\documentclass[aps,prd,nofootinbib]{revtex4} 
\usepackage[utf8]{inputenc} 
\usepackage{lmodern}  
\usepackage{amsmath} 
\usepackage{amsfonts}  
\usepackage{amssymb}  
\usepackage{tabularx} 
\usepackage{graphicx} 
\usepackage{todonotes} 
\usepackage[english]{babel} 
\usepackage[caption=false]{subfig} 
\usepackage[pdfauthor="Lisa Glaser",pdftitle="The Ising model coupled to 2d orders"]{hyperref}

\graphicspath{ {figures/}}
\newcommand{\Sa}{\mathcal{S}}
\newcommand{\jc}[1]{j^{#1}_{c}}
\newcommand{\av}[1]{\langle #1 \rangle}
\newcommand{\C}{\mathcal{C}}

\begin{document}
\title{The Ising model coupled to 2d orders}
\date{\today}
\author{Lisa Glaser}
\affiliation{Radboud University, Nijmegen}

\begin{abstract}
In this article we make first steps in coupling matter to causal set theory in the path integral.
We explore the case of the Ising model coupled to the $2$d discrete Einstein Hilbert action, restricted to the $2$d orders.
We probe the phase diagram in terms of the Wick rotation parameter $\beta$ and the Ising coupling $j$ and find that the matter and the causal sets together give rise to an interesting phase structure.
The couplings give rise to five different phases.
The causal sets take on random or crystalline characteristics as described in~\cite{surya_evidence_2012} and the Ising model can be correlated or uncorrelated on the random orders and correlated, uncorrelated or anti-correlated on the crystalline orders.
We find that at least one new phase transition arises, in which the Ising spins push the causal set into the crystalline phase.
\end{abstract}
\maketitle

Causal set theory is a theory of discrete quantum gravity, in which space-time is encoded as a discrete partial order, called a causal set because the partial order relations correspond to causal relations~\cite{bombelli_space-time_1987} (see  \cite{surya_directions_2011} for a relatively recent review).
The aim of the theory is to quantise gravity by taking the path integral over a suitable class of causal sets.

One approach to quantise causal set theory is to sum over the class of all causal sets.
However, it is unclear if this approach will work, since the class of all partial orders is dominated by the Kleitmann Rothschild (KR) orders~\cite{kleitman_phase_1979,kleitman_asymptotic_1975}.
These are a particular class of three layer orders and it has been proven that in the $N\to \infty$ limit any random partial order is almost certainly of KR type.
In practice, simulations exploring the state space over all causal sets give rise to the expectation that the KR orders will start to dominate the path integral for $N> 100$~\cite{henson_onset_2015}.
This entropic dominance of the KR orders could possibly be broken by weighting causal sets with an action that strongly suppresses this type of orders.
It is not clear whether the current causal set action of choice, the Benincasa-Dowker (BD) action, which was introduced in ~\cite{benincasa_scalar_2010} for $2$ and $4$ dimensions and later generalised to any dimension~\cite{dowker_causal_2013,glaser_closed_2014}, can achieve this.
However, recent work shows that it does suppress another class of pathological orders, the so called bilayer orders~\cite{loomis_suppression_2017}.
In addition, summing over all causal sets does not fix the dimension of the spaces summed over, this leads to the question which dimension to chose for the BD-action.
In an ideal world, the choice of action might influence the dimension of space-time, so using a $2$d action could create a path integral dominated by $2$d causal sets; however, reality is rarely ideal.

The $2$d orders are a class of causal sets that has proven very useful for study through computer simulations.
While the ultimate goal will be to simulate a path integral over a wider class of causal sets, the $2$d orders are an interesting model that gives us a fixed dimension and comes with embedding coordinates, which are useful to visualise the system.
Another interesting property of the $2$d orders is that they entropically favour states that correspond to $2$d flat space-times~\cite{brightwell_2d_2008}.
In the first computational study of the $2$d orders two phases, one random, entropy dominated and one crystalline, action dominated, were found~\cite{surya_evidence_2012}.
The transition between these states was later found to be a first order phase transition~\cite{glaser_finite_2017}.
In this work we also found that the location of the phase transition scales inverse linearly in the system size and inverse quadratically in the non-locality parameter $\epsilon$.
We also found that the action exhibits different scaling behaviour before and after the phase transition, with the scaling before the phase transition pointing at the possibility of a dynamically generated cosmological constant.

An important point in any theory of quantum gravity is how to couple it to matter.
There are two different approaches to study matter on a causal set.
The first is to use the d'Alembertian operator~\cite{sorkin_does_2007}.
This approach has led to the Benincasa-Dowker action~\cite{benincasa_scalar_2010}, and been generalised to work on causal sets in any dimension~\cite{dowker_causal_2013,glaser_closed_2014}.
It can be generalised even further to include additional layers~\cite{aslanbeigi_generalized_2014}.
Studying the continuum limit of scalar fields coupled to a fixed causal set like this gives rise to interesting non-local phenomenology~\cite{belenchia_nonlocal_2015}, which could be explored using optical experiments~\cite{belenchia_tests_2016} and has even been proposed as an explanation for dark matter~\cite{saravani_off-shell_2017}.

The other approach for coupling a scalar field to a causal set is to use the Feynman propagator~\cite{johnston_particle_2008}.
This method has been described for massless fields on causal sets in $2$ and $4$ dimensions in~\cite{johnston_particle_2008} and generalised to massive fields and also to $3$ dimensions in~\cite{ahmed_scalar_2017}.
The continuum limit of the Feynman propagator on causal sets leads to a unique choice of vacuum for quantum field theory, the Sorkin-Johnston vacuum~\cite{afshordi_distinguished_2012}.

Using either the d'Alembertian or the Feynman propagator it is then possible to couple scalar fields to dynamical causal sets in the path integral.
Another way of coupling matter is to chose a statistical physics matter model to couple to the causal set elements.
In this work we add spin degrees of freedom to the elements of the causal set and thous couple the causal set to an Ising model.
The Ising spins are only coupled to their nearest neighbours, which in a causal set are the linked elements, physically this corresponds to an Ising model coupled along the light-cones.

The ordinary Ising model corresponds, in the continuum limit, to a conformal field theory with charge $\frac{1}{2}$.
This model has been studied in many different guises and approaches to quantum gravity.
In~\cite{kazakov_exact_1986} they find that they can solve the Ising model coupled to arbitrary planar random lattices with spherical topology, using a two matrix model.
In this model of euclidean quantum gravity they find a continuous phase transition at which the Ising spins develop long range couplings.
So the Ising model coupled to euclidean quantum gravity leads to a strong coupling in which the geometry changes drastically under the influence of the Ising spins.

Coupling the Ising model to causal fluctuating geometries has been studied in the context of CDT~\cite{ambjorn_new_1999}.
They use the high temperature expansion and Monte Carlo simulations to study the finite size scaling at the single phase transition of the Ising model coupled to $2$d CDT.
They find that, contrary to the results for the euclidean model, the critical exponents in this model are not different from those coupled to flat space and the critical exponents of the geometry remain the same as without the spins.

The Ising model coupled to causal set theory has one important difference from the Ising model on the random two matrix model or in CDT, which is that in our model all nearest neighbours are time-like, while in the random matrix description and the CDT model all edges are space-like.
In~\cite{ambjorn_new_1999} they postulate that the difference between the Ising model coupled to CDT and the Ising model coupled to dynamical triangulations arises through the restriction on fluctuations of geometry that arise from the causality conditions.
While the $2$d orders are Lorentzian by construction they still allow for wild fluctuations, including non-manifold-like causal sets, such as the crystal orders, which makes us hope to find a strong interaction between the Ising model and the causal set.


While the ordinary Ising model can be solved analytically on a lattice in two dimensions, results in higher dimensions depend on computer simulations and different approximations.
One approximation that works particularly for the Ising model in dimensions higher than $4$ is mean field theory.
In mean field theory the interaction between the spins is simplified to just the behaviour of a single spin in a mean, background field generated through its nearest neighbours.
This approximation works particularly well in higher dimensions, since regular lattices there have higher valency nodes.
A higher valency means more nearest neighbours, hence the influence of all neighbours is better approximated through an average and thus mean field theory works better.
The causal set graphs that describe continuous manifolds are also of very high valency~\cite{bombelli_discreteness_2009}, which is a reason to expect mean field theory to be useful in our system.
We will test this hypothesis and find that, despite additional approximations that are due to the randomness of causal sets, mean field theory can give us a good estimate of some of the critical points of our system.

To study this model we will proceed as follows.
First, in section 1, we will give a short introduction into causal set theory and the observables we plan to use, then, in section 2, we will will study the Ising model on fixed $2$d orders taken from either the random phase or the crystalline phase.
To do so we will explore first order observables, variances and the fourth order cumulant for behaviour characteristic of phase transitions.
After thus establishing a baseline for the behaviour of the model, in section 3 we will explore the Ising model on dynamical causal sets in the same manner, trying to sketch the phase diagram of the coupled system and to characterise the phases arising.
In the 4th section we will explore how well mean field theory can explore the location of those phase transitions that are characterised through a jump in the magnetisation.
Section 5 is a short conclusion and outlook for future work.

\section{Introduction}
\subsection{Some details about causal sets}
Causal set theory describes space time as a partially ordered set $\C$.
If two elements $x,y \in C$ are related we write this as $x \prec y$. In addition a causal set is
\begin{itemize}
\item {\bf transitive} for all $x,y,z \in \C$, if $x \prec y$ and $y \prec z$ then $x \prec z$
\item {\bf antisymmetric} if $x,y \in \C$ then $x \prec y \prec x$ is not possible
\item {\bf locally finite} for all $x,y \in \C$  $| I(x,y)| < \infty $
\end{itemize}
If there is no intervening element $z$ s.th. $x \prec z \prec y$ we say $x$ is linked to $y$ which we will here denote as $x \prec_l y$.
A set of elements $\{x_0, ... , x_n\}$ so that $x_0 \prec x_1 \prec ... \prec x_n$ is called a chain, and if all relations are links $x_0 \prec_l x_1 \prec_l ... \prec_l x_n$ it is called a path.
A causal interval are all elements $z$ such that $x\prec z \prec y$, the number of causal intervals of size $i$ is denoted as $N_i$ and the abundance of intervals can be used to measure the manifold likeness of a causal set and its dimension~\cite{glaser_towards_2013}.
A convenient way to encode the causal set is through the matrices
\begin{align}
  A_{ik}&= \delta_{i\prec k} & L_{ik}&= \delta_{i\prec_l k} \;,
\end{align}
where $A_{ik}$ is called the adjacency, or relation matrix while $L_{ik}$ is the link matrix.

The simplest action for a $2$d causal set is
\begin{equation*}
\Sa_{BD} = 4 \left(N- 2  N_0 +  4 N_1 - 2 N_2 \right)\;,
\end{equation*}
however, this fluctuates strongly~\cite{sorkin_does_2007}, so to suppress fluctuations Sorkin introduced an intermediate length scale $\frac{l_{Pl}^2}{l^2}=\epsilon$
\begin{align*}
\Sa_{BD}(\epsilon)&= 4 \epsilon \left( N- \sum_{n} N_n f(n,\epsilon) \right) \\
f(n,\epsilon)&= (1-\epsilon)^i \left(1- \frac{2 n \epsilon}{1-\epsilon}+ \frac{n (n-1)\epsilon^2}{2 (1-\epsilon)^2} \right)\;.
\end{align*}
In the continuum limit this action recovers the Einstein-Hilbert action of general relativity~\cite{benincasa_scalar_2010}.
The action for the Ising model on the causal set can be written using the spin vector $s_i$ with entries $\pm 1$ denoting the spin of the $i$-th element
\begin{align}
  \Sa_{I}(j)= j \sum_{i,k} s_i s_k L_{i k}
\end{align}
The $2$d orders $\Omega_{2d}$ can be defined as the union of two total orders.
For a given set $S = (1, . . . , N )$ let $U = (u_1 , u_2 , . . . ,u_N )$ and $ V = (v_1 , v_2 , . . ., v_N )$, such that $u_i,v_i\in S$, with $u_i=u_k \Rightarrow i=k$, and $v_i=v_k\Rightarrow i=k$.
Then $U$ and $V$ have total orders induced by the integers.
We can create a $2$d order $C \equiv U \cap V$, in which $e_i\equiv (u_i,v_i)\in C $ with the induced partial order $e_i \prec e_k$ in $C$ iff $u_i < u_k$ and $v_i < v_k$~\cite{el-zahar_asymptotic_1988,winkler_random_1990,brightwell_2d_2008}.

Using the $2$d orders we can define the path sum over this restricted class as
\begin{align}
  \mathcal{Z}_{\Omega_{2d}} = \sum_{\mathcal{C} \in \Omega_{2d}} e^{- \frac{i}{\hbar} \Sa_{BD}(\epsilon)} \;.
\end{align}
To explore the system on the computer requires a Wick rotation.
Since the causal sets themselves can not be Wick rotated, we introduce a Wick rotation parameter $\beta$ and analytically continue it from $\beta \to -i \beta$, leading to a path sum
\begin{align}
  \mathcal{Z}_{\Omega_{2d}} = \sum_{\mathcal{C} \in \Omega_{2d}} e^{- \frac{\beta}{\hbar} \Sa_{BD}(\epsilon)} \;.
\end{align}
The parameter $\beta$ acts like an inverse temperature, and hence a free parameter, when we treat the system as purely thermodynamic.
In higher dimensions, where the prefactor of the action is dimensionful, the parameter $\beta$ would also be dimensionful, in this case the value of $\beta_c$ might tell us something about the Planck length of the system.
However this is outside the scope of the simple $2$d model explored here, we can hence assume $\hbar=1$ for the remainder of this article.

\begin{figure}
  \includegraphics[width=0.49\textwidth]{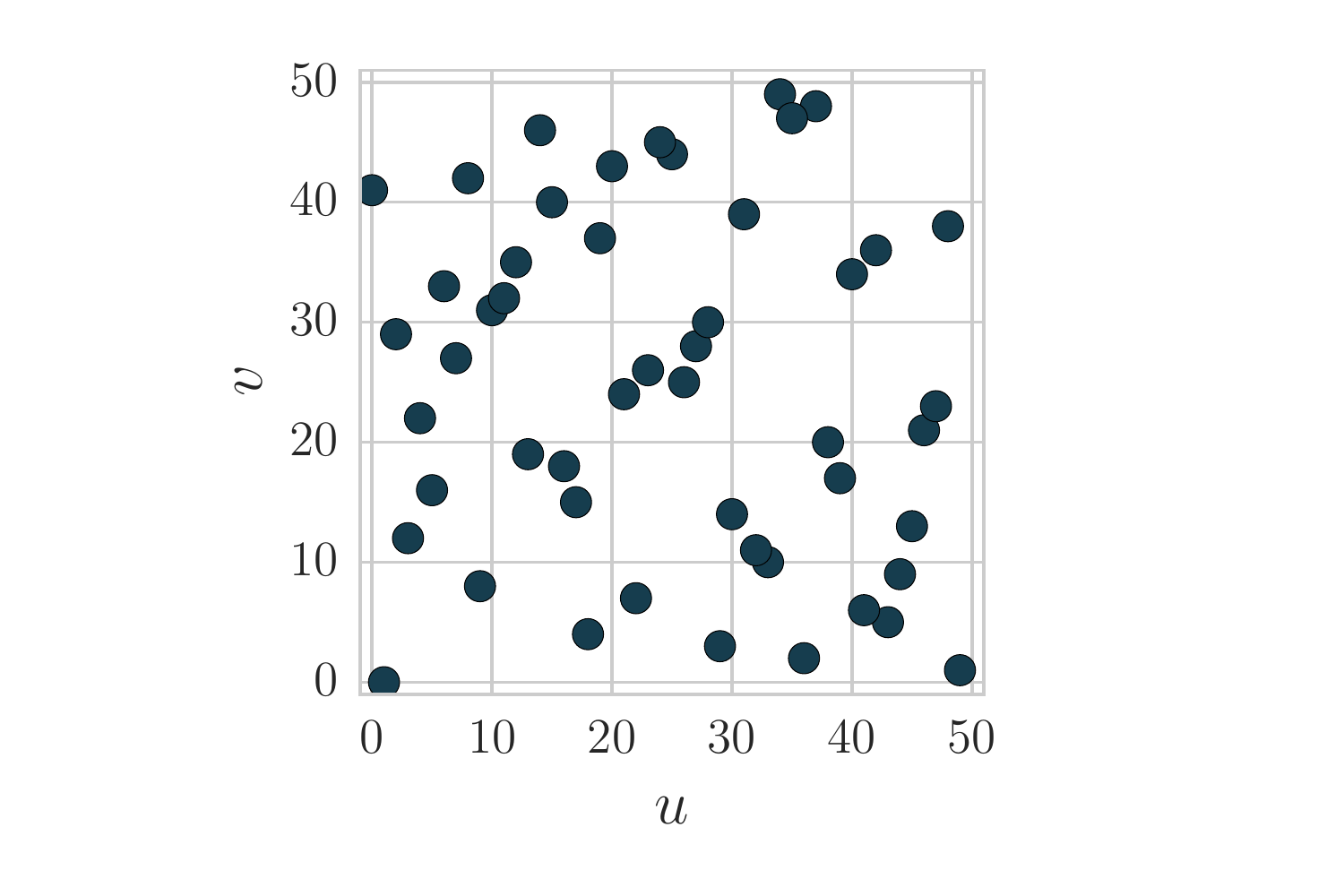}
  \includegraphics[width=0.49\textwidth]{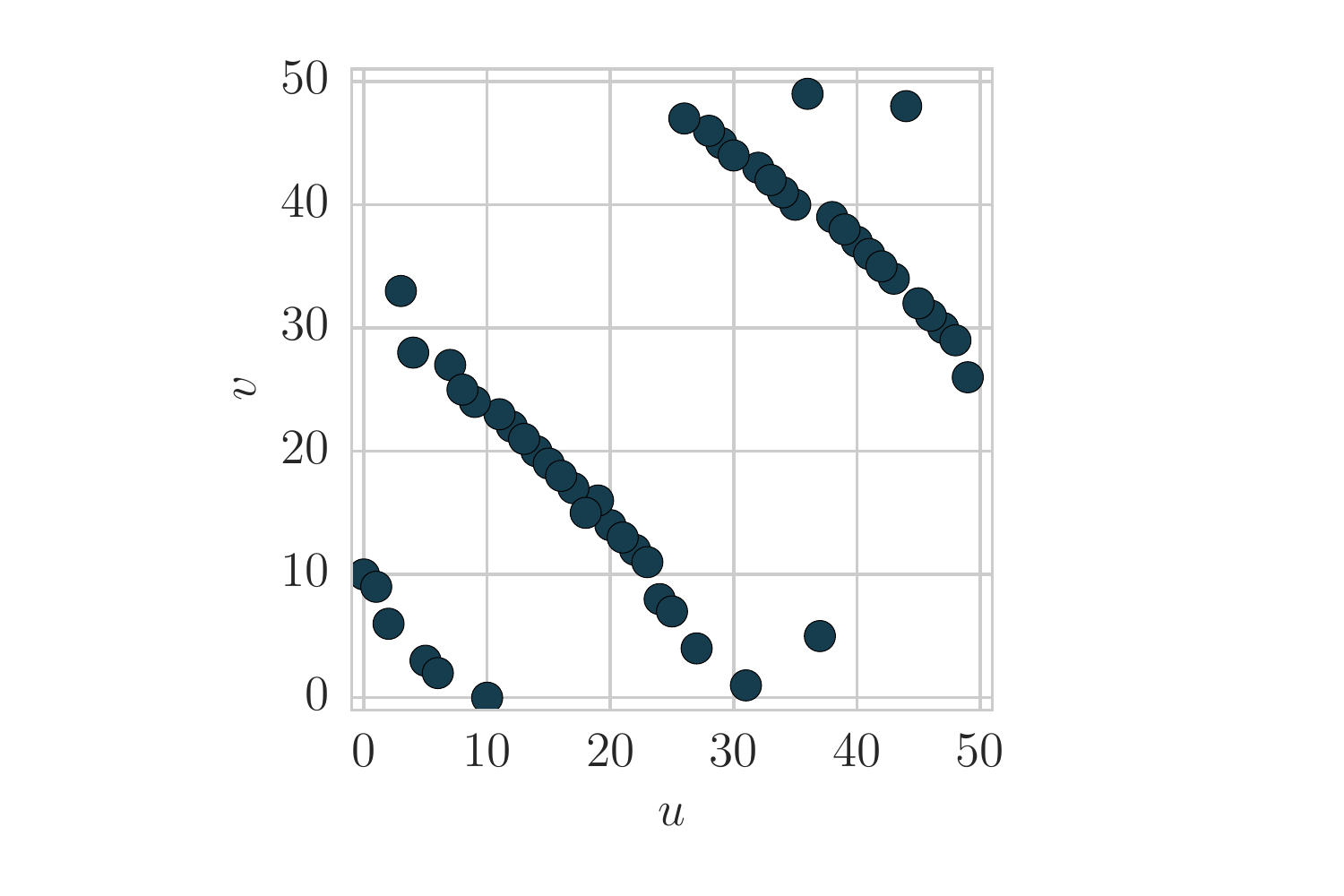}
  \caption{Examples of a typical random $2$d order and of a crystal $2$d order.}
 \label{fig:R+Ccauset}
\end{figure}

This system was first explored in~\cite{surya_evidence_2012}, where Surya found a phase transition at $\beta_c$ that changes with the non-locality parameter $\epsilon$.
The transition was between disordered, random causal sets and crystalline causal sets which show a layered structure, illustrated in Figure~\ref{fig:R+Ccauset}.
In \cite{glaser_finite_2017}, we explored this phase transition in more detail, and established that it is of first order.
We also found that for a given $N,\epsilon$ value the phase transition $\beta_c$ is given by
\begin{align}
	\beta_c(N,\epsilon) \approx \frac{1.66}{N\epsilon^2} + \mathcal{O}(\frac{1}{N^2})\;.
\end{align}
These results build the backdrop upon which to explore the $2$d orders coupled to the Ising model.

\subsection{Observables we will examine}
In \cite{surya_evidence_2012} Surya used many different properties of the causal sets, e.g. their height, the Interval abundance~\cite{glaser_towards_2013}, their ordering fraction and their action to examine the structure of the causal sets on either side of the phase transition.
In \cite{glaser_finite_2017} on the other hand, we focused on the action and its specific heat $C$ to locate the phase transition.
To observe how the causal set coupled to the Ising model behaves we need observables that are sensitive to the states of the Ising model.
The most obvious such observable is the overall magnetisation of the state
\begin{align}
  M= \frac{1}{N} \sum_i s_i \;.
\end{align}
In practice the magnetisation in a totally ordered state can be $+1$ or $-1$ and the system can jump between these during a Monte Carlo run.
To avoid this, which could average the magnetisation to zero for long runs even if the system spend most of its time in a totally ordered state we use a modified version, in which we add an absolute value to calculate the average over a Monte Carlo chain $c$ of length $|c|$ containing causal sets $\C$,
\begin{align}
  \av{M}= \frac{1}{|c|} \sum_{C \in c}  \left| \frac{1}{N} \sum_{i\in \C} s_i \right|\;.
\end{align}
The magnetisation is the order parameter for the Ising model when it jumps from a magnetised, ordered phase to a random, non magnetised phase.
Another frequent observable in the Ising model is the spin spin correlator
\begin{align}
  S_{ik}= s_i s_k L_{i k}\;.
\end{align}
We are here using element wise notation for matrix and vector elements and will write summations explicitly when they occur.
This can easily be measured on a fixed causal set, however if the partial order relations change the matrix $S_{ik}$ becomes meaningless, a more sensible observable is the sum of this over the causal set
\begin{align}
  S= \sum_{i,k} s_i s_k L_{i k}
\end{align}
this is the action divided by $j$, so we do not need to measure it independently and can instead explore the action $\Sa_I$ alone.
The above spin spin correlator depends on links, however to explore if the model also exhibits longer ranged interaction we can also measure the correlator for elements that are related,
\begin{align}
  R_{ik}= s_i s_k \delta_{i \prec k} \\
  R= \sum_{i,k} s_i s_k A_{i k} \;.
\end{align}
In principle we can also restrict the correlator to any length of path between the elements, this requires some more computational effort, and will thus only be computed for areas of the phase diagram we want to study in detail
\begin{align}
  C^{n}_{ik} = s_i s_k L^{n}_{ik} \;.
\end{align}
Again, the correlator between any pair $C^n_{ik}$ is meaningless if the causal set is changing, so we calculate the average $\av{C^n}=\sum_{i,k} s_i s_k L^{n}_{ik}$.
We can calculate this for all $n$ to see how the correlation falls off.

We will thus measure $\av{M},\av{R}$ in addition to $\av{\Sa_I},\av{\Sa_{BD}}$ and supplement this with the path correlators $\av{C^n}$ for select points, with these quantities we can hope to understand the behaviour of our system.

Throughout the article we will calculate the average, the variance and the $4$th order cumulant of these observables.
The statistical errors on these quantities are estimated using bootstrapping, where we have taken into account the autocorrelation time of the data.

\section{Ising model on fixed causal sets}
Before jumping into the analysis of the entire system we will explore how the Ising model behaves when coupled to fixed causal sets.
This allows us to explore the behaviour of our observables, and gives us a baseline for the behaviour of the system.
The partition function simulated for this is
\begin{align}
\mathcal{Z}_{I} = \sum_{I \text{on }\mathcal{C}} e^{- \beta \Sa_I(j) }
\end{align}
with $I$ the spin states on a fixed causal set $\C$.
In the simulations on a fixed causal set we attempt $20\,000 N$ spin flips for each data point.

\subsection{A fixed random causal set}
We will first analyse the Ising model on a fixed random causal set.
When the lattice is fixed we effectively only have one coupling constant, either $\beta$ or $j$, since they multiply both terms.
We have thus decided to only vary $j$, keeping $\beta$ fixed at $0.1$, for no reason at all.

We have run the analysis for $20$ random causal sets of $N=50$ elements generated at $\beta=0$ with the algorithm described in~\cite{surya_evidence_2012}.

All sets explored show $2$ phase transitions, one at negative $j,\jc{-}$ and one at positive $j, \jc{+}$.
The transition at $\jc{-}$ is accompanied with a jump in the absolute magnetisation and hence is a transition between an ordered phase in which all spins point in the same direction and a disordered phase with order parameter $M$.
This transition is also clearly visible in peaks of both the action $\Sa_{I}$ and the magnetisation $M$.
The transition at $\jc{+}$ is only visible as a soft peak in $Var(\Sa_{I})$ and we have not been able to identify a good order parameter for it.

We found the phase transition location to vary with the causal set, see Table \ref{tab:PTrandom}.
\begin{figure}
\includegraphics[width=\textwidth]{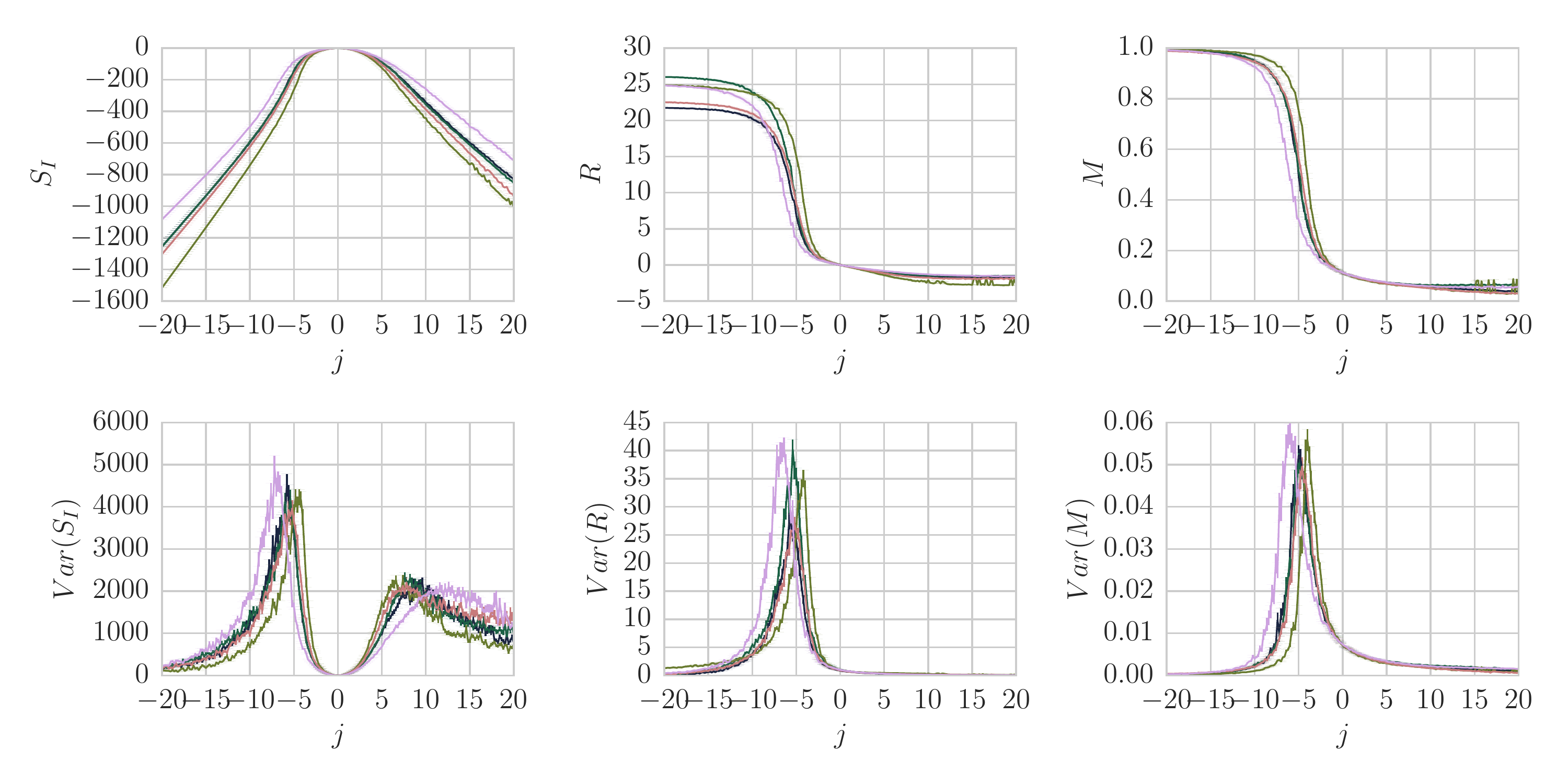}
  \caption{Comparing the observables on $5$ of the $20$ different random causal sets we see that the qualitative features remain the same.}
  \label{fig:compRCausets}
\end{figure}
 \begin{table}
  \caption{Phase transition points for the Ising model on different random causal sets}
  \label{tab:PTrandom}
  \begin{tabular}{l r r r r r r r r r r} 
    \toprule
    & 1&2&3&4&5&6&7&8&9&10\\
    \colrule
    $\jc{-}$&  $ -5.2 \pm 0.4 $ & $ -5.0 \pm 0.4 $ &  $-6.4 \pm 0.6 $ & $ -4.2 \pm 0.8 $ &  $ -5.6 \pm 0.6 $ & $ -5.2 \pm 0.6 $ & $ -6.0 \pm 0.6 $
    & $ -5.4 \pm 0.6 $ &
     $ -5.8 \pm 0.6 $& $ -5.6 \pm 0.8 $  \\
    $\jc{+}$& $ 10.4 \pm 1.4 $& $ 7.4 \pm 1.2 $ & $ 8.0 \pm 1.6 $ & $ 11.6 \pm 2.0 $& $ 7.0 \pm 1.0 $ & $ 8.4 \pm 1.6 $ & $ 9.8 \pm 2.6 $ &
     $ 9.0 \pm 1.2 $ & $ 10.6 \pm 1.2 $ &
    $ 7.2 \pm 1.2 $ \\[10pt]
    \toprule
     &11&12&13&14&15&16&17&18&19&20\\
     \colrule
    $\jc{-}$& $ -5.8 \pm 0.6 $ & $ -5.8 \pm 0.6 $ & $ -5.4 \pm 0.4 $ & $ -6.2 \pm 0.3 $ &  $ -5.4 \pm 0.4 $ &  $ -5.4 \pm 0.4 $ & $ -5.0 \pm 0.4 $ &
     $ -4.8 \pm 0.6 $ & $ -5.6 \pm 0.4 $ & $ -5.2 \pm 0.4 $     \\
    $\jc{+}$& $ 9.8 \pm 1.8 $ &$ 8.6 \pm 1.4 $& $ 8.8 \pm 1.6 $& $ 10.2 \pm 1.2 $&$ 8.2 \pm 1.6 $& $ 8.8 \pm 1.4 $& $ 8.6 \pm 1.2 $&
    $ 7.4 \pm 1.0 $& $ 8.8 \pm 1.2 $& $ 9.6 \pm 1.4 $    \\ \colrule
  \end{tabular}
\end{table}
In Figure~\ref{fig:compRCausets} we show the observables for $5$ of the causal sets, and we can see that the behaviour of the system on the different random causal sets is qualitatively the same, despite the changing phase transition temperatures.
In addition to changing phase transition location, the maximum and minimum value of the spin correlation and magnetisation show some dependence on the underlying causal set, however the existence of $2$ phase transitions at $\jc{\pm}$ is clear for all $20$ cases.

\begin{figure}
\includegraphics[width=\textwidth]{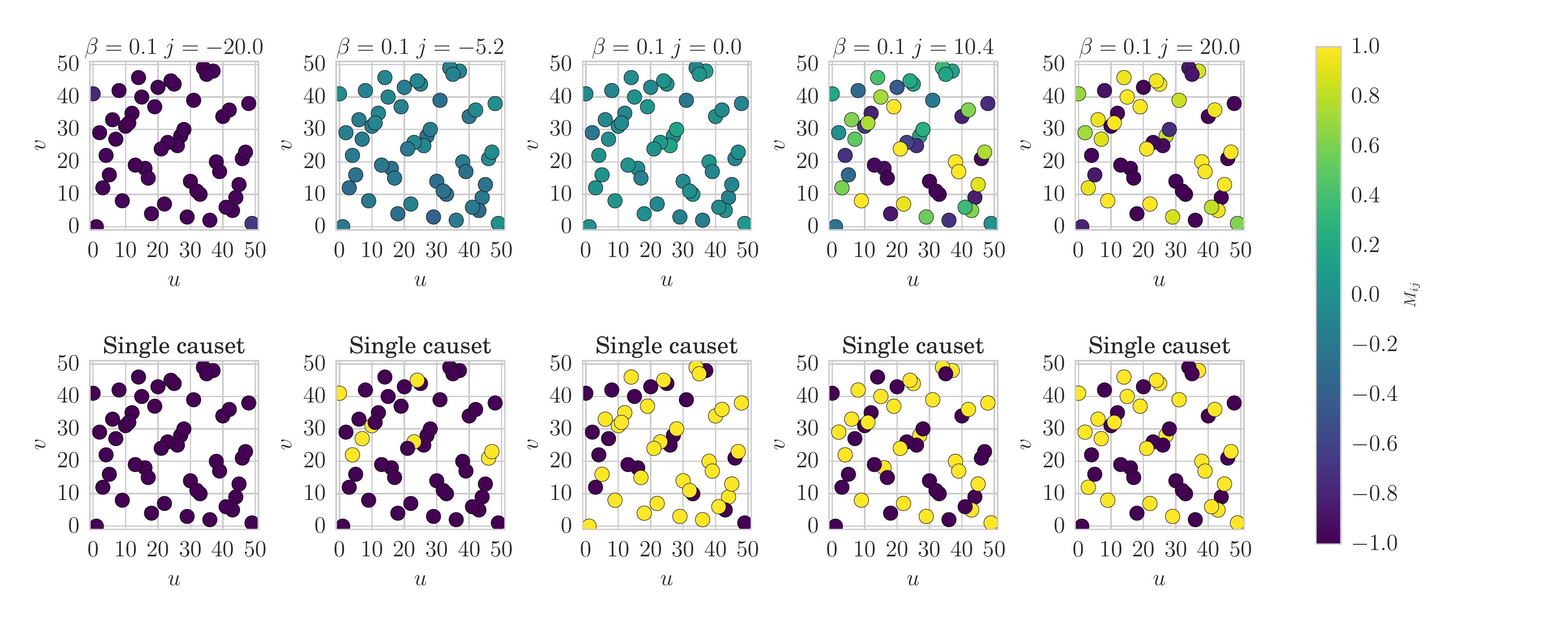}
  \caption{The causal set counted as the first random causal set in our collection. In the upper plot the elements are coloured by their average magnetisation $M_{ik}$ while the lower plots show example causal sets for the given parameter values.}
  \label{fig:Mij_random}
\end{figure}

One qualitative way to understand the behaviour of the Ising model in the different phases on the fixed causal set is to plot the causal set and colour the elements by their average magnetisation $M_{ik}$.
In this definition of the magnetisation we are not using the absolute value, since it would mask the anticorrelated phase.
In Figure \ref{fig:Mij_random} we do so for the causal set with index $5$ in our collection of random causal sets, the data is averaged over $200$ states of the causal sets.
We plot the causal set for $j=-20.0,-5.2,0.0,10.4,20.0$ which are the minimal and maximal value, as well as the two phase transition candidates and the neutral element.
The colour makes it clear that at negative $j$ all spins are very highly magnetised. There is some thermal movement left, but the spins are overwhelmingly aligned.

At the phase transition from the ordered to the disordered phase the spins appear to, however looking at an example causal set from the given run we see that the spins are mostly aligned.
The overall magnetisation of $0$ arises because the spins fluctuate wildly, taking on mostly aligned states in either direction or completely random states.

\begin{figure}
  \centering
  \includegraphics[width=0.6\textwidth]{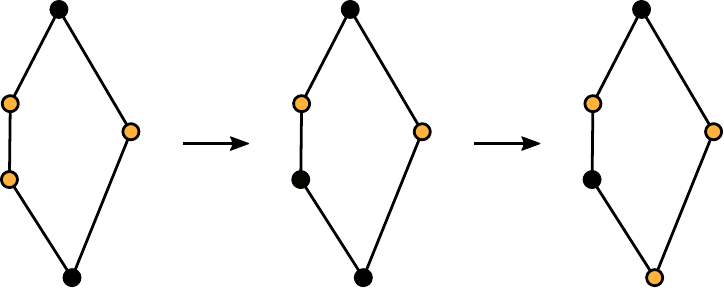}
  \caption{If $j>0$ the transition between consecutive states in this diagram is energy neutral.}
  \label{fig:randomAntiExample}
\end{figure}

At the neutral point the spins average to $0$ magnetisation, an example state with these parameters shows that the Ising spins are equally distributed between up and down states.
The phase transition at positive $j$ and the state at large positive $j$ show an interesting behaviour, some elements of the causal set have a magnetisation of $-1$ while others have a magnetisation of $1$, which indicates  that these spins are fixing each other in anticorrelated positions.
A single given state for these parameter values can not be distinguished from the $j=0.0$ state.
Interestingly in the random $2$d orders a consistent optimally anti correlated state might not exist, instead there will be several states that are of same energy and can thus be switched between without penalty.
A simple example of such a circle is shown in Figure \ref{fig:randomAntiExample}, all changes shown there are energy neutral.
In larger causal sets, cycles like this will also influence the state of the entire causal set, and there will be overlapping circles of constant energy.
This then muddies the phase transition at positive $j$ for the random causal sets and makes it very causal set dependent.
This is likely part of the reason why we have not been able to identify a good order parameter for the uncorrelated to anti correlated phase transition on random orders.

The next step after establishing that the phase transition seems independent of the representative of the class of causal sets is to establish the order of the phase transition.
We do this by exploring the fourth order cumulant, often called Binder cumulant, of the phase transition
\begin{align}\label{eq:binder}
  B_{\mathcal{O}}= 1 - \frac{1}{3}\frac{\av{\left(\mathcal{O}-\av{\mathcal{O}}\right)^4}}{\av{\left(\mathcal{O}-\av{\mathcal{O}}\right)^2}^2}\;.
\end{align}
The forth order cumulant is useful to detect phase transitions, and to understand the order of a phase transition.
It was first introduced in~\cite{binder_finite-size_1984}, and a very good explanation of the properties we require is given in~\cite{tsai_fourth-order_1998}.
We calculate the Binder cumulant for the magnetisation and the energy and hope to determine the order of the phase transition from there.
In the Binder cumulant of the magnetisation all terms $\av{M^n}$ with $n$ odd are $0$ in an ideal system.
However, due to the long autocorrelation time of the system in the ordered state we can not observe this.
A flip of all spins from $+M$ to $-M$ will not occur with sufficiently frequency to lead to $\av{M}=0$, instead we decide to set these terms to $0$ by hand and use the definition
\begin{align}
B_{M}= 1 - \frac{1}{3}\frac{\av{M^4}}{\av{M^2}^2}\;,
\end{align}
for the magnetisation, while still using equation~\eqref{eq:binder} for all other observables.
Using this definition we expect the Binder coefficient to be $2/3$ in the ordered phase and $0$ in the disordered phase.
For a second order phase transition, with order parameter $M$, the cumulant will smoothly interpolate between the two values with the interpolation region getting smaller for larger $N$, for a first order transition on the other hand the jump should be sharp for any size of system $N$ and the value of the Binder cumulant should dip below $0$ at the phase transition.

The forth order cumulant of the Ising action should be $0$ away from the phase transition.
Using a simple Gaussian approximation it should not change at the phase transition either, however in~\cite{tsai_fourth-order_1998} the plots for the second order phase transition show a characteristic up down behaviour, in which the cumulant peaks to an above $0$ value on the ordered side of the phase transition and dips to a below $0$ value on the disordered side.
These dips get sharper for larger $N$.
For a first order transition the forth order cumulant of the Ising action should only dip below $0$ at the phase transition and not show any further fluctuations.

\begin{figure}
\subfloat[][$B_M$]{\includegraphics[width=0.5\textwidth]{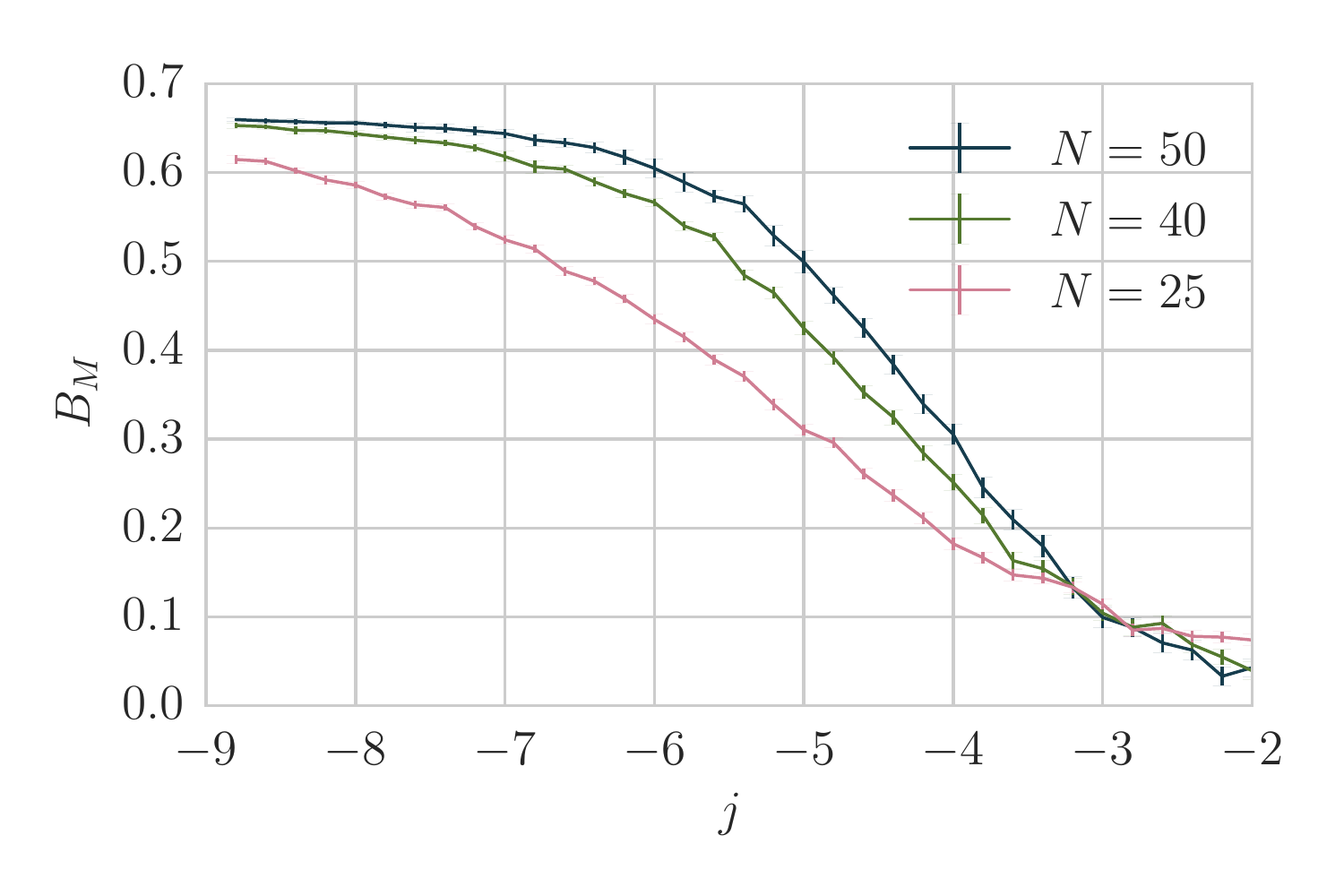}}
\subfloat[][$B_{\Sa_I}$]{\includegraphics[width=0.5\textwidth]{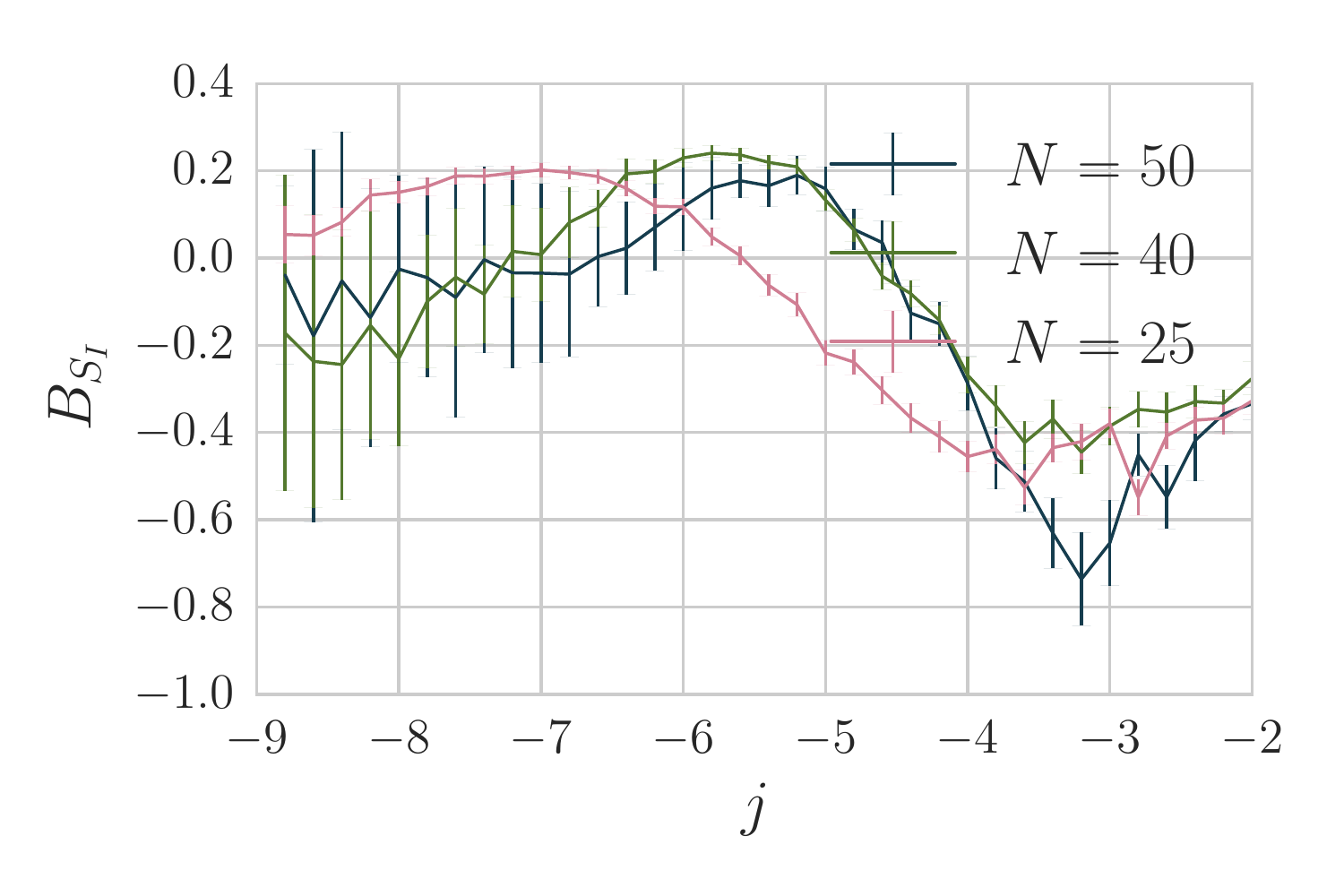}}
  \caption{Binder cumulant for the Ising model averaged over fixed random $2$d orders around $\jc{-}$.}
  \label{fig:binderR}
\end{figure}
To explore the cumulant for our systems we calculate the average of the fourth order cumulant over the $20$ orders we explored above, and over $20$ orders each of $N=25,40$ to have a comparison for how the behaviour changes with size.
We plot the results in Figure~\ref{fig:binderR}, and find them to be just as we described for a continuous phase transition.
This leads us to conclude that the transition at $\jc{-}$ is a continuous phase transition, in agreement with other Ising model results.

We can also use the fourth order cumulant to approximate the phase transitions location for infinite system size.
The cumulant of the magnetisation measured for different system sizes should intersect in a single point, which corresponds to the location of the phase transition for infinite system size.
Looking at Figure~\ref{fig:binderR} we can read off an intersection point at $\jc{-}=-3.2$ which is quite far removed from the transition points of the individual systems.
This is an estimate of the crossing value at infinite system size, and since the sets examined here are very small it is likely that finite size effects lead to a much larger phase transition point for them.

\begin{figure}
  \centering
  \includegraphics[width=0.5\textwidth]{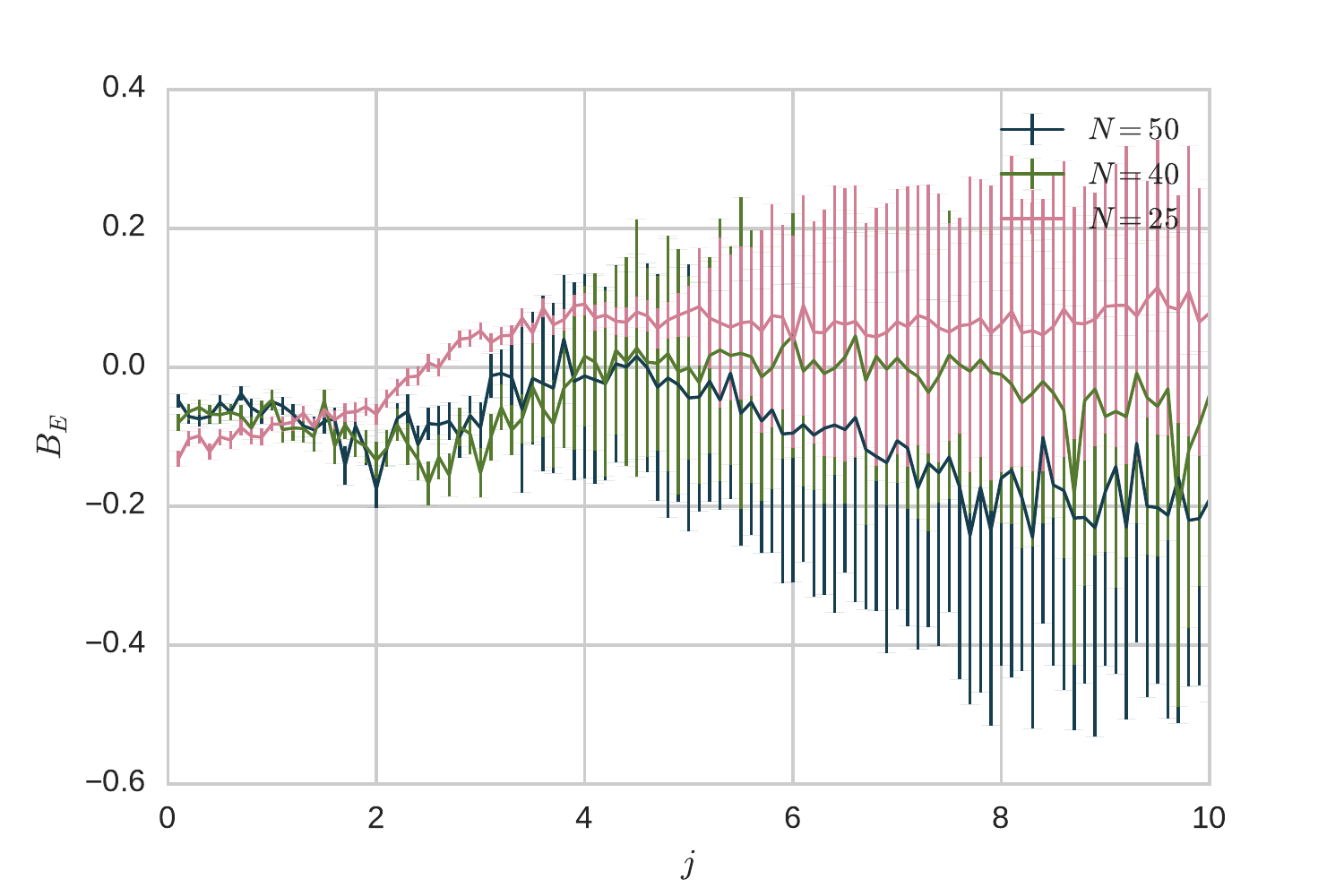}
  \caption{The fourth order cumulant for the transitions $\jc{+}$ on random causal sets.}
  \label{fig:BMj+R}
\end{figure}

Exploring the $\jc{+}$ transition we find that the fourth order cumulant does not help us.
For the magnetisation we never expected it to, since the magnetisation along this transition remains $0$, however even the forth order cumulant of the action is useless.
Looking at Figure~\ref{fig:BMj+R}, we see that it shows very large errorbars at large $j$ which might be due to fluctuations between equivalent energy states.
It is clear that we can not learn anything about this transition using the observables we currently have.

The random $2$d orders are causal sets that are well approximated by flat space, hence one might wonder how the system here compares to the flat spaced $2$d Ising model.
While we have not done any quantitative checks for this, it is clear that there is a qualitative difference, due to the structure of the $2$d orders.
In the random $2$d orders each element has a relatively high valency, and the valency can differ strongly between different elements.
As we will see in section \ref{sec:meanfield} this high valency makes mean field theory an effective tool in studying this model.

\subsection{Ising model on fixed crystal causal set}
Prior study of the $2$d orders found two phases, a random, manifold-like one and a crystalline one.
In the previous section we showed that the Ising model coupled to fixed random $2$d orders has two phase transitions and thus shows a correlated, a disordered and one other, likely anti-correlated phase which we could not characterise well.
When we couple the Ising model to dynamic causal sets we expect that at some temperature the causal sets will transition to the crystalline phase.
Hence to understand the results it will be helpful to also study the Ising model coupled to a fixed crystalline causal set.

\begin{figure}
  \includegraphics[width=\textwidth]{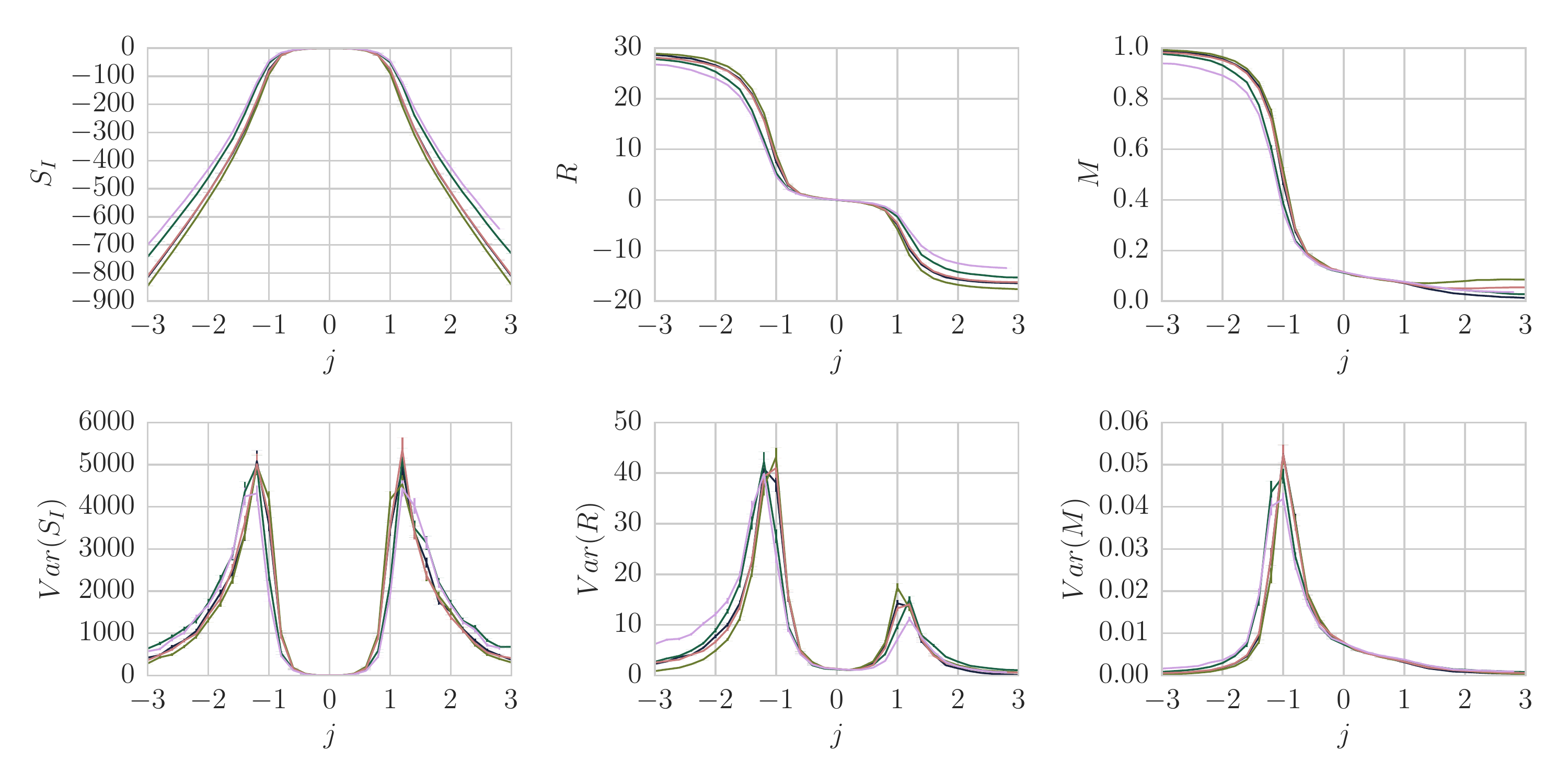}
  \caption{Comparing the observables on $5$ different crystal causal sets we see that the qualitative features remain the same.}
  \label{fig:compCCauset}
\end{figure}

\begin{figure}
\subfloat[][$B_M$]{\includegraphics[width=0.5\textwidth]{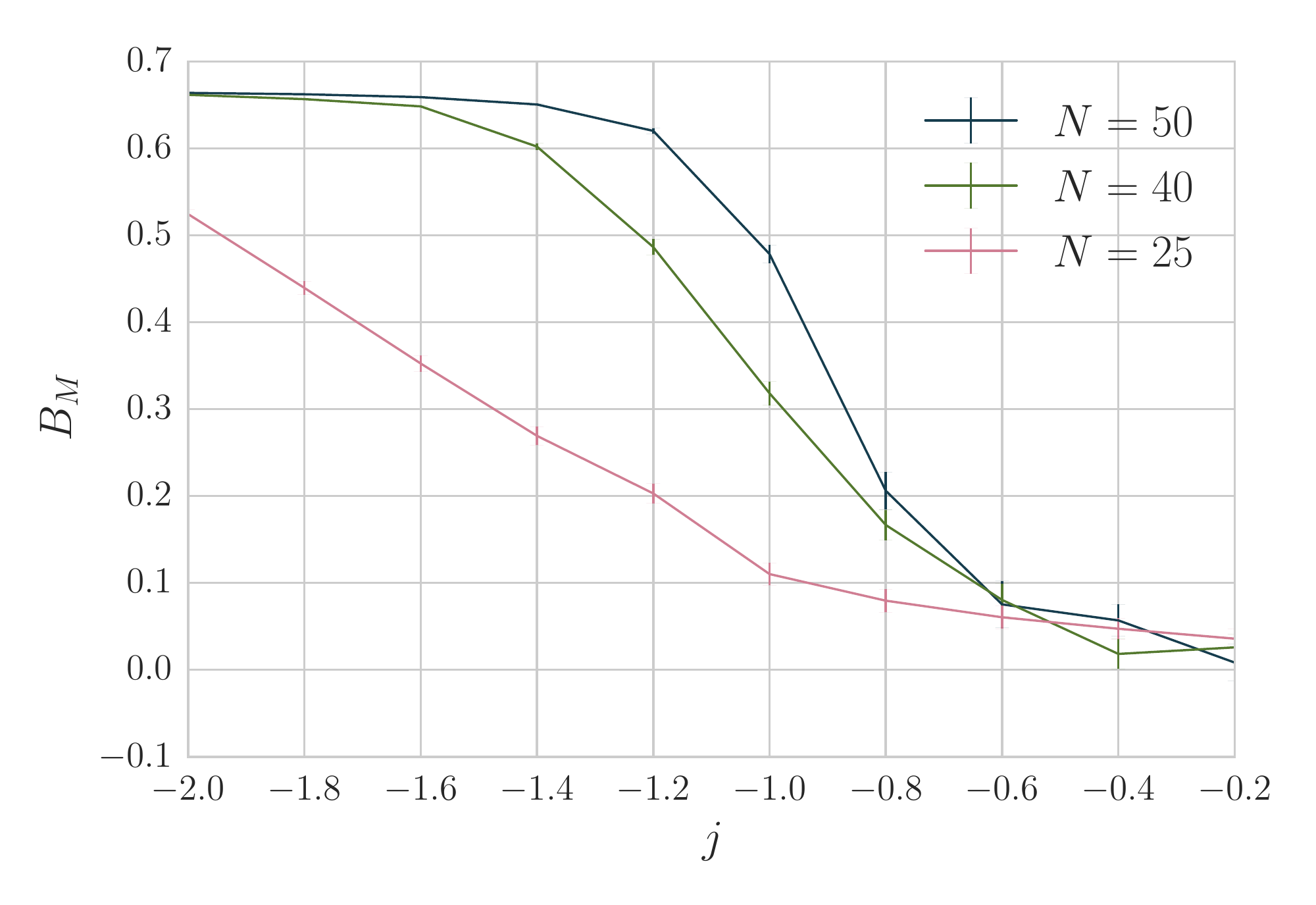}}
\subfloat[][$B_E$]{\includegraphics[width=0.5\textwidth]{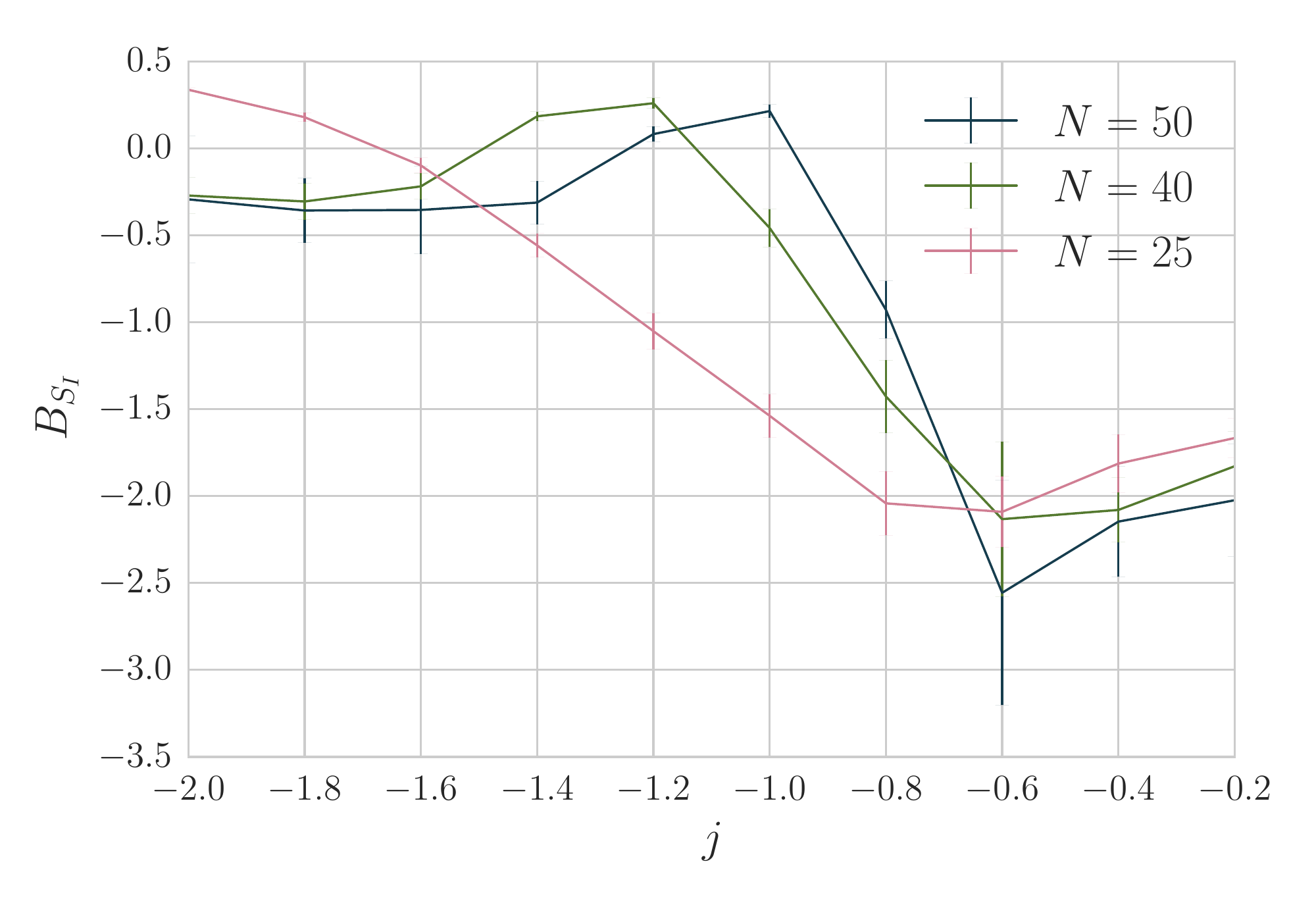}}
  \caption{Binder cumulant for the Ising model averaged over fixed crystalline $2$d orders.}
  \label{fig:BinderC}
\end{figure}

Again we can compare the behaviour on different fixed crystal causal sets for $N=50$, as before we simulated $20$ different causal sets, but only plot $5$ of these to make the plots easier to interpret.
The fixed crystalline causal sets are obtained using the algorithm from~\cite{surya_evidence_2012} at $\epsilon=0.21$ and $\beta=1.5$ which is well within the crystalline phase.
The behaviour is consistent, and we see in Figure~\ref{fig:compCCauset} that in the case of the crystal causal sets the location of the phase transition seems almost independent of the choice of causal set.
This is due to the fact that crystal orders are much more structurally constrained than the random orders,  which leads to more similar behaviour.
The absolute spin correlation is much larger than for the random $2$d orders, because of the higher number of links in the crystal orders.
\begin{figure}
  \includegraphics[width=\textwidth]{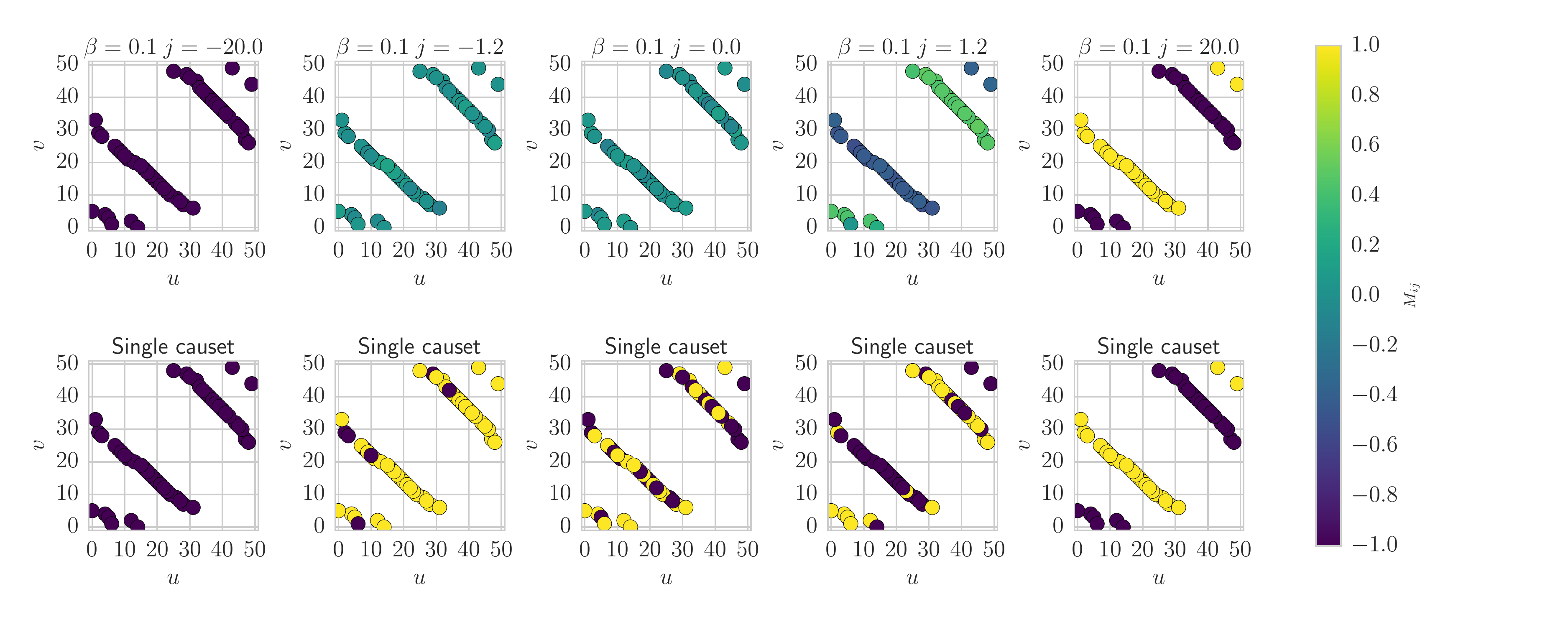}
  \caption{The causal set counted as the first crystal causal set in our collection. In the upper plot the elements are coloured by their average magnetisation $M_{ik}$ while the lower plots show example causal sets for the given parameter values.}
  \label{fig:magC}
\end{figure}
It is much easier to detect the anticorrelated phase transition in the crystal phase, since it leads to opposing orientations on consecutive layers, which we can see in Figure~\ref{fig:magC}, where we again averaged over 200 states.
As for the system on random $2$d orders we can also explore the Binder cumulant for the average over crystalline orders.
We show this in Figure~\ref{fig:BinderC}, the cumulants show the same indicators of a second order transition as for the random causal sets.
The crossing of $B_M$ for different system sizes is not as clearly located as for the random causal sets, it does lead to an estimate of $\jc{-}=-0.1,\dots,-0.3$.

\begin{figure}
  \includegraphics[width=0.5\textwidth]{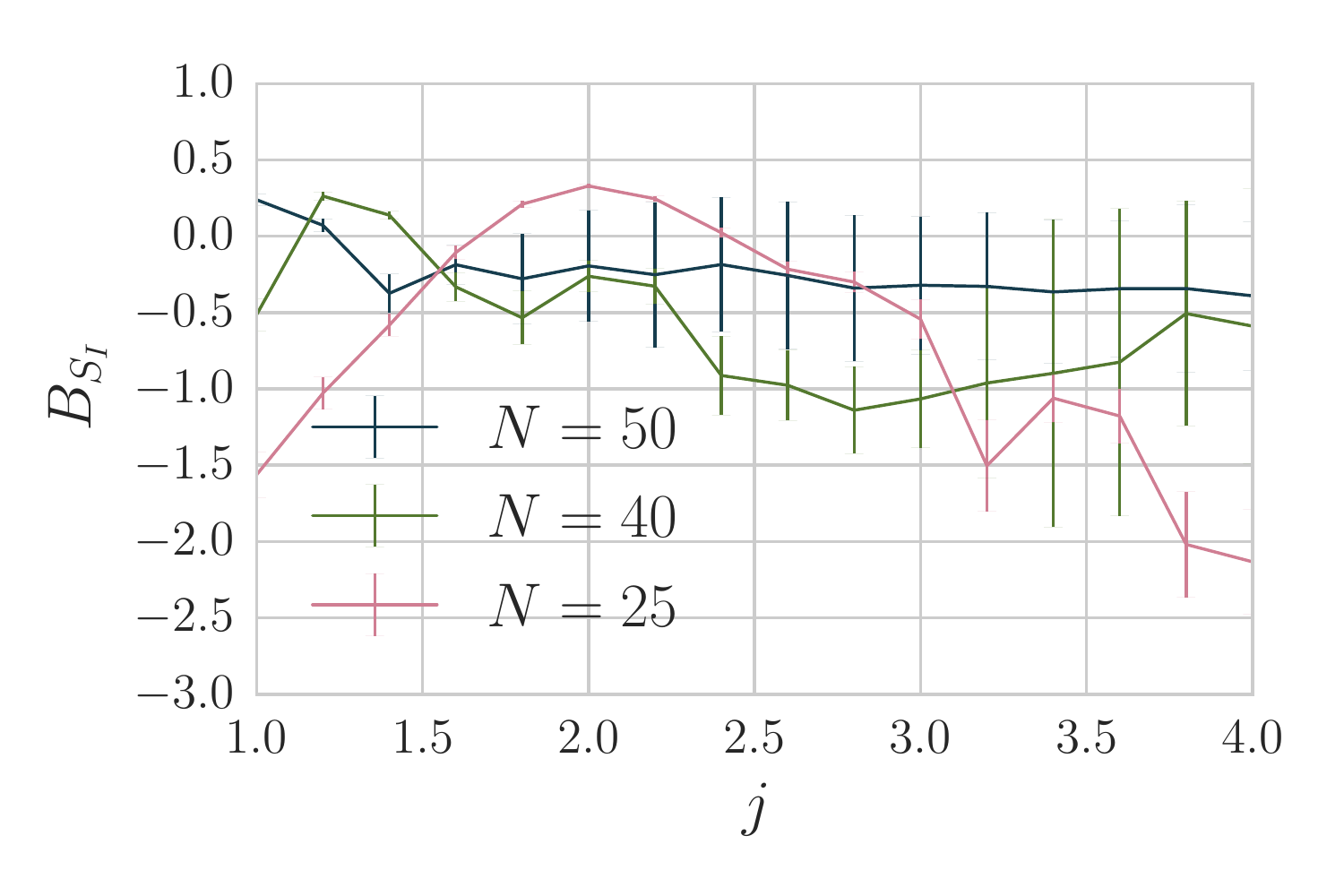}
  \caption{The fourth order cumulant for the transitions $\jc{+}$ on crystal causal sets.}
  \label{fig:binderCp}
\end{figure}

Since the magnetisation is not part of the $\jc{+}$ transition its fourth order cumulant at this transition can not help us understand the system, we can look at the cumulant for $S_{I}$, since the variance of $S_{I}$ shows some signs for the phase transition.
We show it for different system sizes in Figure~\ref{fig:binderCp}.
While the errorbars are considerably smaller than for the random causal sets, the behaviour does not show any uniformity between the system sizes, or any clear indicators of behaviours we could interpret.
It is thus clear that we are not currently equipped to explore phase transitions to anticorrelated states.

\section{Ising model on varying causal set}
We found that the Ising model on a fixed causal set undergoes a continuous phase transition, between a state in which the spins are completely correlated and a state in which the spins are disordered and a second, less understood transition between the disordered state and an anticorrelated state.
The next step to explore this is to also vary the causal set in the path integral, and simulate the partition function
\begin{align}
\mathcal{Z}_{\Omega_{2d},I} = \sum_{\mathcal{C} \in \Omega_{2d}} \sum_{I \text{on} \C} e^{- \beta \left( \Sa_{BD}(\epsilon)+ \Sa_I(j) \right)} \;,
\end{align}
in which we vary the causal set $\C$ as well as the spin state $I$.
To simulate the $2$d orders we use a modified version of the code presented in~\cite{cunningham_causal_2017}.
The code generates $2$d orders, using the same algorithm as described in~\cite{surya_evidence_2012}, and we couple the Ising model to this by assigning each element a spin variable which can take on values $\pm 1$.
In the Monte Carlo step we evolve both the causal set and the spin state.
In one sweep we first attempt $r=N(N-1)/2$ causal set moves, and then attempt $N$ spin flips.
In the computer the simulated $2$d orders are labelled, as explained in~\cite{surya_evidence_2012} we use this labelling to fix the spin degrees of freedom to given causal set elements.

Since we saw in~\cite{glaser_finite_2017} that the behaviour of the system does not change qualitatively for a wide range of the parameter $\epsilon$ we fix $\epsilon=0.21$ and concentrate on varying $\beta,j$.
To explore this we ran a grid of points $j \in [-1.0, 0.9]$ with step size $0.1$ and $\beta \in [-1.0,1.0]$ with step size of $0.1$.
This first run showed the beginning of a new phase in the corner $\beta=-1.0, j=0.9$, so we extended the region and added points in the grid $\beta \in [-1.0,0.0]$ and $j\in [1.0,2.0]$ with the same step sizes.

\begin{figure}
  \subfloat[][$\Sa_{Ising}$]{\includegraphics[width=0.5\textwidth]{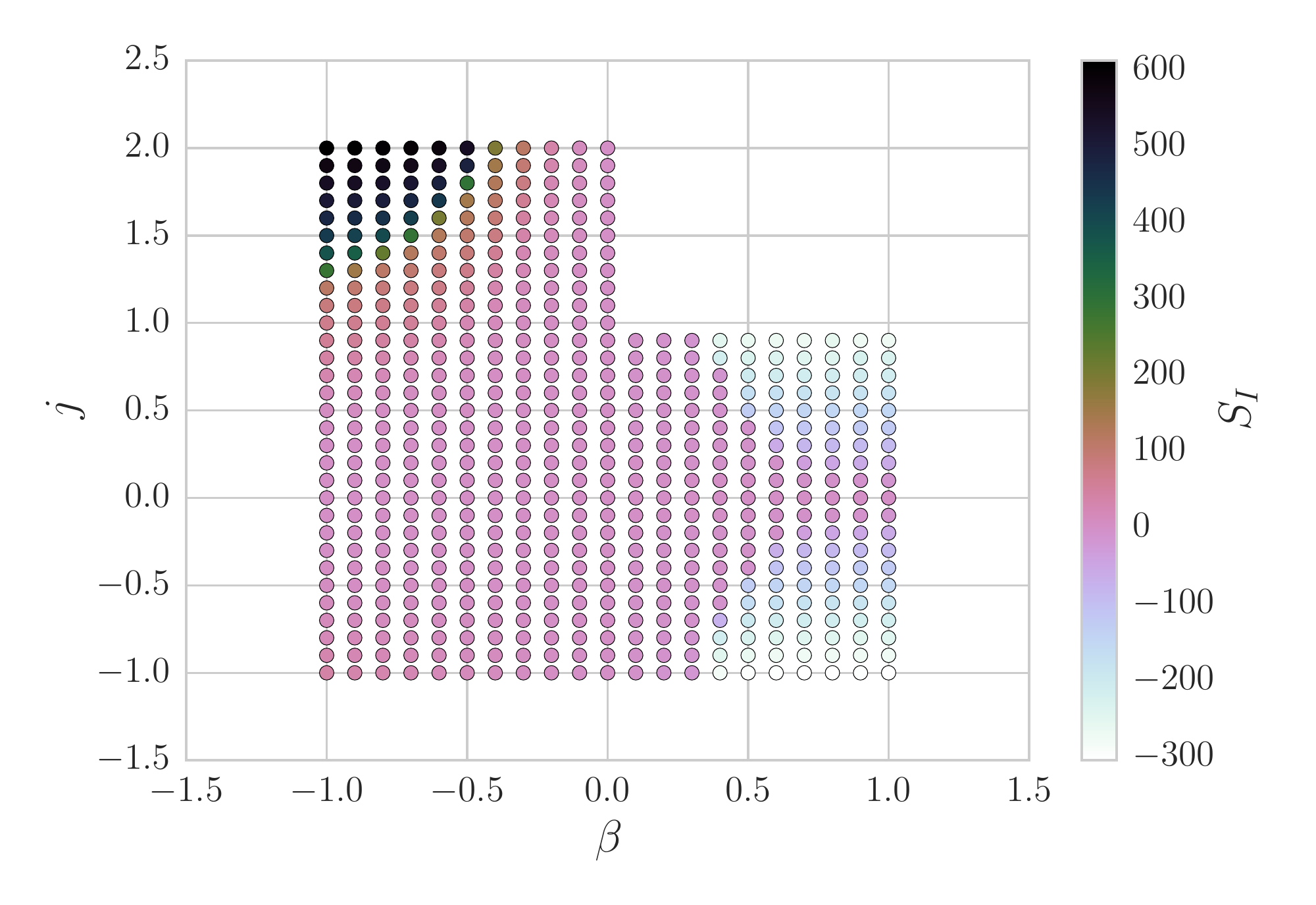}}
  \subfloat[][$\Sa_{BD}$]{\includegraphics[width=0.5\textwidth]{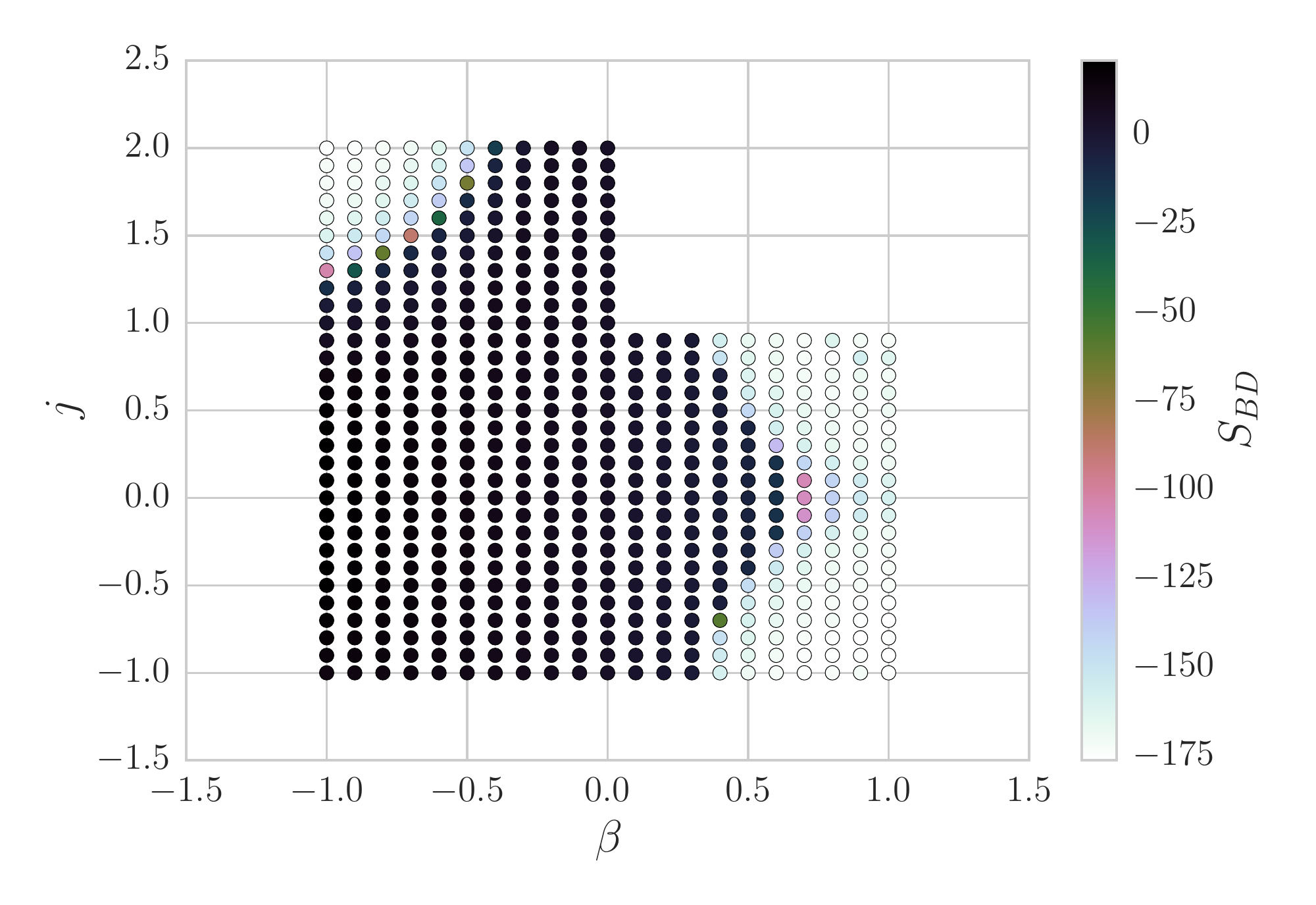}}

  \subfloat[][$M$]{\includegraphics[width=0.5\textwidth]{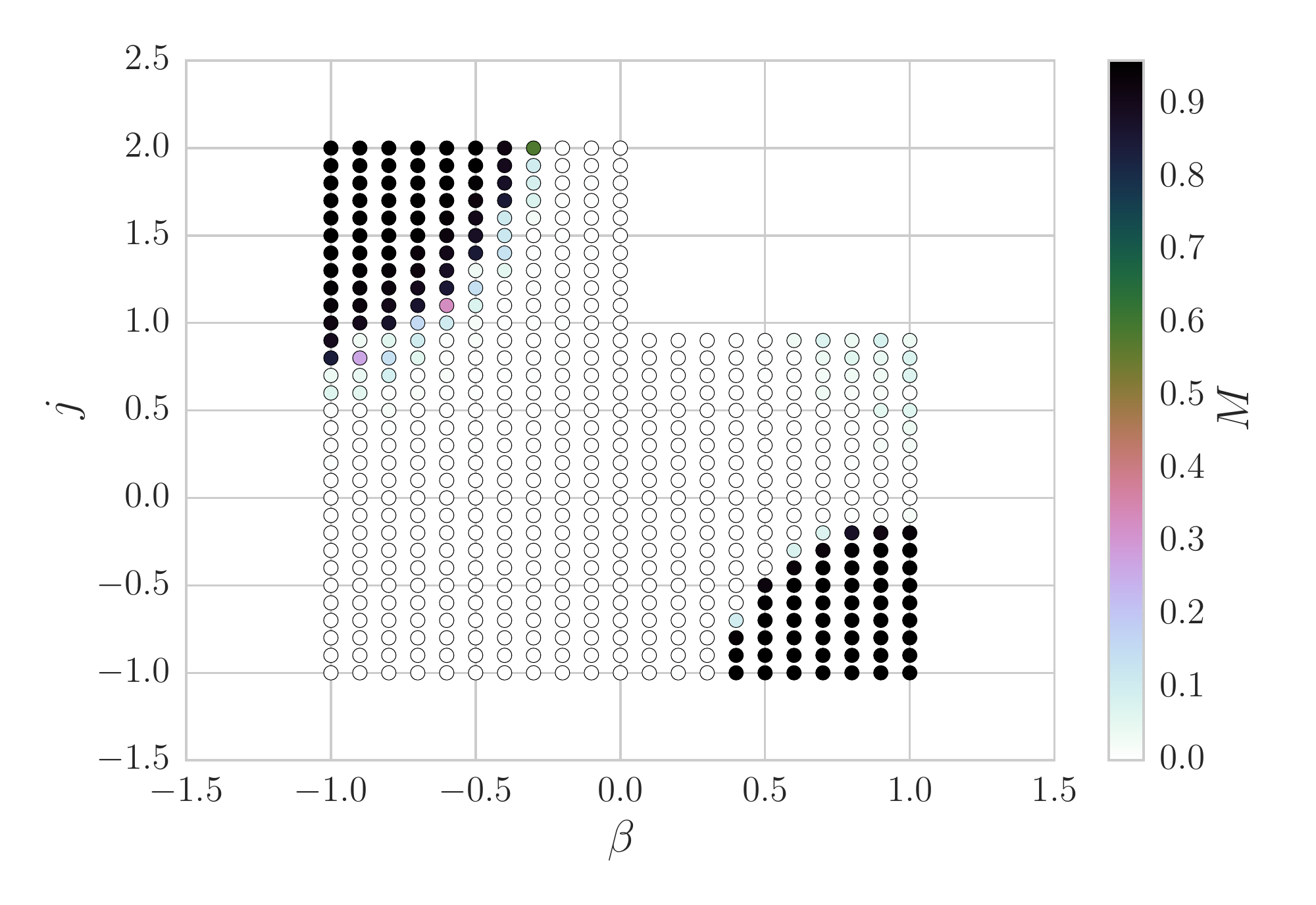}}
  \subfloat[][$R$]{\includegraphics[width=0.5\textwidth]{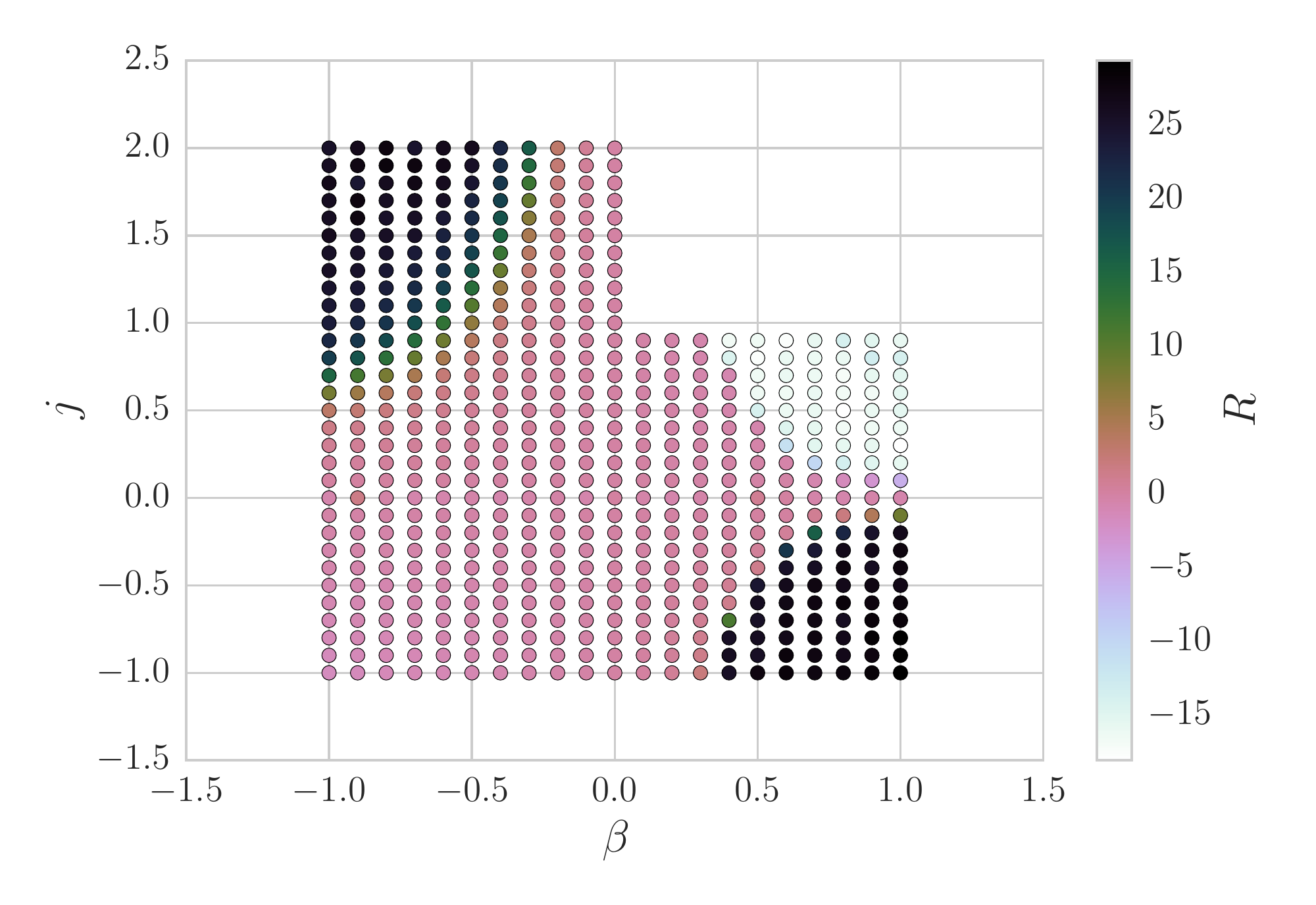}}
  \caption{Scatter plots of the different observables tested for the varying causal set. Combining the information contained in these four images we can distinguish the phases.}
  \label{fig:VarCScatter}
\end{figure}

The complete grid can be seen in Figure \ref{fig:VarCScatter}.
Comparing the colouring of the grid for different observables we find five different states.
The region connected to $\beta=0,j=0$ is the disordered phase, neither the causal set, nor the Ising model show any dominant structure in this region.
At positive $\beta$ we find an ordered phase.
We know from~\cite{glaser_finite_2017}, that the causal set without the Ising model should have a first order phase transition at $\beta_c\approx \frac{1.66}{N \epsilon^2}$ which for our values of $N,\epsilon$ is $\beta_c \approx 0.75$, agreeing with our results, as plotted in Figure \ref{fig:VarCScatter} b).
For $j=0$ these are the crystalline $2$d orders, and the spins are decoupled.
Small $j$ in either direction still leaves the spins disordered, only when the coupling becomes strong enough the spins align (for negative $j$) or anti align (for positive $j$) which we can see as the black (aligned) and blueish/ white (anti aligned) region in the relation correlation in Figure \ref{fig:VarCScatter} d).
For non zero $j$ the transition to the crystal orders happens earlier.
The last phase is at positive $j$ and negative $\beta$, it exhibits a two stage behaviour.
This is particularly interesting, first the interaction of the Ising model aligns the spins and then it forces the causal set elements into a crystal order.

To explore the states more quantitatively, we can compare them with those on the fixed causal sets.
We find that we can characterise each state using two letters, the first one to indicate the state of the causal set, $R$ for random or $C$ for crystalline, and the second to describe the state of the spins, $C$ for correlated, $A$ for anticorrelated and $R$ for random.
The two states in which the causal sets are random $2$d orders are:
\begin{itemize}
  \item Thermal state $RR$; both the causal set and the Ising spins are random. We chose to examine the point $\beta=0.1,j=-0.2$ for this
  \item Ordered spins on random causal sets $RC$; an example point in this state is $\beta=-1,j=1.0$.
\end{itemize}
The three states on the crystal causal sets are:
\begin{itemize}
  \item Disordered crystal set $CR$; we chose the point $\beta=1,j=0.1$ for this state\footnote{This point is not part of the original grid, however we include it in this analysis to have a good example of a $CR$ state.}.
  \item Correlated crystal set $CC$; a good example point is $\beta=0.5,j=-1.0$
  \item Anti correlated crystal set $CA$; the example point for this state will be $\beta=0.5, j=0.9$.
\end{itemize}
Having chosen these points we can compare the values of the observables for them and also compare them to the states on the fixed causal sets.
We will denote the states on the fixed causal sets in a similar manner as above but with an added $f$ in front, so $fCC$ is a correlated state on a fixed crystal causal set.
We will explore the average over $20$ fixed causal sets of the same class for the fixed case.

\begin{figure}
\subfloat[][$\Sa_I$]{\includegraphics[width=0.33\textwidth]{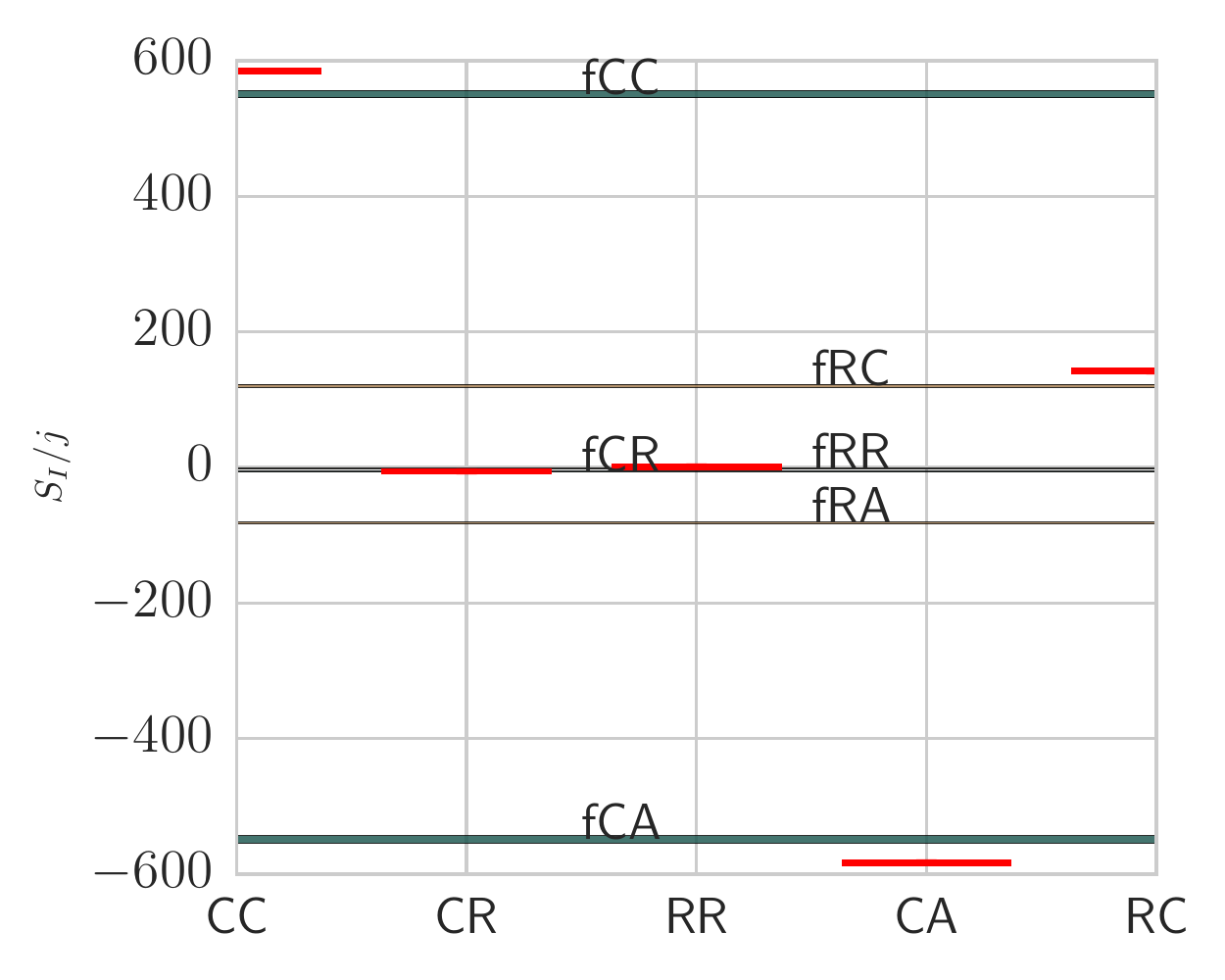}}
\subfloat[][$M$]{\includegraphics[width=0.33\textwidth]{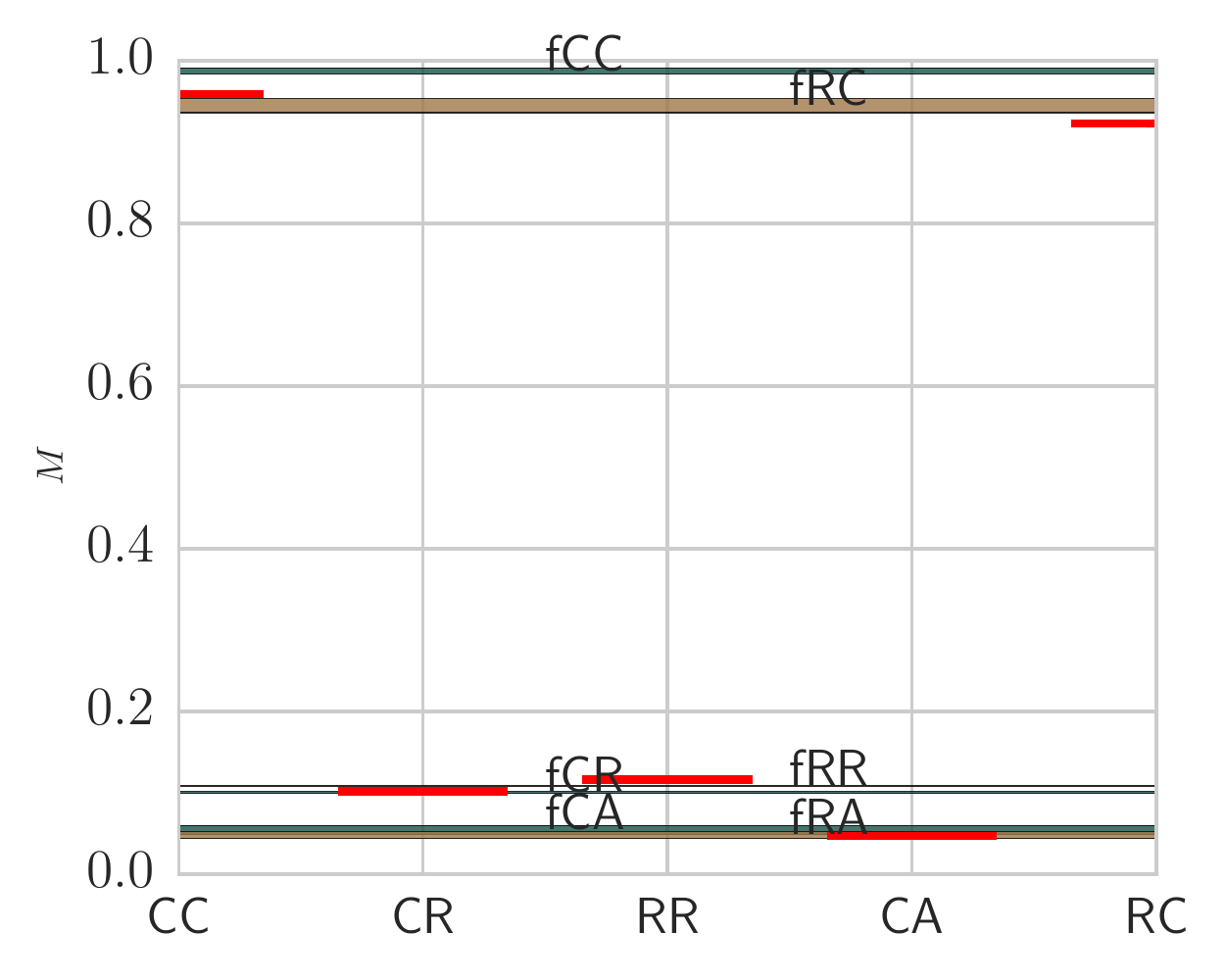}}
\subfloat[][$R$]{\includegraphics[width=0.33\textwidth]{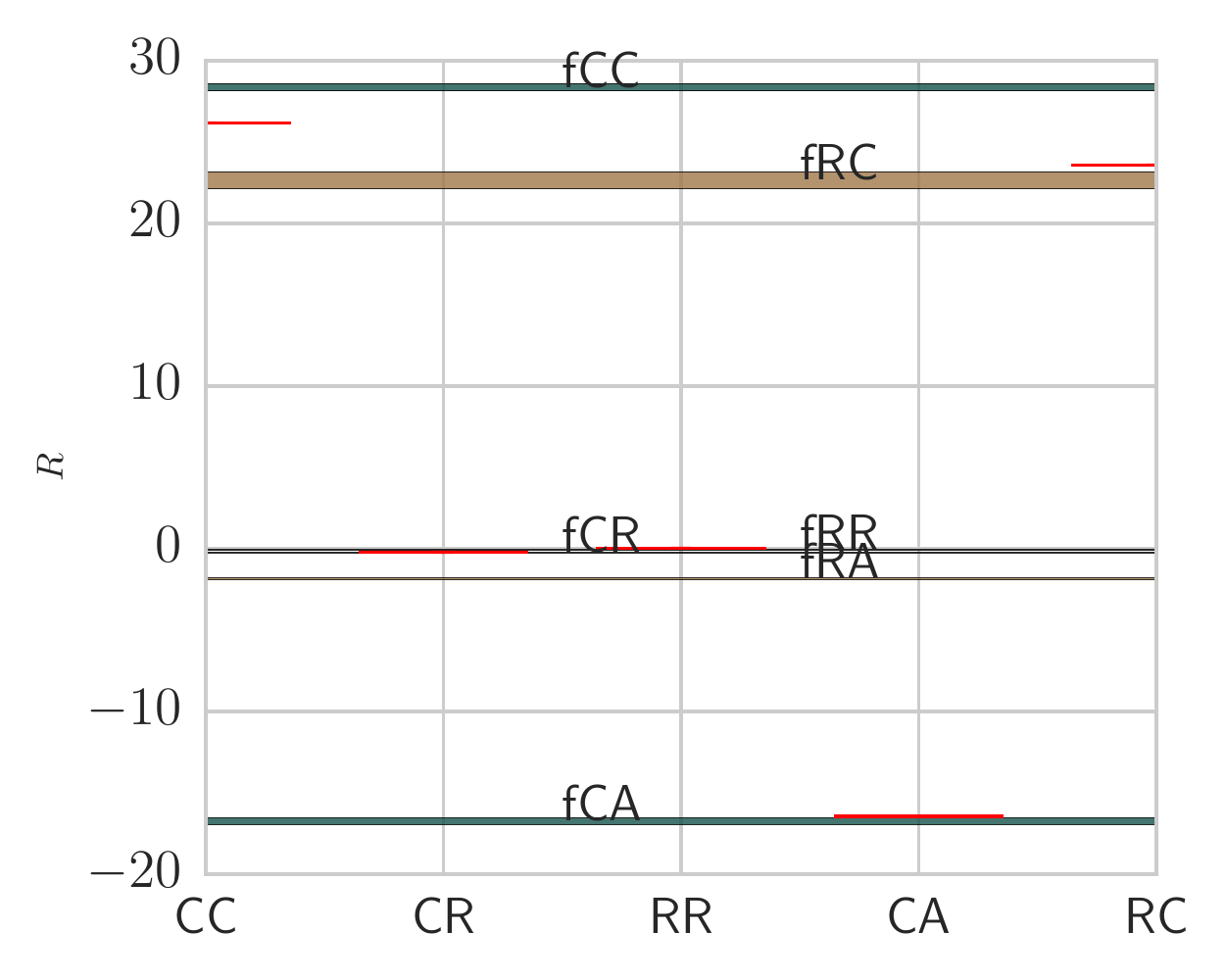}}
  \caption{Comparing the values of observables on the varying causal sets with those on the fixed causal sets. The $x$-axis denotes the different states considered, while the $y$ axis shows the observables. The shorter, red lines indicate the value for the states on varying causal sets, while the long horizontal lines in dark blue/ yellow indicate the average value on the fixed crystal/ fixed random causal sets.}
  \label{fig:statecomp}
\end{figure}

To visualise this data we plot the five states for varying causal sets along the $x$-axis, and plot the six states on the fixed causal sets as unbroken horizontal lines with their width given by their errors, they $y$-axis then denotes the values of different observables.
We can see this for $\Sa_I,M,R$ in figure \ref{fig:statecomp}.
The comparison shows that the states are very similar indeed, $CC$ corresponds to $fCC$, $CA$ to $fCA$ and $RC$ to $fRC$.
One interesting observation is that $CR$ and $RR$ have similar values to $fCR, fRR$, and can not be told apart from the observables we use here.
This is because their difference lies in the different structure of the causal sets, which is not probed by the Ising model when the Ising spins are not coupled strongly.

With this qualitative understanding of the different phases, we can try and explore them by running run more detailed simulations along some lines in the parameter space that cross the phase boundaries.
These lines are given in Table \ref{tab:lines} and visualised in the image next to it. 
\begin{table}
  \begin{minipage}{0.5\textwidth}
  \caption{Table of added lines and  a sketch of the different phase regions and the lines along which we will examine the causal sets in more detail.}
  \begin{tabular}{l r r }
    \toprule
    fixed value & range & step size \\
        \colrule
    $\beta=1$ & $j \in [-0.3,0.3]$& $0.02$ \\
    $\beta=-1$ & $j \in [0.3,1.5]$ & $ 0.02$\\
    $j=-1.0$ & $ \beta \in [0.3,0.5]$ & $0.01$\\
    $j=0.9$ & $ \beta \in [0.3, 0.5]$ & $0.01$\\
    $j=2.0$ & $\beta \in [-0.2,-0.6] $ & $0.01$\\
    \toprule
  \end{tabular}
  \label{tab:lines}
\end{minipage}\begin{minipage}{0.5\textwidth}
\includegraphics[width=\textwidth]{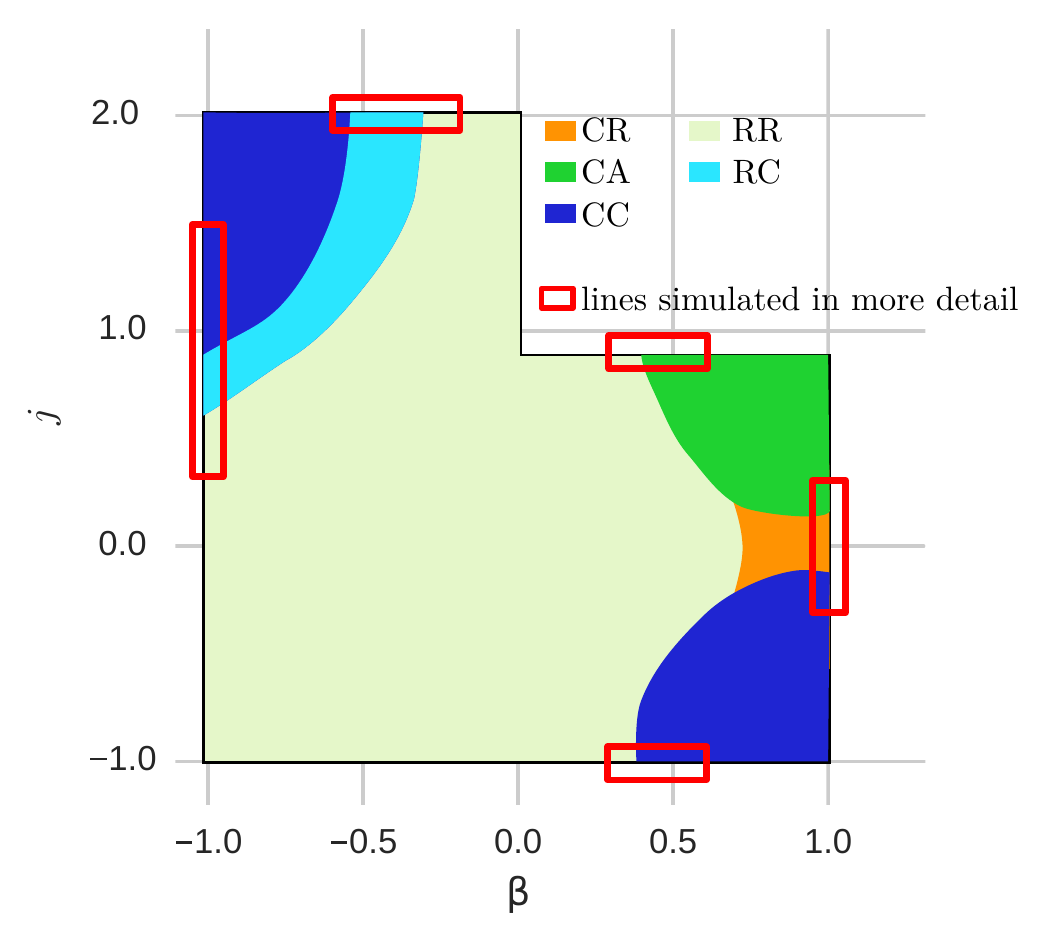}
\end{minipage}
\end{table}
They allow us to explore the phase transitions in more detail.

\subsection{The correlated to disordered to anticorrelated transitions $\beta=1$ }
\begin{figure}
  \centering
  \includegraphics[width=0.9\textwidth]{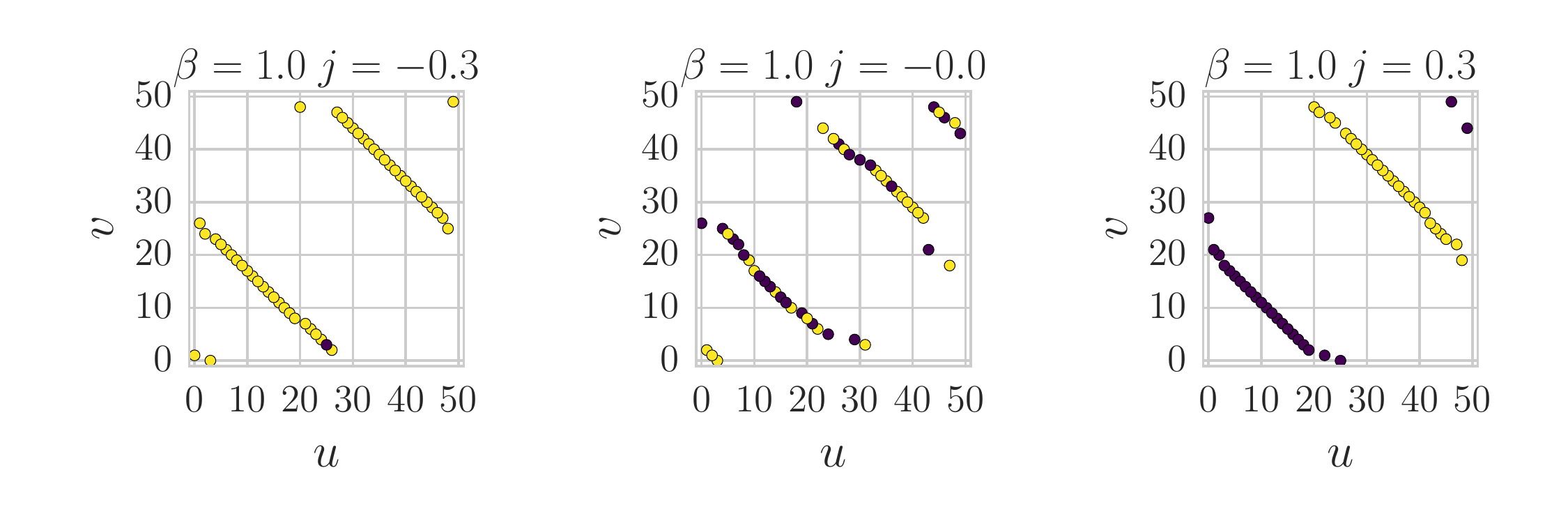}
  \caption{Causal sets taken along $\beta=1$. With $j$ changing from negative to positive the spins change from correlated to anti-correlated with a small uncorrelated region around $j=0$.}
  \label{fig:b1Causets}
\end{figure}

\begin{figure}
  \includegraphics[width=0.49\textwidth]{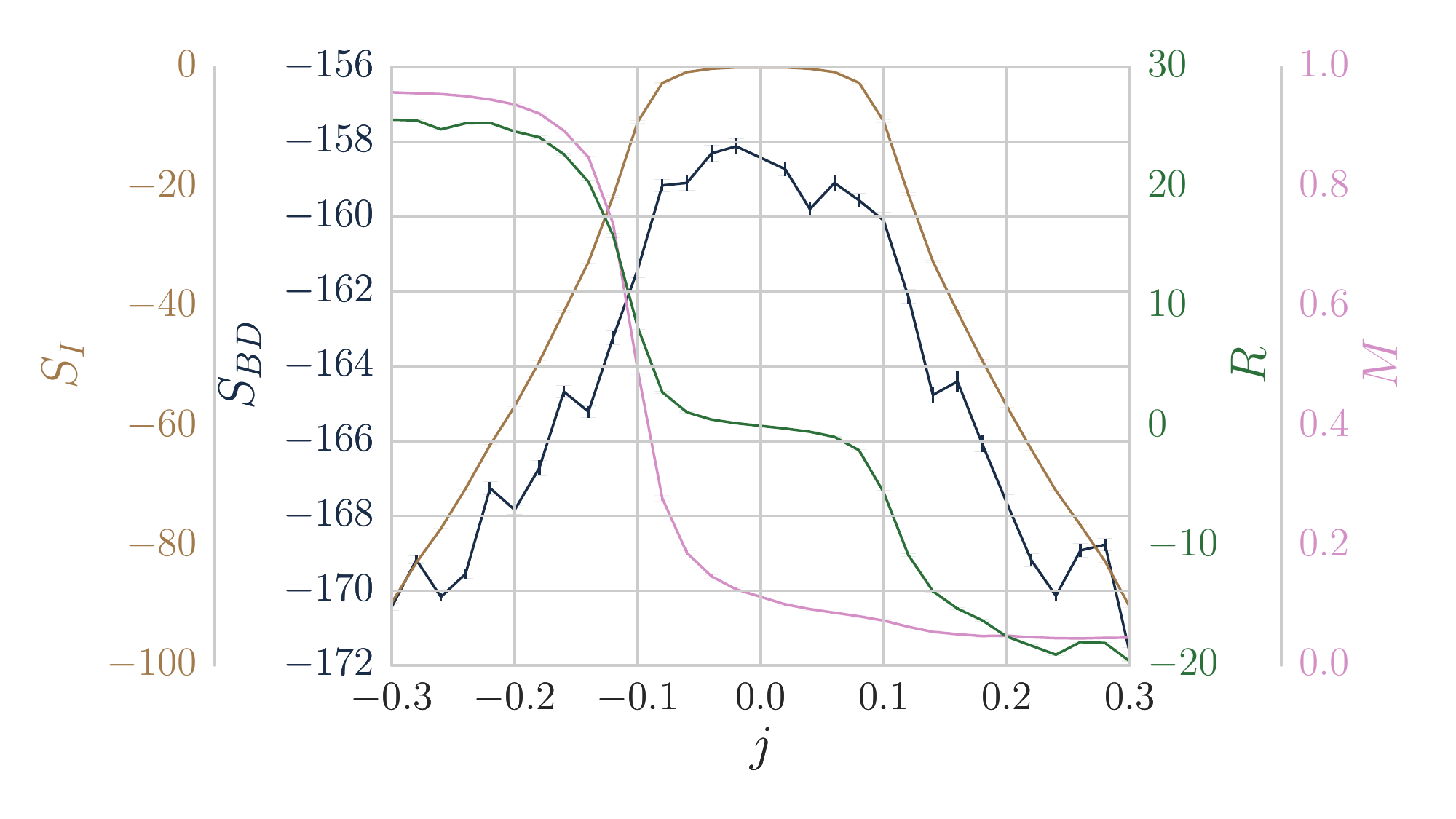}
  \includegraphics[width=0.49\textwidth]{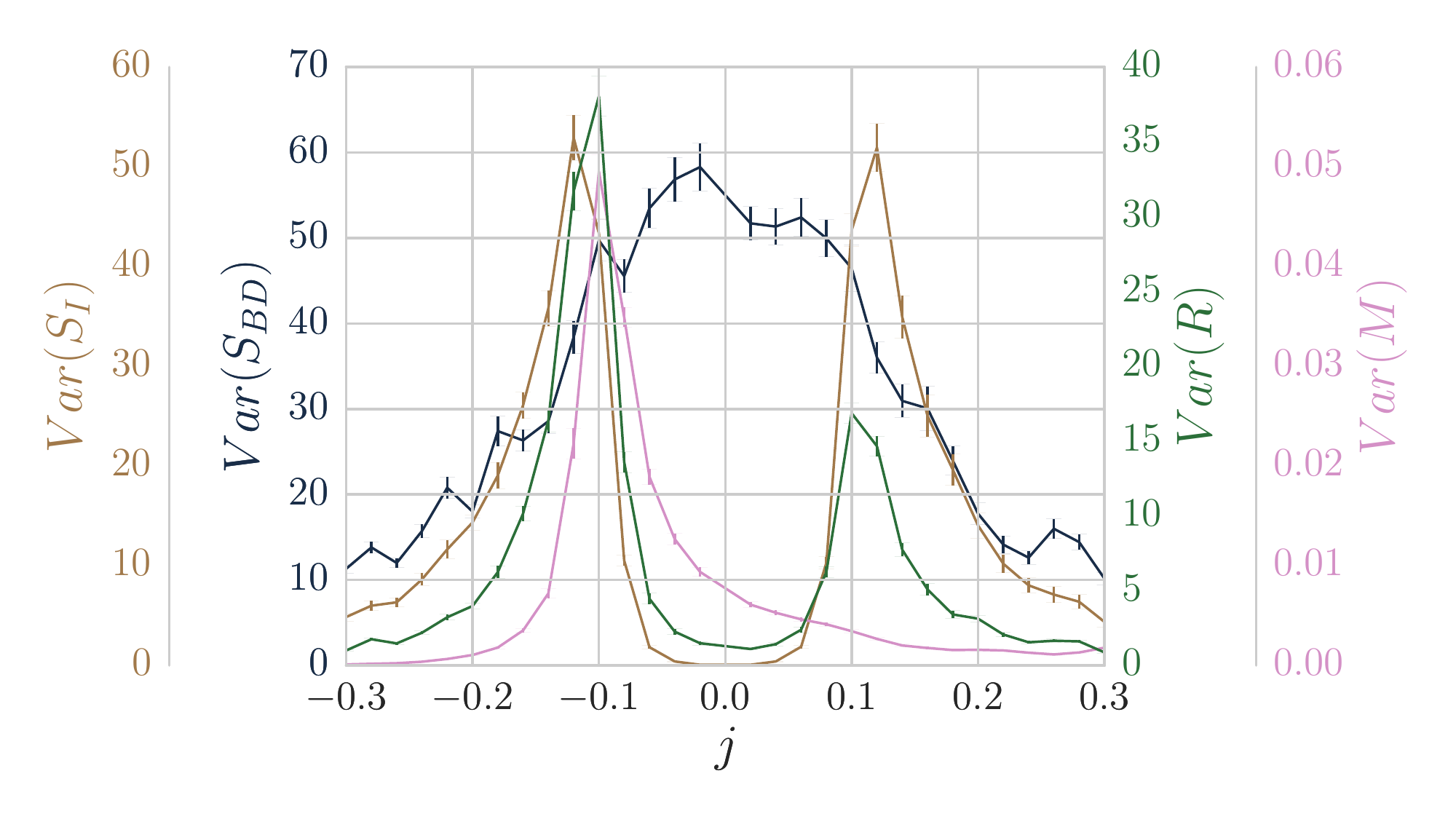}

    \includegraphics[width=0.49\textwidth]{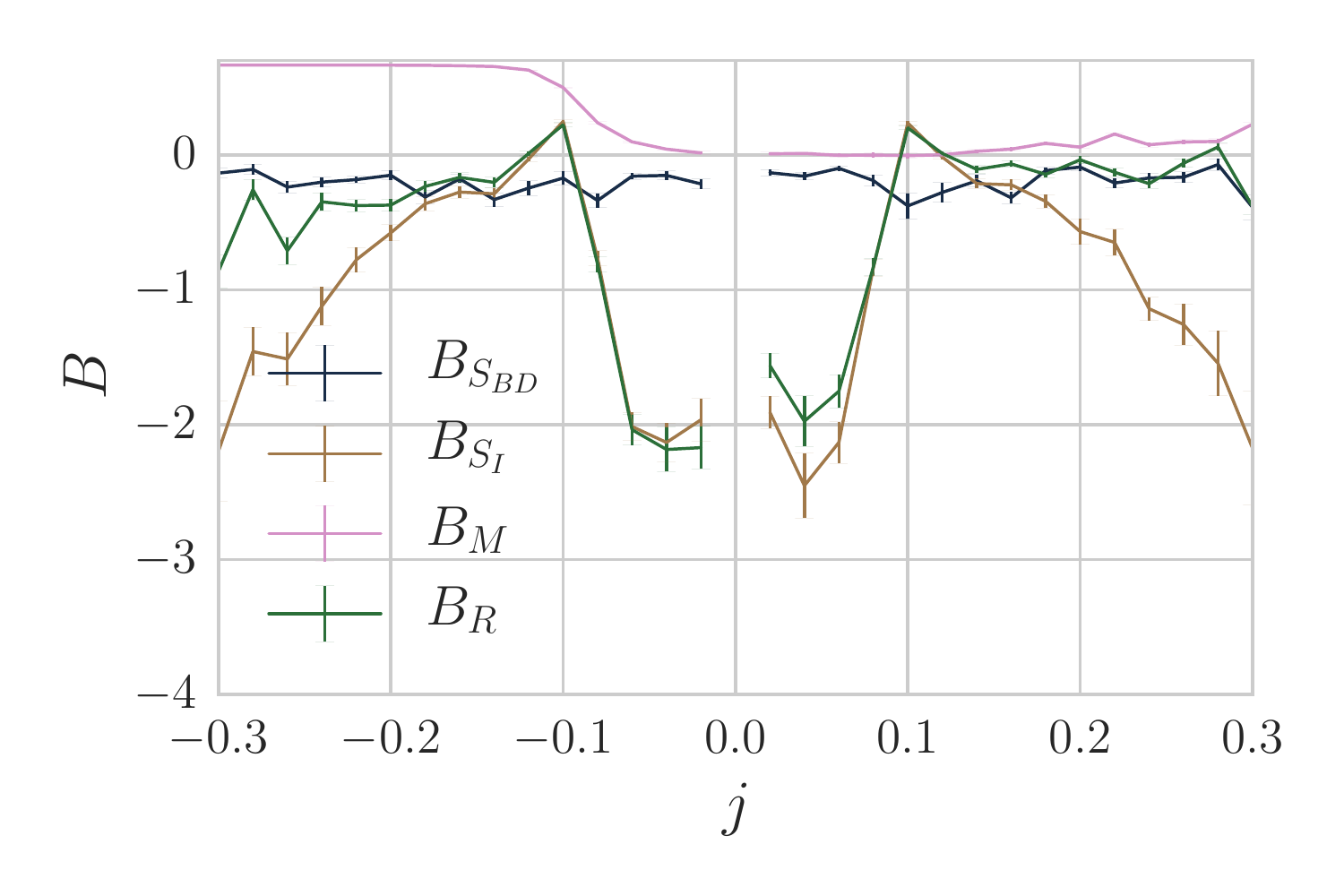}
  \caption{$\beta=1$ phase transition line, from negative to positive $j$ the line crosses from a crystalline causal set with correlated spins to a crystalline causal set with disordered spins, to a crystalline causal set with anticorrelated spins.}
  \label{fig:b1PTline}
\end{figure}

This line is the only one that crosses two different phase transitions $CC \to CR \to CA$, the behaviour of the different observables and variances as it does so is shown in Figure \ref{fig:b1PTline}, and three example causal sets for the different regions are shown in Figure \ref{fig:b1Causets}.
The causal sets in this region are always crystalline, so the spin alignment is the main property that changes.
We can see this from the fact that the BD-action, the dark blue line in Figure \ref{fig:b1PTline}, remains almost constant (it varies by less than $10\%$).
The variance of the BD-action also changes slowly with a slight increase around $j=0$.
Since the Ising model acts as an additional force pulling the causal set into a more ordered state, as we can surmise from the fact that the phase transition starts at lower $\beta$ for larger $j$, it is likely that larger $j$ suppresses fluctuations and thus leads to lower $\Sa_{BD}$ and a smaller variance.

The phase transitions are not visible in $\Sa_{BD}$ or its variance, they are however very clear in the variance of the Ising action, the yellow line in Figure \ref{fig:b1PTline}.
We can also see the correlated spin to disordered spin transition in the magnetisation and its variance (pink lines in Figure \ref{fig:b1PTline}).
The transition to the anticorrelated phase is not visible in the magnetisation, since it has close to zero magnetisation, but we can see it in $R$ and the variance of $R$ (green line in Figure \ref{fig:b1PTline}).

\begin{figure}
  \includegraphics[width=0.49\textwidth]{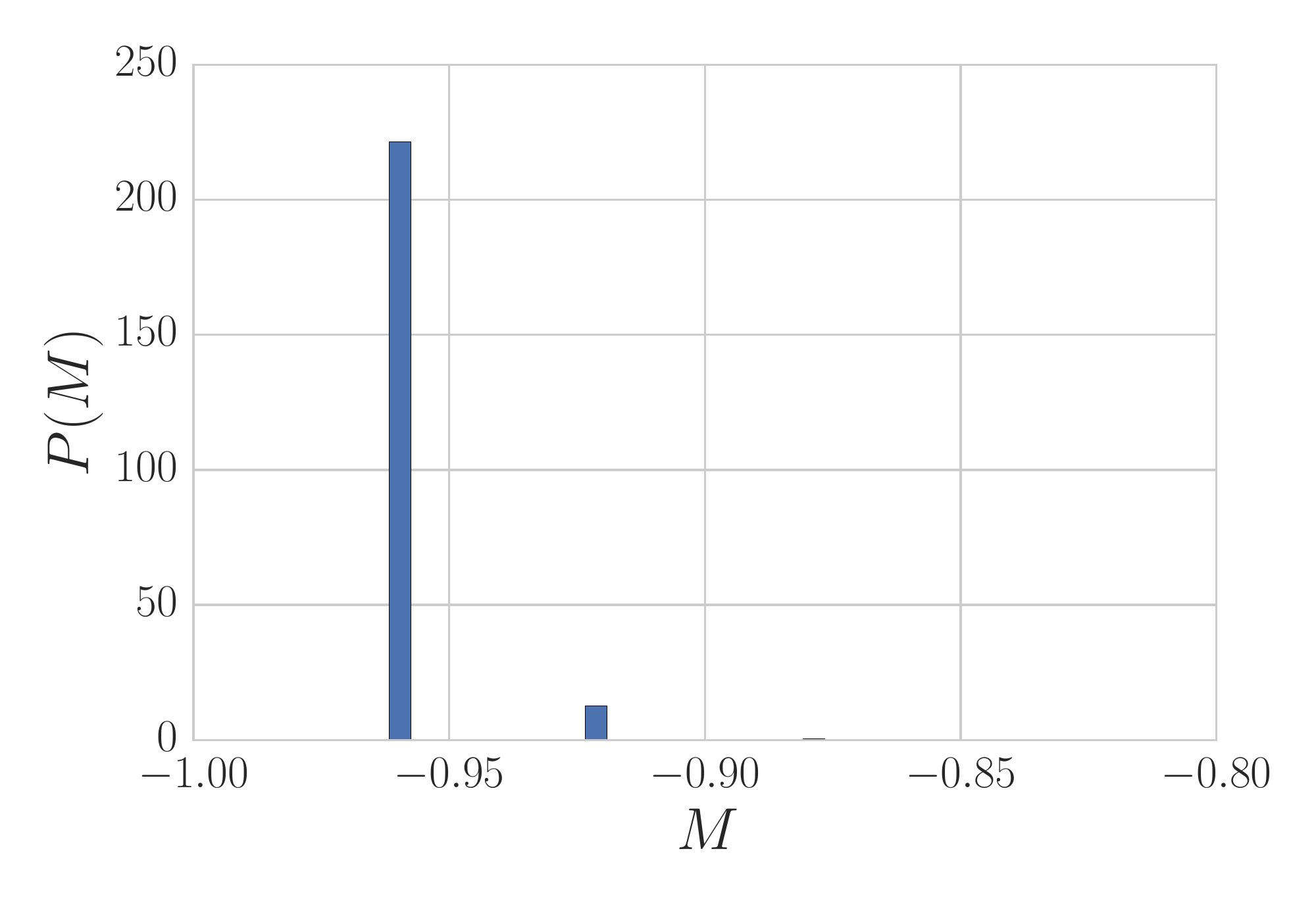}
  \includegraphics[width=0.49\textwidth]{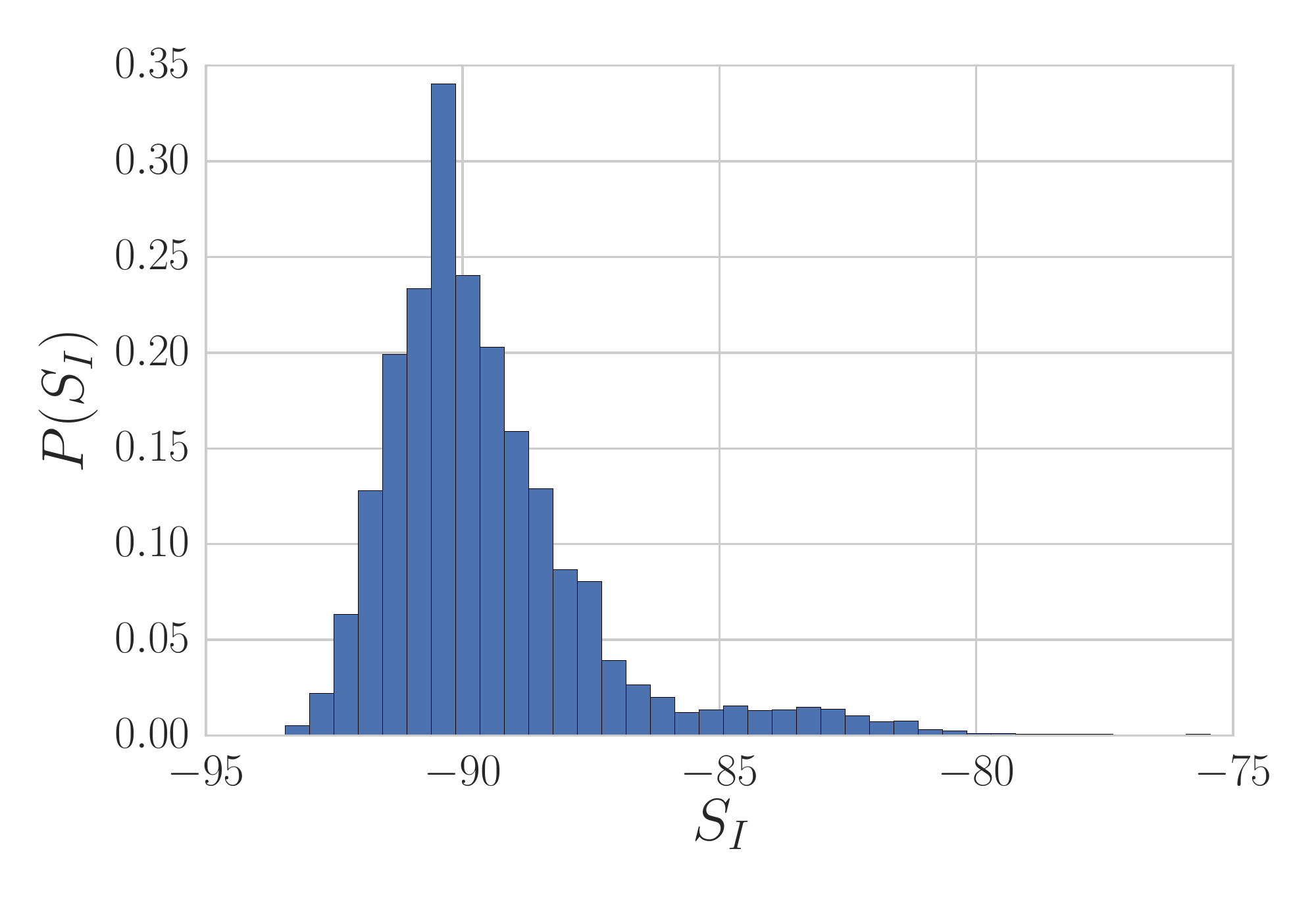}

  \caption{Histogram to show the frequency distribution of the magnetisation states in the Ising model for $j=-0.02$ and $j=-0.3$.}
  \label{fig:fluctuation}
\end{figure}

The Binder cumulant of the magnetisation shows that the phase transition at negative $j$ is a second order phase transition again, while the cumulant of $\Sa_{BD}$ makes it clear that the causal sets are not undergoing a qualitative change in states along this line.
For the cumulant $B_{\Sa_I}$ the fluctuations above and below $0$ coinciding with the correlated and anticorrelated phase transitions indicate that both of the transitions are higher order.
The falloff to negative values for very low or large $j$ can be explained through the distribution.
The values for the fourth order cumulant assume that the observables are Gaussian distributed around the mean value.
This is a good approximation in the thermal region of the phase diagram, however at large $\beta$ this breaks down.
We show the distribution of $\Sa_I$ and $M$ in Figure~\ref{fig:fluctuation}, and see that they are both far from Gaussian. 
In the left hand figure we show the distribution of $M$, almost all states are in an almost completely aligned state, with very few fluctuations.
Combining this with the smearing of the $\Sa_I$ states that arises through the fluctuations of the causal set leads to the distribution of $\Sa_I$ shown in the right hand figure.
Here we can clearly see that the distribution has two roughly Gaussian peaks.
The second, lower peak, arises when the Ising model fluctuates into a less aligned state.
This behaviour is less important closer to the phase transition, where the spins move more, so that we are confident that the fourth order cumulant there is still a good observable.
It also does not influence either the cumulant of $\Sa_{BD}$ or $M$, however it does lead to the negative values for $B_{S_I}$.

\begin{figure}
  \centering
  \includegraphics[width=0.5\textwidth]{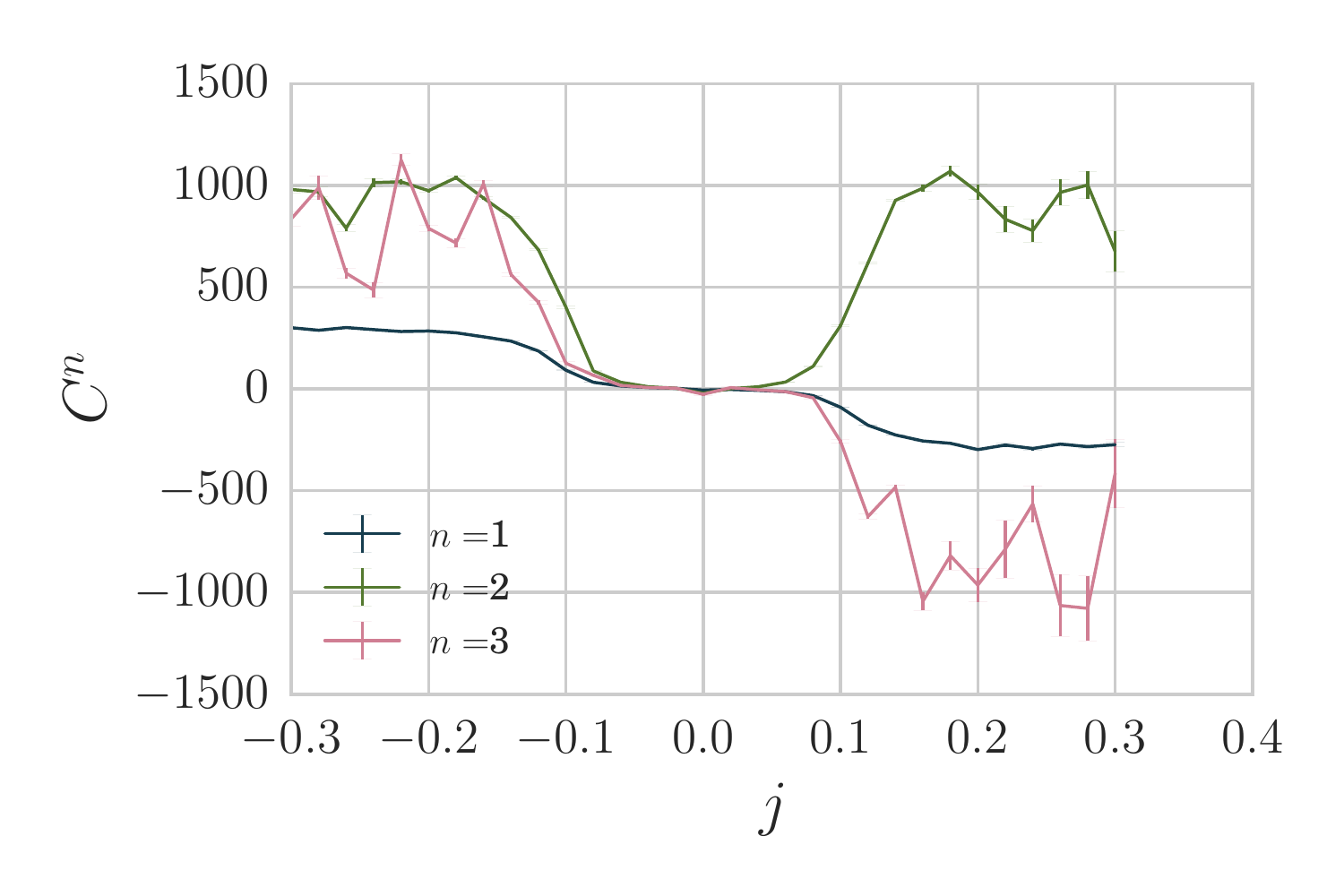}
  \caption{Path correlators along the $\beta=1$ line. The correlated and anticorrelated phases are distinguished by the value of $C^{1,3}$.}
  \label{fig:b1Nchain}
\end{figure}

We can gather a little more information about the phase transition by exploring the $n$ path correlator $C^n$.
This correlator shows us how strongly elements at path distance $n$ are correlated to each other.
Looking at Figure \ref{fig:b1Nchain} we can clearly see how the system changes from correlated to anticorrelated when traversing the $j=0$ line.
For negative $j$ all elements are correlated, independent of their path distance.
The correlation grows smaller as $j \to 0$ and disappears in the region around $j=0$.
Then as $j$ becomes positive, elements at path distance $1$ grow correlated again, while linked elements and elements at path distance $2$ grow anticorrelated.
Due to the crystalline structure, the causal sets explored here are rather flat and rarely contain paths of more than length $2$, which is why it is not useful to plot longer path distances.

\subsection{Random state to anticorrelated spins on a crystal order $j=0.9$}
\begin{figure}
  \includegraphics[width=\textwidth]{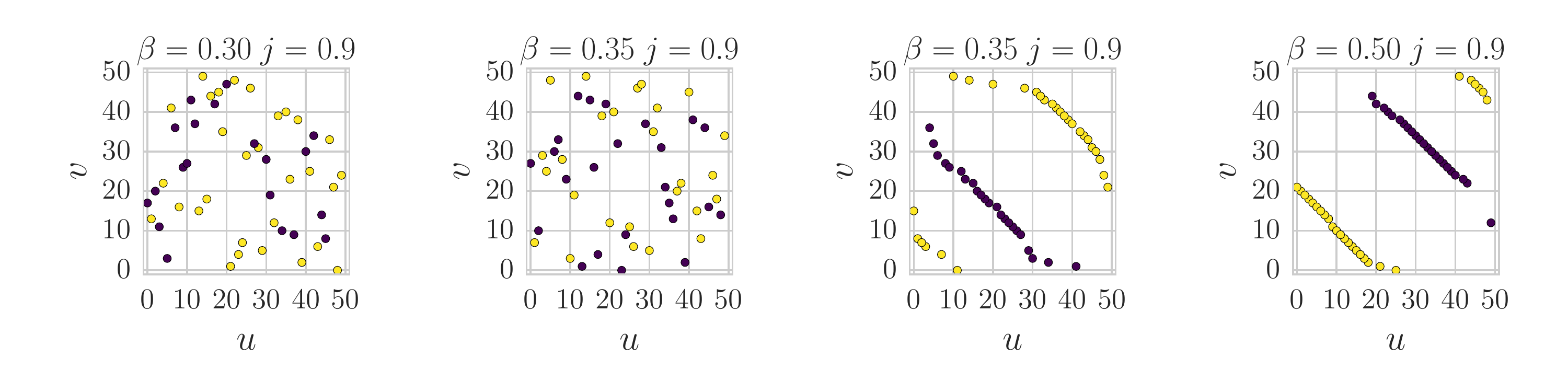}
  \caption{Three causal sets in the different regions along the $j=0.9$ lines.}
  \label{fig:j45Causets}
\end{figure}

Along this line the system undergoes a single phase transition $RR \to CA$, at $\beta=0.35$ the system transitions from a random order with uncorrelated spins to a crystal order with anti-correlated spins, this is illustrated nicely in the four causal sets in Figure \ref{fig:j45Causets}.
We show two different causal sets at $\beta=0.35$ to illustrate that at the phase transition two different states of the causal set coexist.

\begin{figure}
  \includegraphics[width=0.49\textwidth]{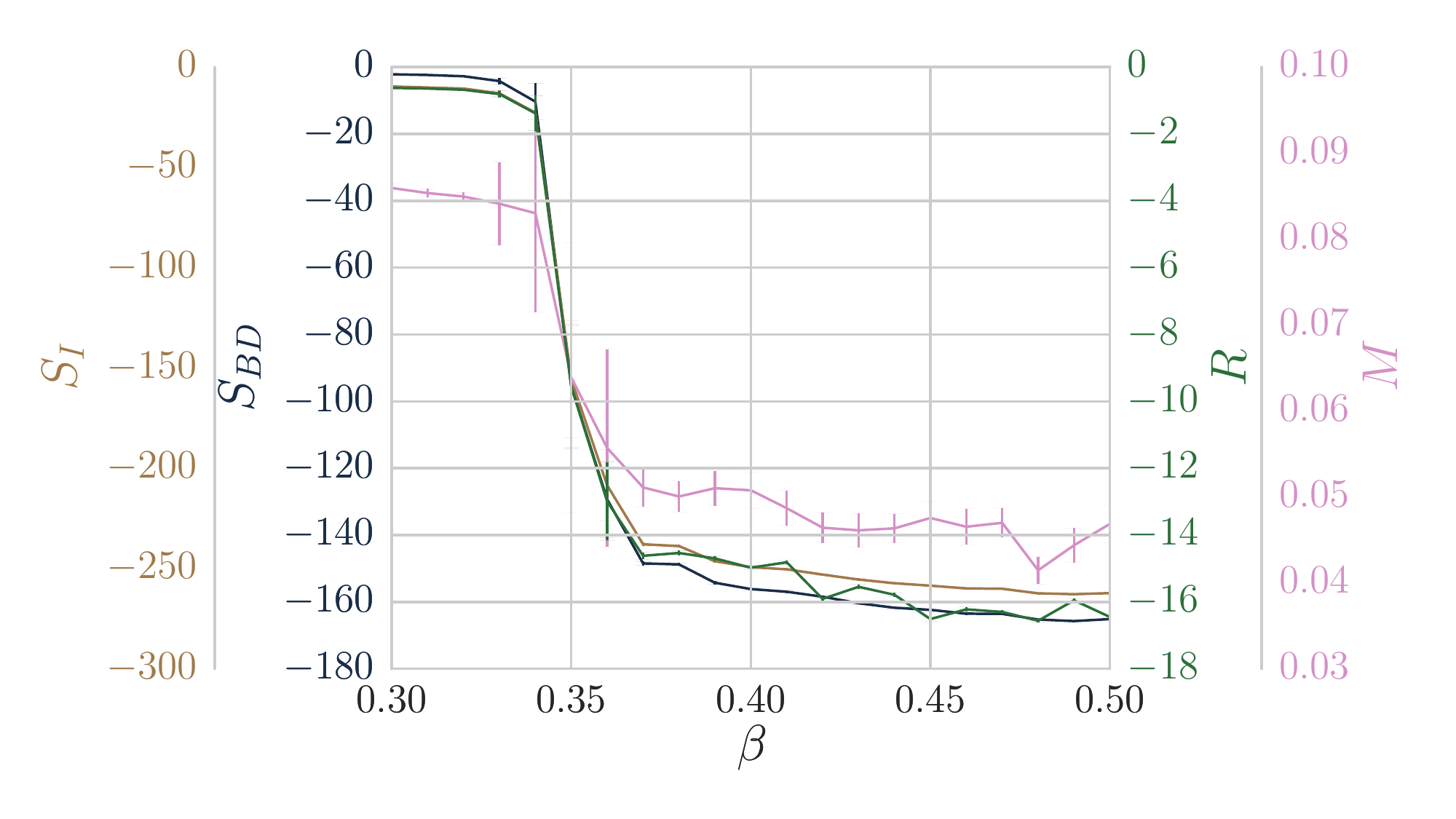}
  \includegraphics[width=0.49\textwidth]{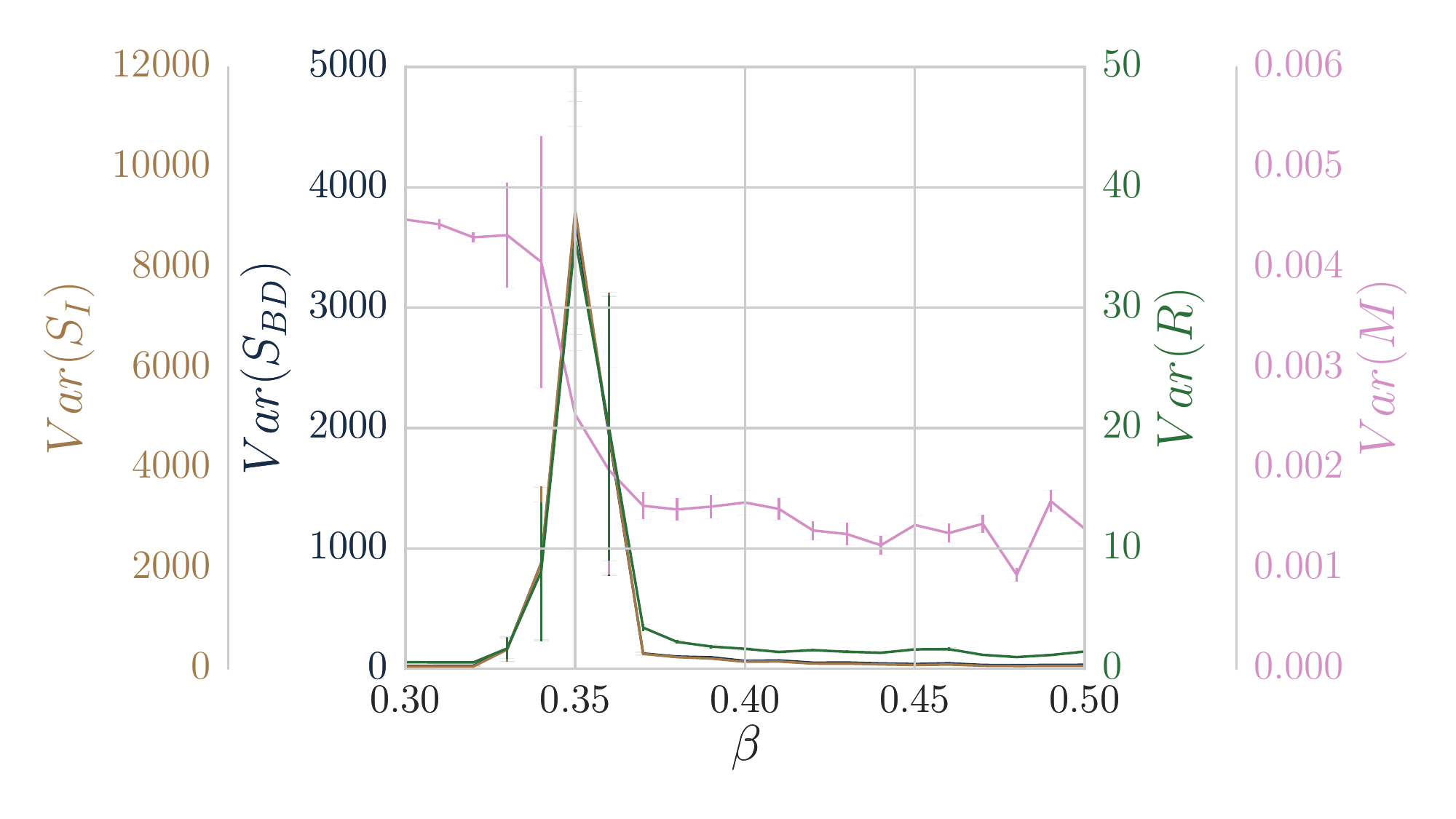}

  \includegraphics[width=0.49\textwidth]{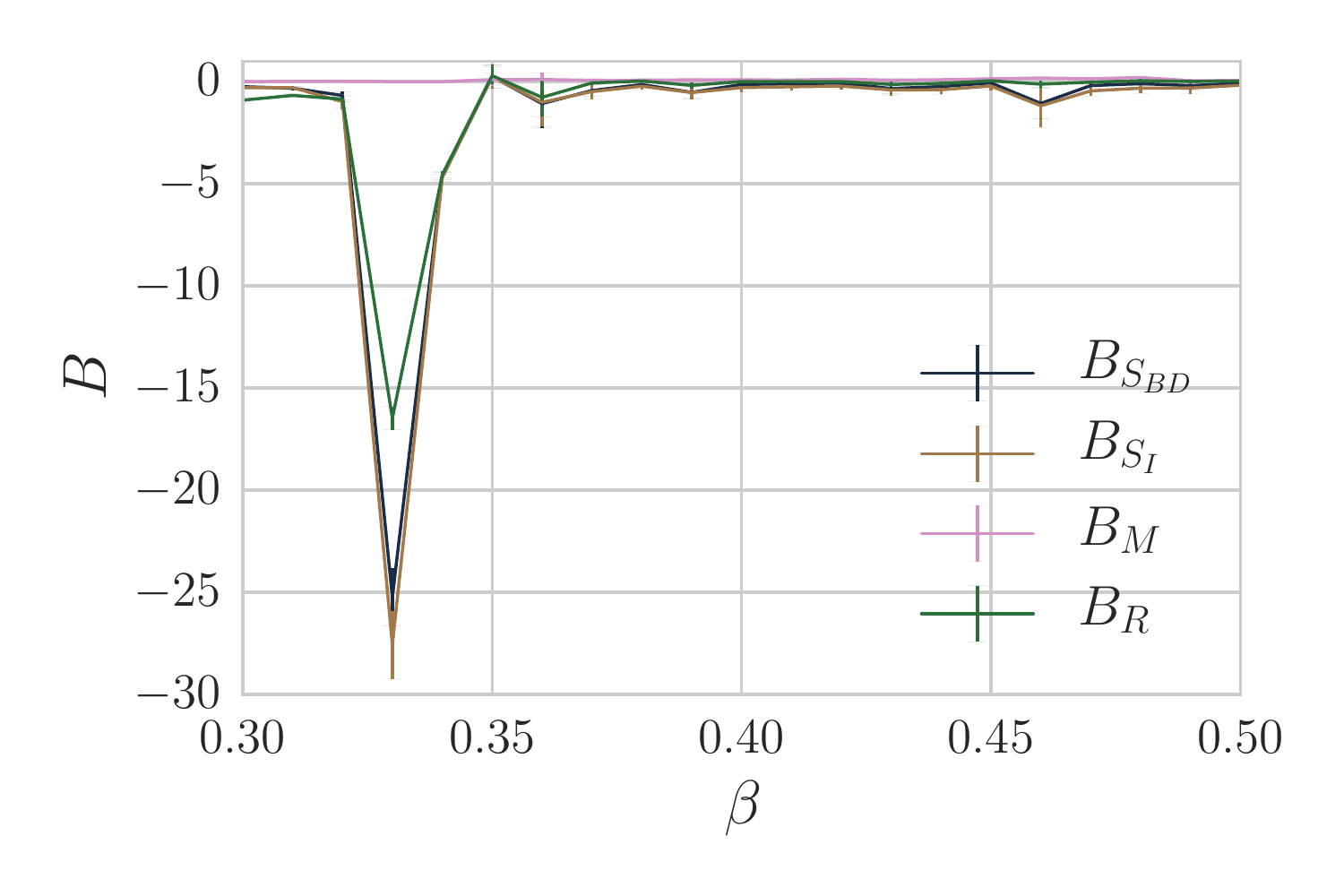}

  \caption{Different observables to characterise the phase transition at $j=0.9$.}
  \label{fig:j45PTlines}
\end{figure}

The phase transition is clearly visible in all four observables in Figure \ref{fig:j45PTlines}, even the magnetisation which is very small on both sides of the transition, but still changes in a way significant within the errorbars.
The variance of the magnetisation is the only variance not to show a clear peak at the phase transition, all three other observables peak at the phase transition, and show very similar shapes in their peaks (although the values of the variances are quite different).

The forth order cumulants also show us that at this phase transition the magnetisation does not participate in the phase transition.
The cumulants of the actions point towards a possible first order transition, which would be one explanation for the coexistence behaviour of the different states of the causal set.
On the other hand coexistence always happens for limited system size, so it is not a definite signal.
Without a clearer understanding of the best order parameter to characterise this transition we are not very confident in this identification.

\subsection{Random state to correlated spins on a crystal causal set $j=-1.0$}
\begin{figure}
  \includegraphics[width=\textwidth]{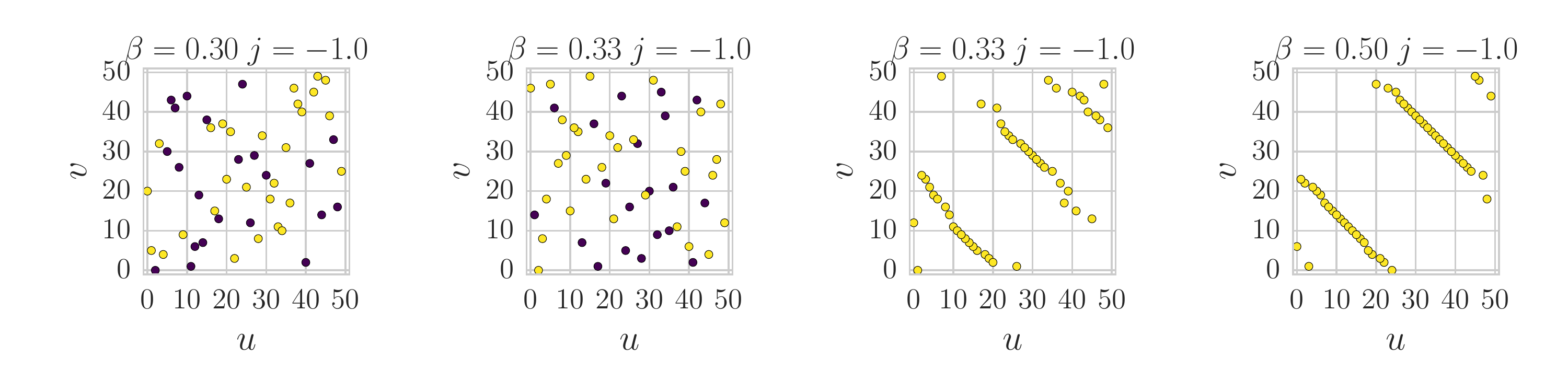}
  \caption{Three causal sets in the different regions along the $j=-1.0$ lines.}
  \label{fig:j-05Causets}
\end{figure}

\begin{figure}
  \includegraphics[width=0.49\textwidth]{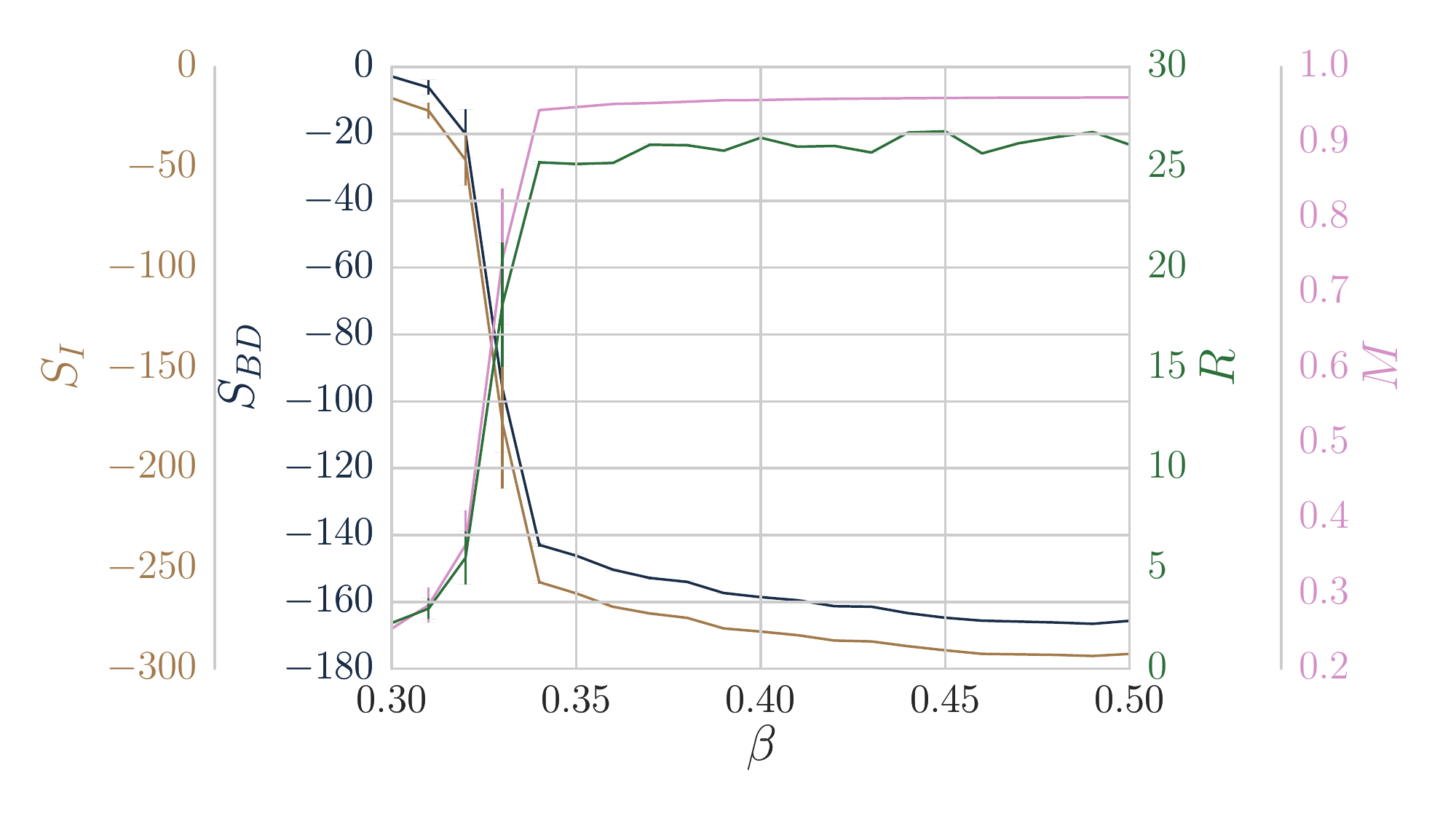}
  \includegraphics[width=0.49\textwidth]{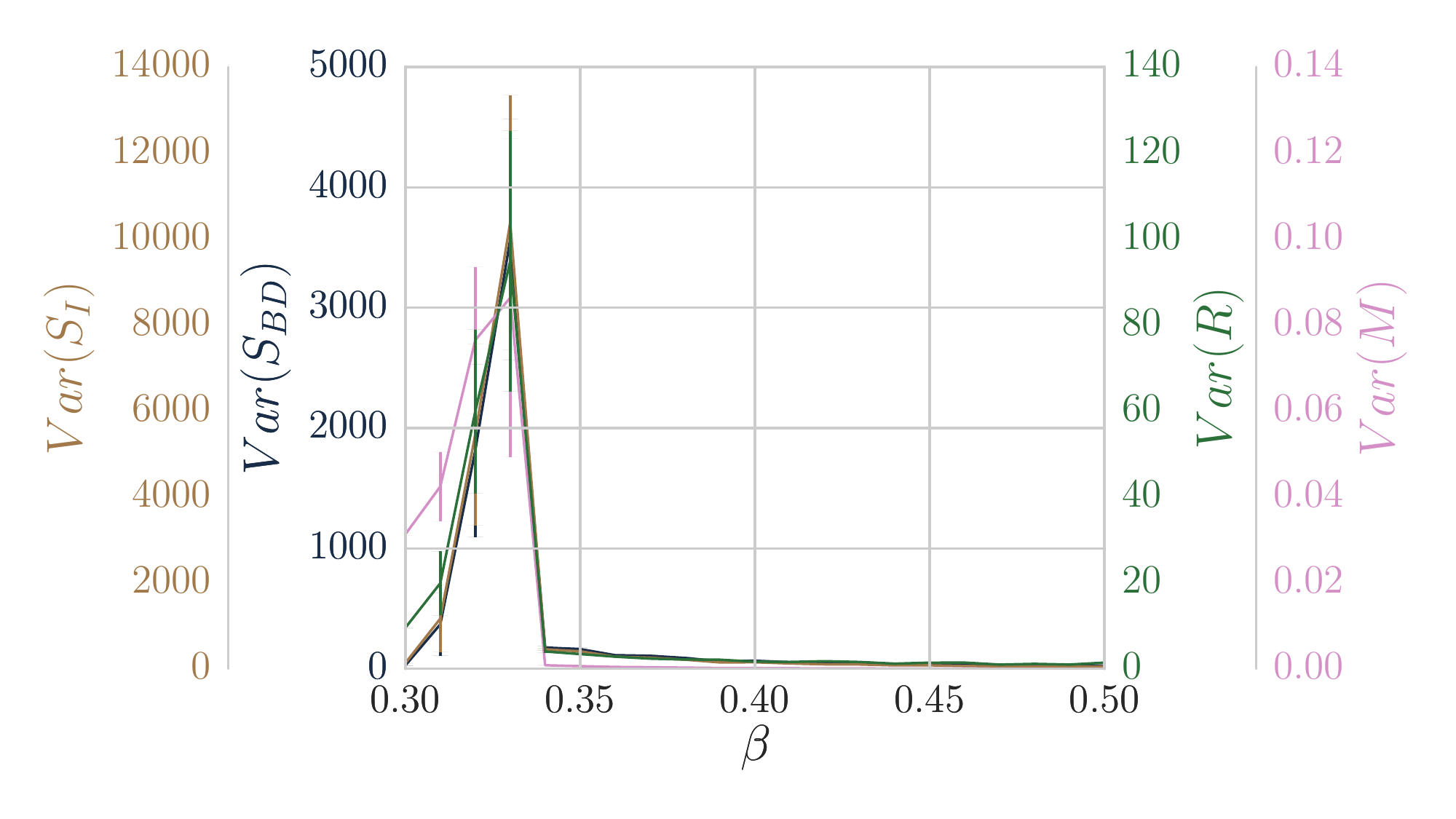}

    \includegraphics[width=0.49\textwidth]{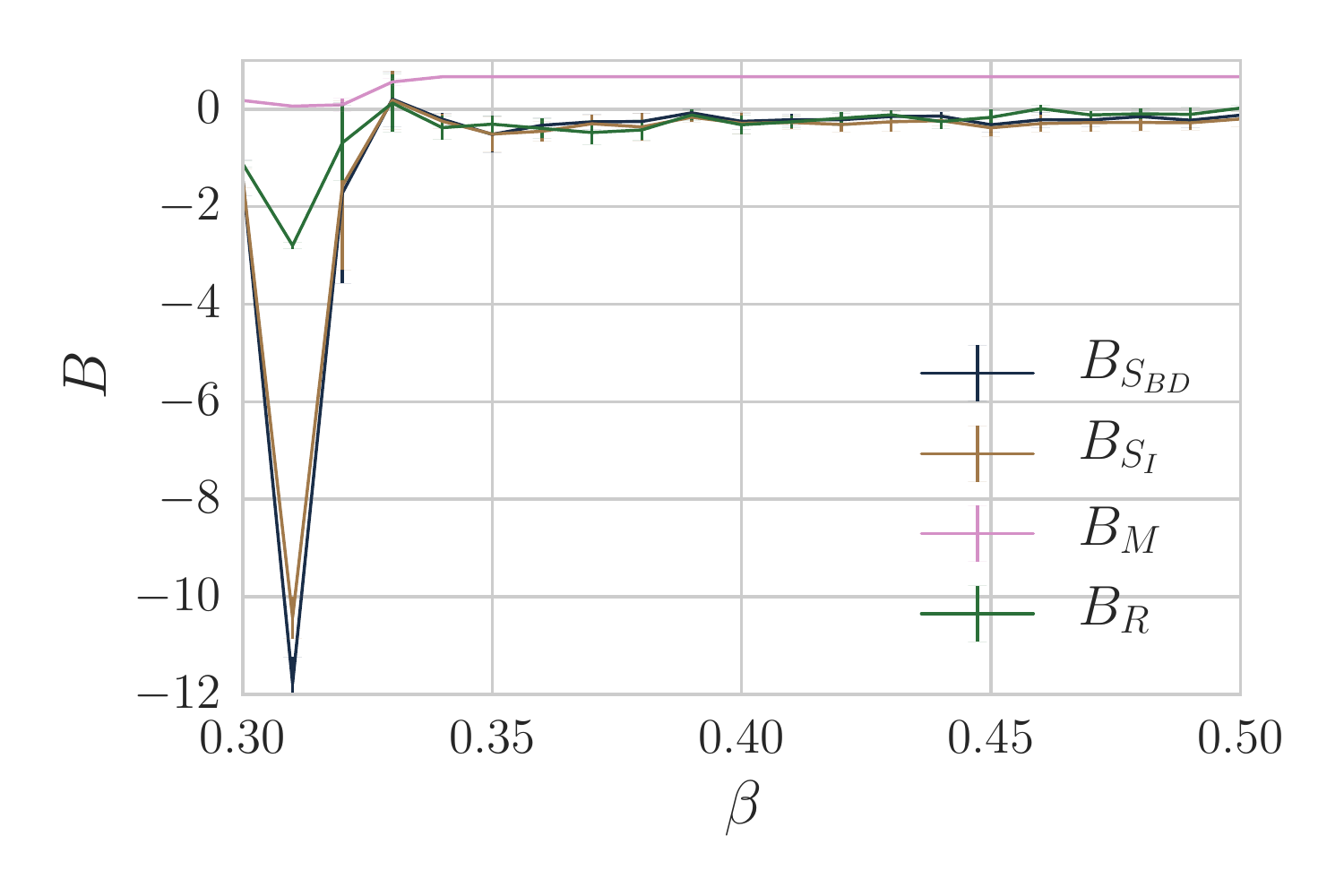}
  \caption{Different observables to characterise the phase transition at $j=-1.0$}
  \label{fig:j-05PTlines}
\end{figure}

This line follows a fixed, negative $j=-1.0$ from a random $2$d order with disordered spin states to a crystalline ordered causal set on which all spins are aligned $RR \to CC$, we show this in Figure \ref{fig:j-05Causets}.
As for the previous line we plot two causal sets at the phase transitions $\beta=0.33$ to illustrate that the phases coexist at this point.
The behaviour of the $2$d orders tracks the behaviour at $j=0$ closely, except that the phase transition arises much earlier, $\beta=0.33$ as compared to $\beta_c\approx 0.75$ as found in \cite{glaser_finite_2017}.
The behaviour of the observables and their variances along this line is very clear, with all observables showing a rapid change around $\beta=0.33$ and all variances peaking at this point, as shown in Figure \ref{fig:j-05PTlines}.
The forth order cumulant of the magnetisation transitions from $0$ to $2/3$ as we expect for a second order phase transition, while the cumulants for both actions dip far below $0$ and then bump up above $0$ before levelling out around $0$.
This makes the order of the transition here unclear.
The behaviour of the fourth cumulant of the magnetisation points towards a higher continuous transition, while the coexistence in Figure~\ref{fig:j-05Causets} and the strong negative dip in $B_{S_{BD}}$ point at a first order transition.

\subsection{From disordered causal sets with uncorrelated spins to disordered causal sets with correlated spins to crystal causal sets with correlated spins $\beta=-1$ and $j=2$}
\label{subsec:doubletrans}
Since the phase transitions along these lines behave very similar we will discuss them together.
We will first focus on the $\beta=-1$ line, and then add some comments on the $j=2$ line.
These lines are particularly interesting since there are two consecutive changes in behaviour.

\begin{figure}
  \includegraphics[width=0.49\textwidth]{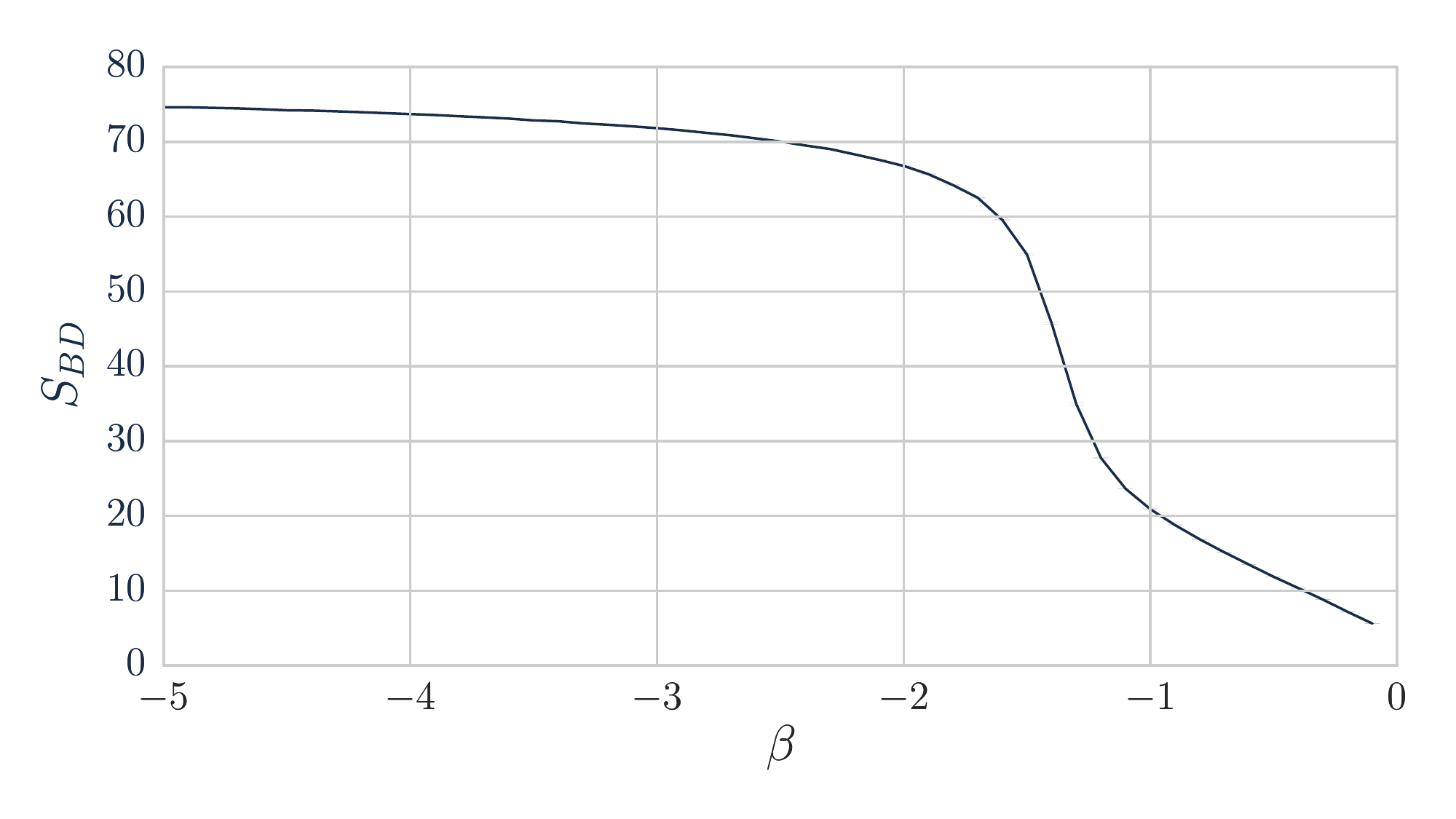}
  \includegraphics[width=0.49\textwidth]{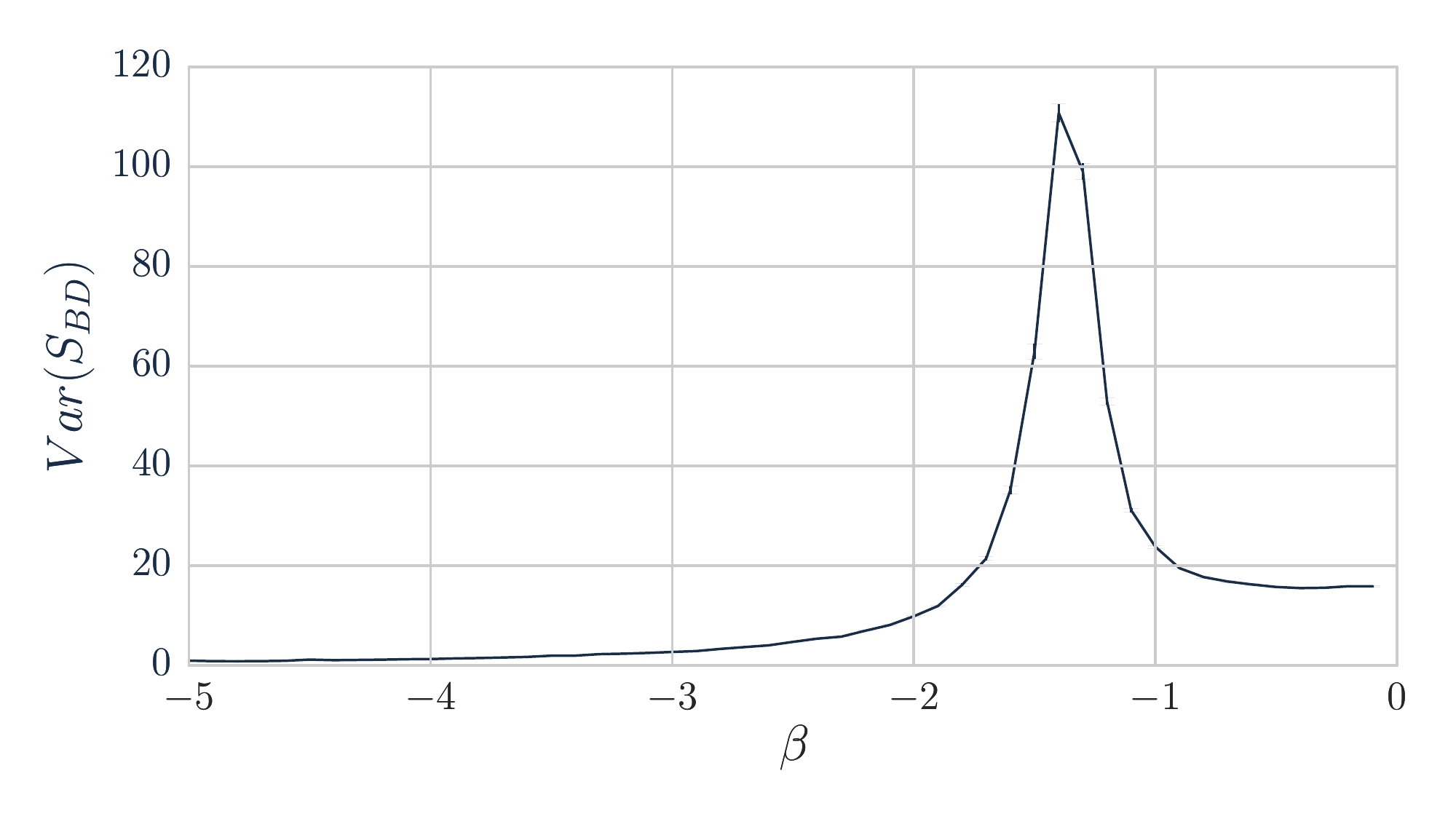}
  \caption{$\Sa_{BD}$ and $Var(\Sa_{BD})$ for $j=0,\beta<0$}
  \label{fig:negb}
\end{figure}

\begin{figure}
  \centering
  \includegraphics[width=0.9\textwidth]{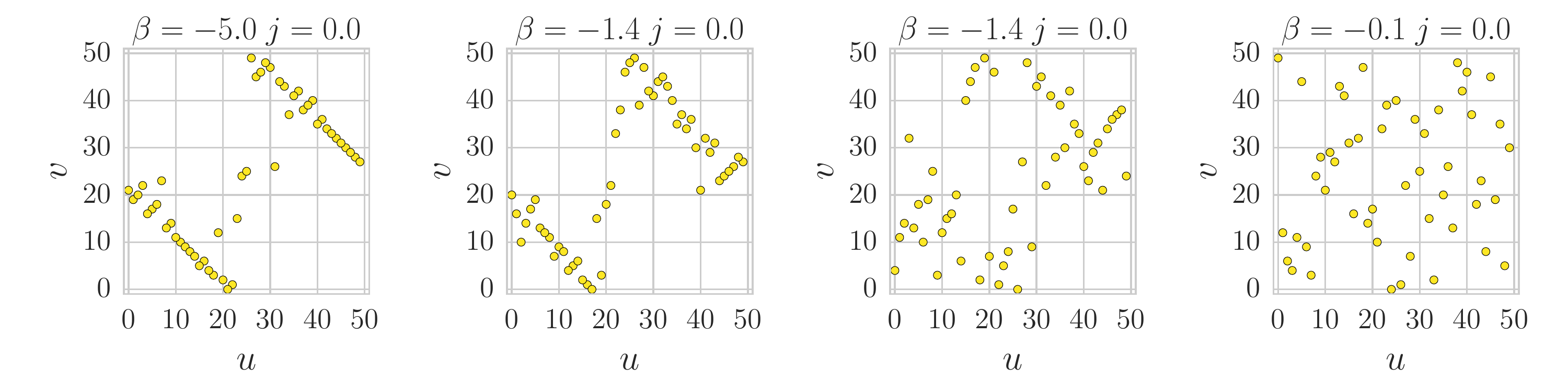}
  \caption{Some examples of the $2$d orders along the phase transition for negative $\beta$.}
  \label{fig:negbvis}
\end{figure}

We first ran some simulations at negative $\beta$ with $j=0$ to have a benchmark against which to compare the behaviour of the model coupled to the Ising spins.
We show these in Figure~\ref{fig:negb}, and see from the jump in the variance that the system undergoes a phase transition between $\beta=-1.3 \dots -1.4$.
For $\beta$ above this critical value the $2$d orders are of a random type, although they are slightly more elongated than the truly random $2$d orders.
We show some examples of the $2$d orders in Figure~\ref{fig:negbvis}, and can see that the maximal action state seems to consist of two layers with a few elements in the middle between them, since this configuration maximises the number of $1$-paths in the causal set.
From these plots it is clear that for $\beta=-1$ the system is still far enough from the phase transition that the orders are of the random type.

\begin{figure}
  \includegraphics[width=\textwidth]{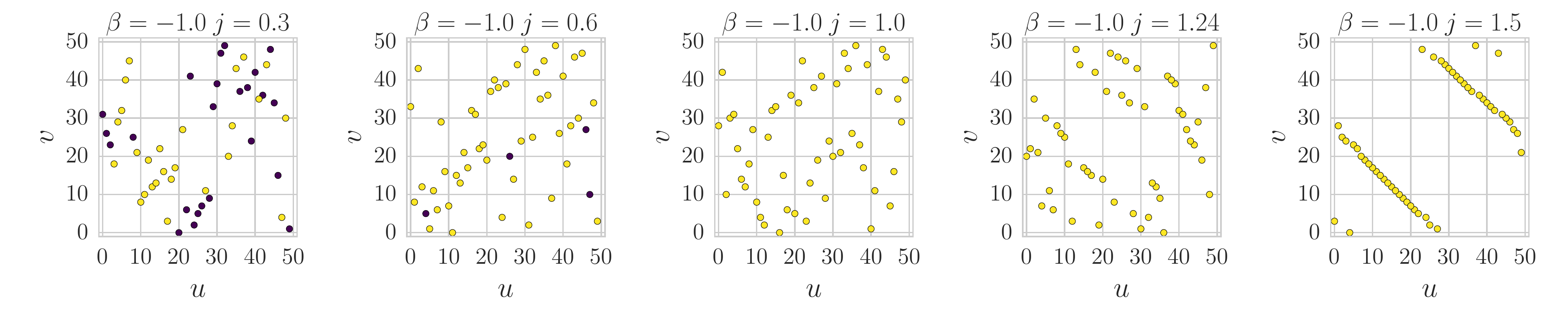}
  \caption{Five causal sets in the different regions along the $\beta=-1$ line and at the two phase transition points.}
  \label{fig:b-1Causets}
\end{figure}

Varying $j$ at fixed $\beta=-1$ takes us from a phase that seems random visually (leftmost image in \ref{fig:b-1Causets}), to a random causal set phase with ordered spins, to a crystalline causal set phase with ordered spins, we illustrate these states in Figure \ref{fig:b-1Causets}.
The orders at large $j$ are clearly of the layered type that minimises $\Sa_{BD}$ and form the crystal causal sets.
It is interesting that by maximising the number of links they minimise $\Sa_{BD}$ and $\Sa_{Ising}$ at the same time\footnote{For negative $j$ a maximal number of links in a totally ordered state minimises the action. For positive $j$ the crystal orders allow for complete anticorrelation between adjacent layers, which again minimises $\Sa_{Ising}$}.
This means that here both actions push the geometries in the same direction, which leads to the earlier phase transition for these geometries.

\begin{figure}
  \includegraphics[width=0.49\textwidth]{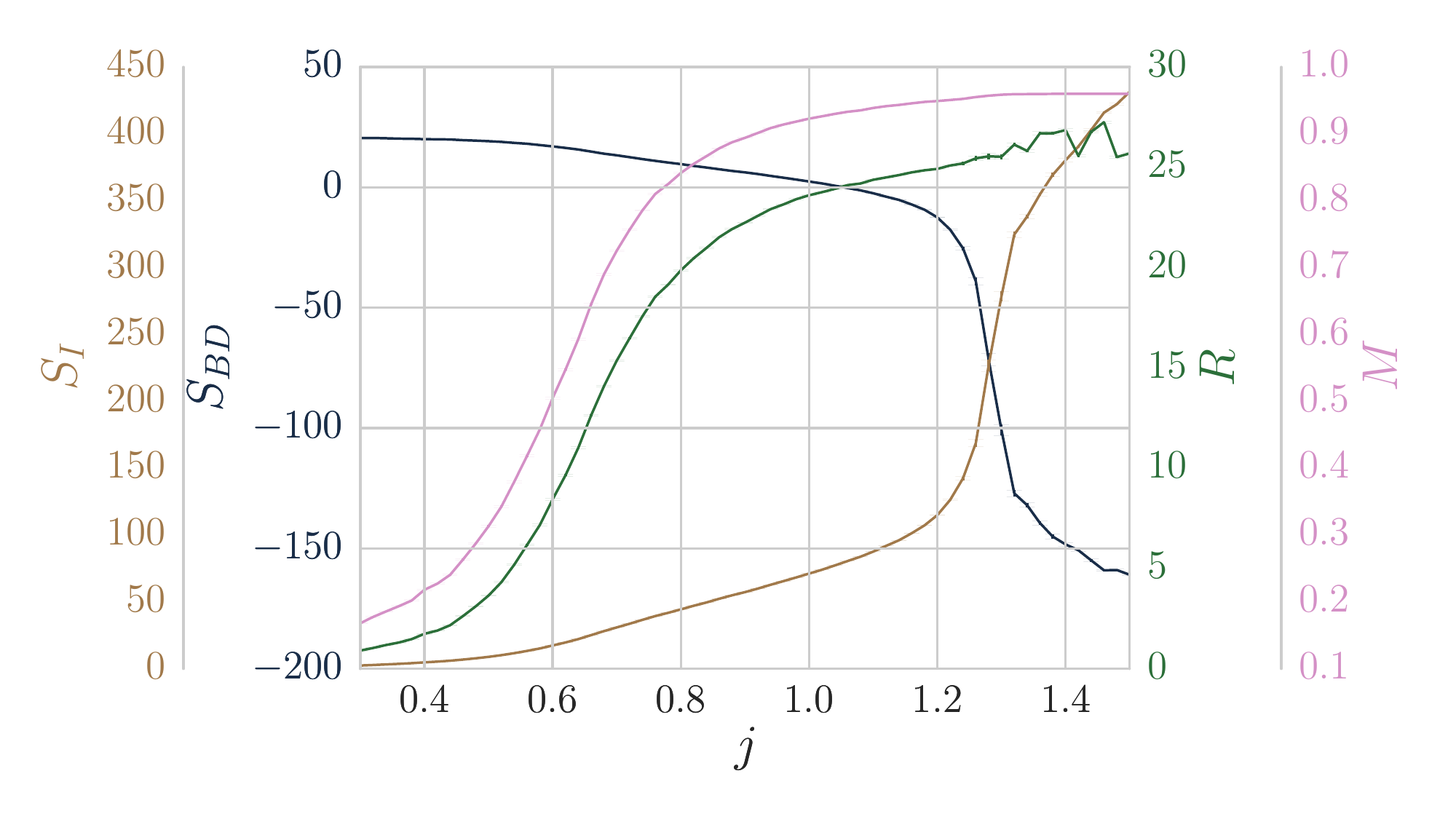}
  \includegraphics[width=0.49\textwidth]{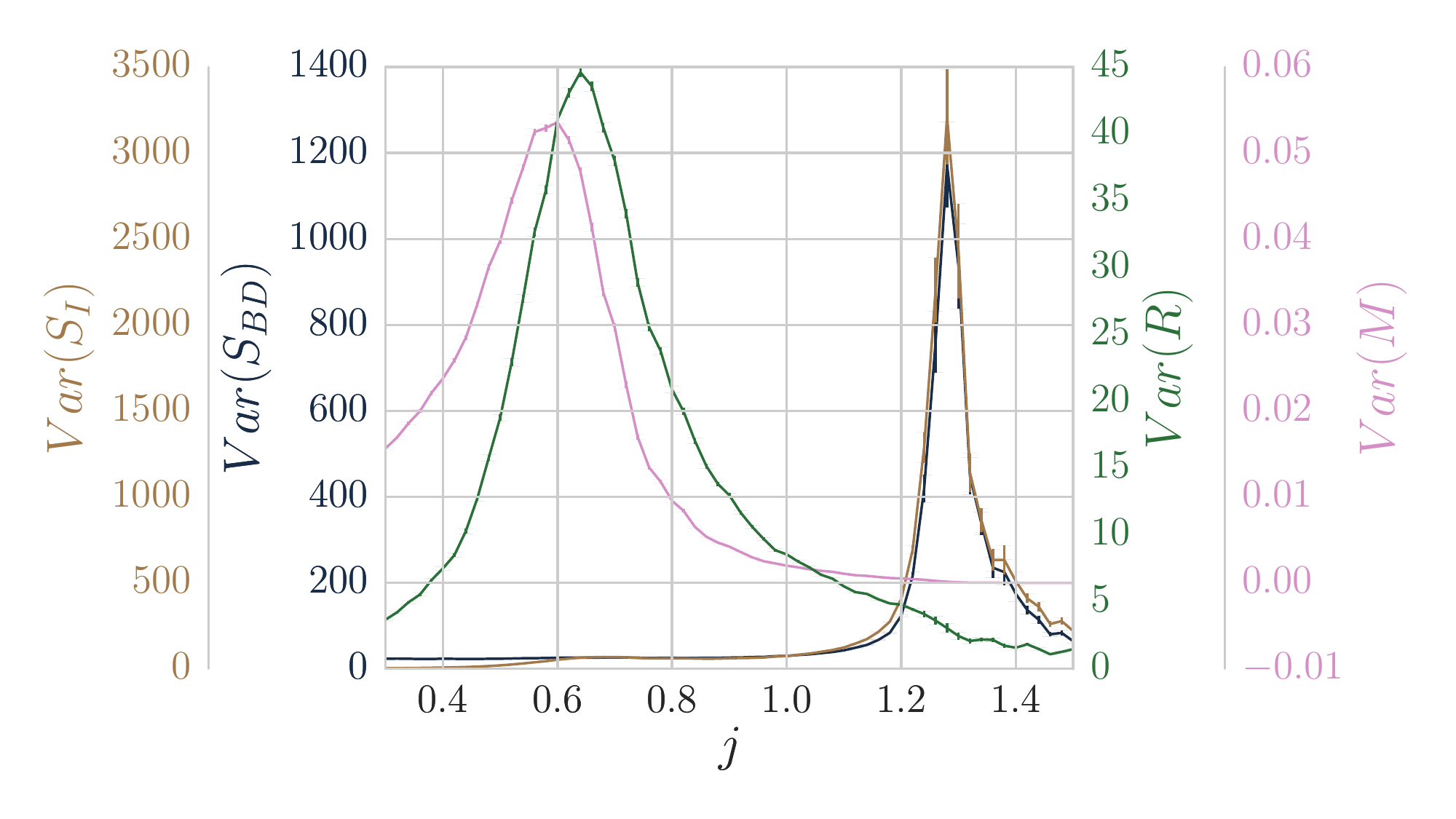}

  \includegraphics[width=0.49\textwidth]{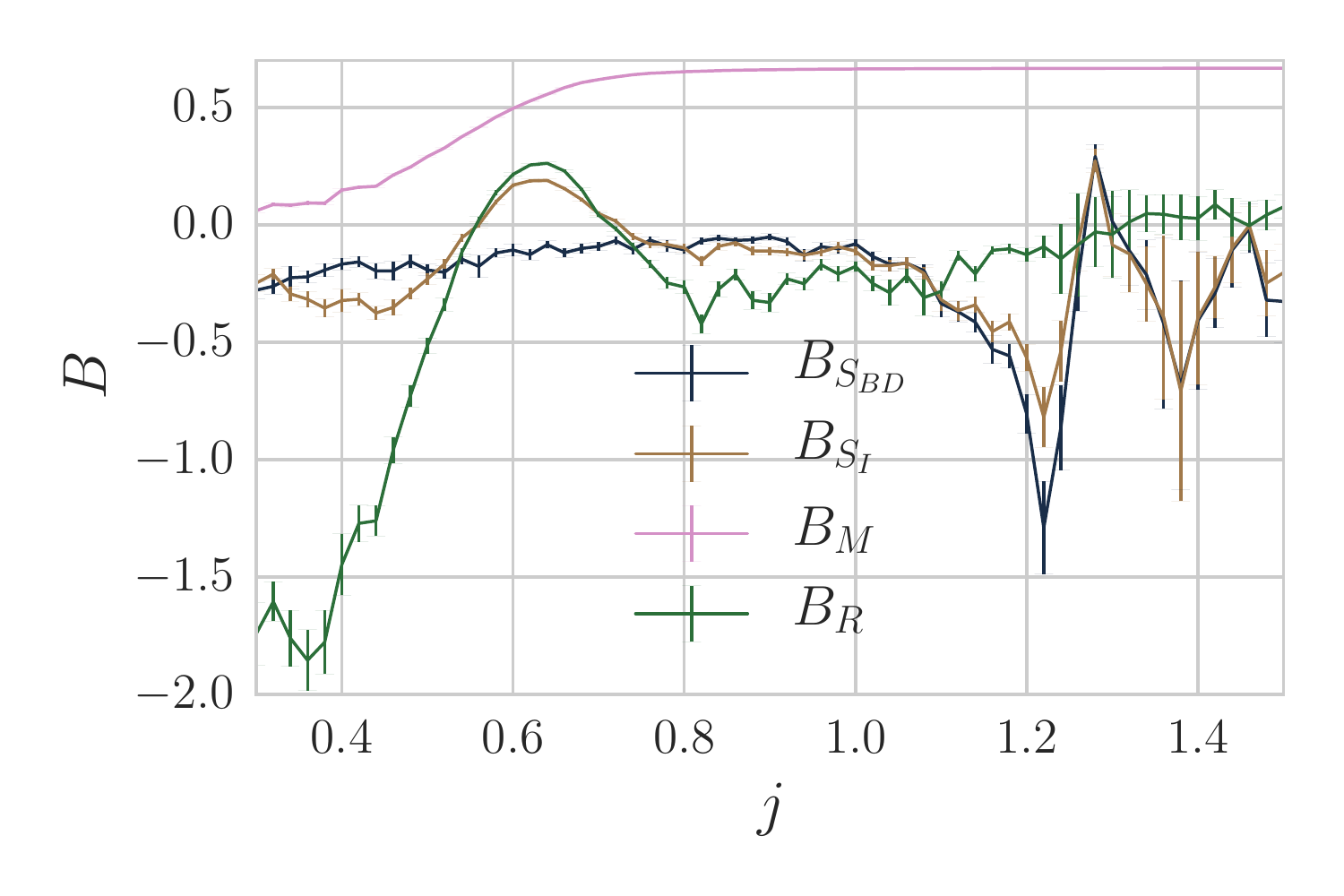}
  \caption{$\beta=-1$ The average observables and their variances show the two phase transitions that happen along this line.}
  \label{fig:b-1PTline}
\end{figure}

At low $j$ we see  $2$d orders that seem random, however the average of the BD-action in this phase is $\sim 20$ as opposed to the value for random $2$d orders which should be $4$, so they are not quite random anymore.
It is possible that this behaviour is comparable to the negative constant action arising before the phase transition found in~\cite{glaser_finite_2017}, however this will require further study.
The fact that the system undergoes two different changes is very clear in the observables and variances shown in Figure \ref{fig:b-1PTline}.
It is particularly interesting, since in this case the Benincasa-Dowker action and the Ising model action are counteracting each other.
The Benincasa Dowker action would be maximised by states such as the leftmost one in Figure~\ref{fig:negbvis}, which for negative $\beta$ is the energetically preferred state.
The Ising action however tries to maximise itself by maximising the number of links, which arises through crystalline causal sets, such as the rightmost one in Figure~\ref{fig:b-1Causets}.
However since the maximal value the Benincasa Dowker action can take on in its energetically optimal state is only about a quarter of the energetically optimal value of the Ising action the Ising model wins and the causal sets take on a crystalline structure.

The transition happens in two steps, first at $j\approx 0.6$ the magnetisation and the correlation measure $R$ increase, and their variances peak (green and pink lines in \ref{fig:b-1PTline}).
Then afterwards at $j=1.24$ both the Benincasa-Dowker action and the Ising action change rapidly, accompanied by a clear peak in their variance.

These two phase transitions are also clear from the forth order cumulant.
The cumulant of the magnetisation shows behaviour indicating a second order phase transition at $j\approx 0.6$, which is the Ising transition of this line.
The cumulant of the Ising model action also shows the up down behaviour at this point, as expected.
However at $j=1.24$ both the Ising model action and the BD-action show particular behaviour.
They dip to negative values, then shoot up to positive values to dip below $0$ again at which point our data ends.
This could indicate a higher order transition, and will warrant further study in the future.
This is interesting since it is a transition of the geometries, not the Ising model, yet it still shows behaviour that is not easily explained as first order, it opens the possibility at having found a higher order transition of the geometry.
Another hint that this transition might be of higher order is that the second causal set from the right in Figure~\ref{fig:b-1Causets} shows some characteristics of both phases and is not clearly part of either.

\begin{figure}
  \includegraphics[width=0.49\textwidth]{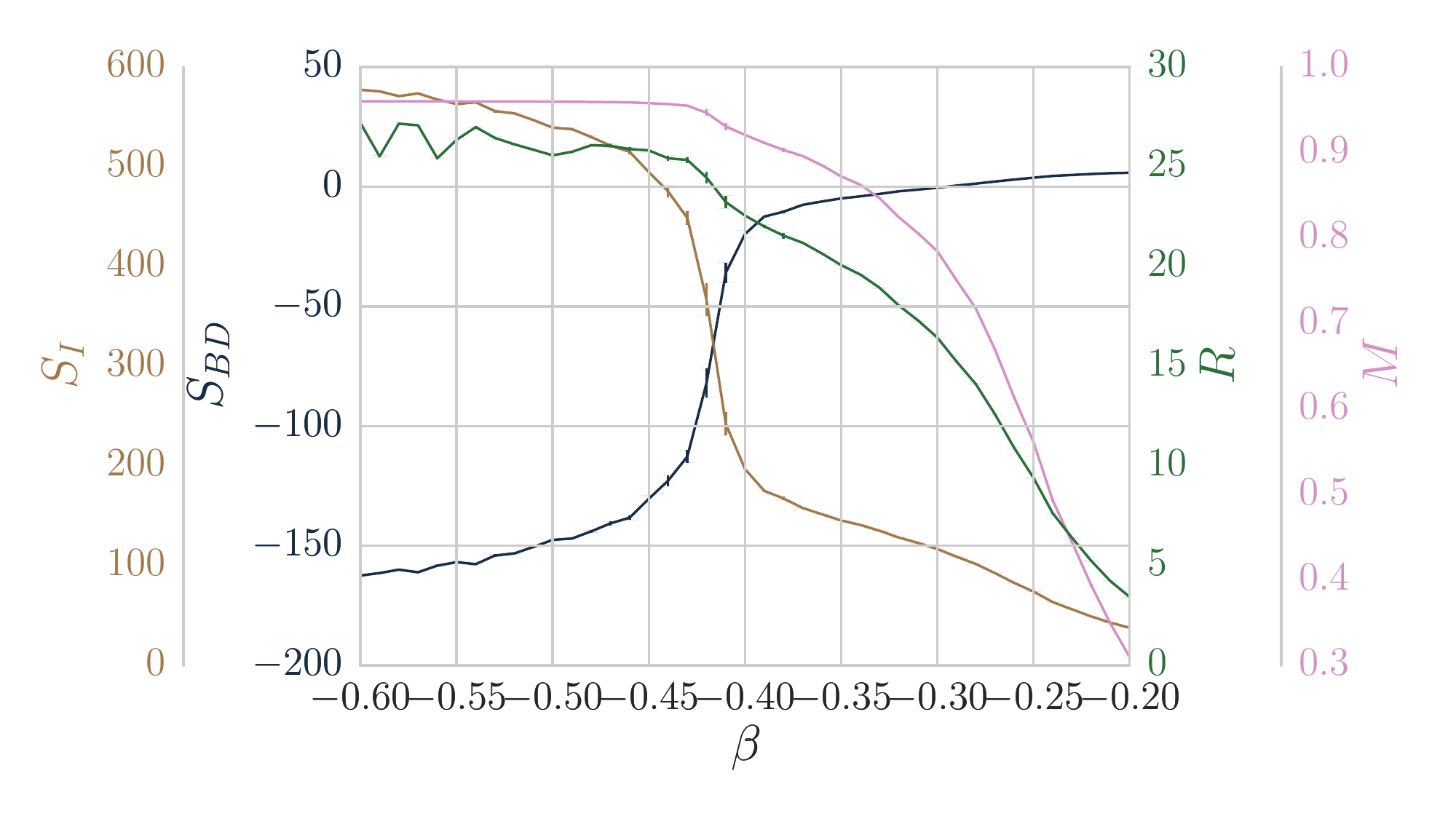}
  \includegraphics[width=0.49\textwidth]{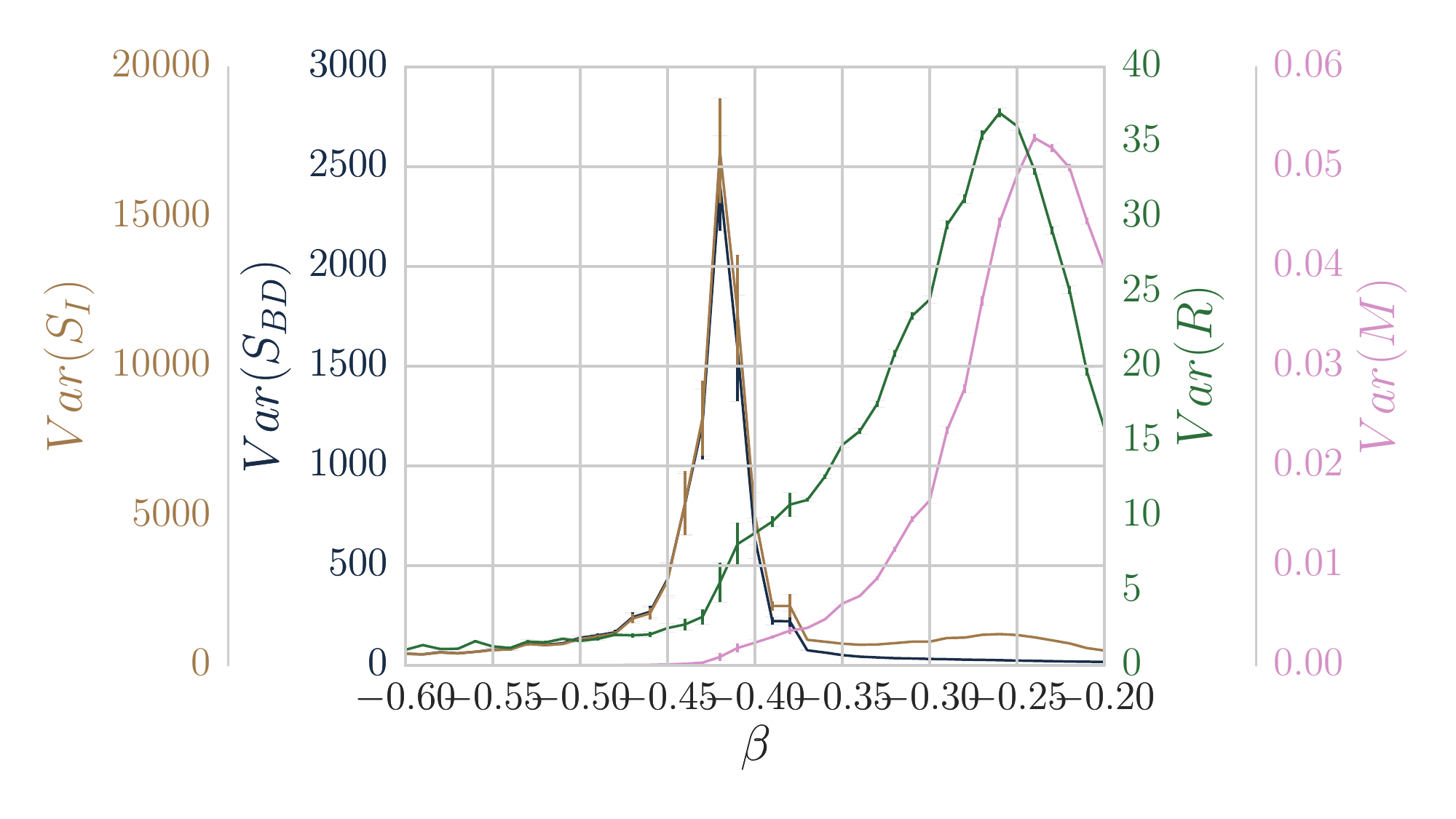}

  \includegraphics[width=0.49\textwidth]{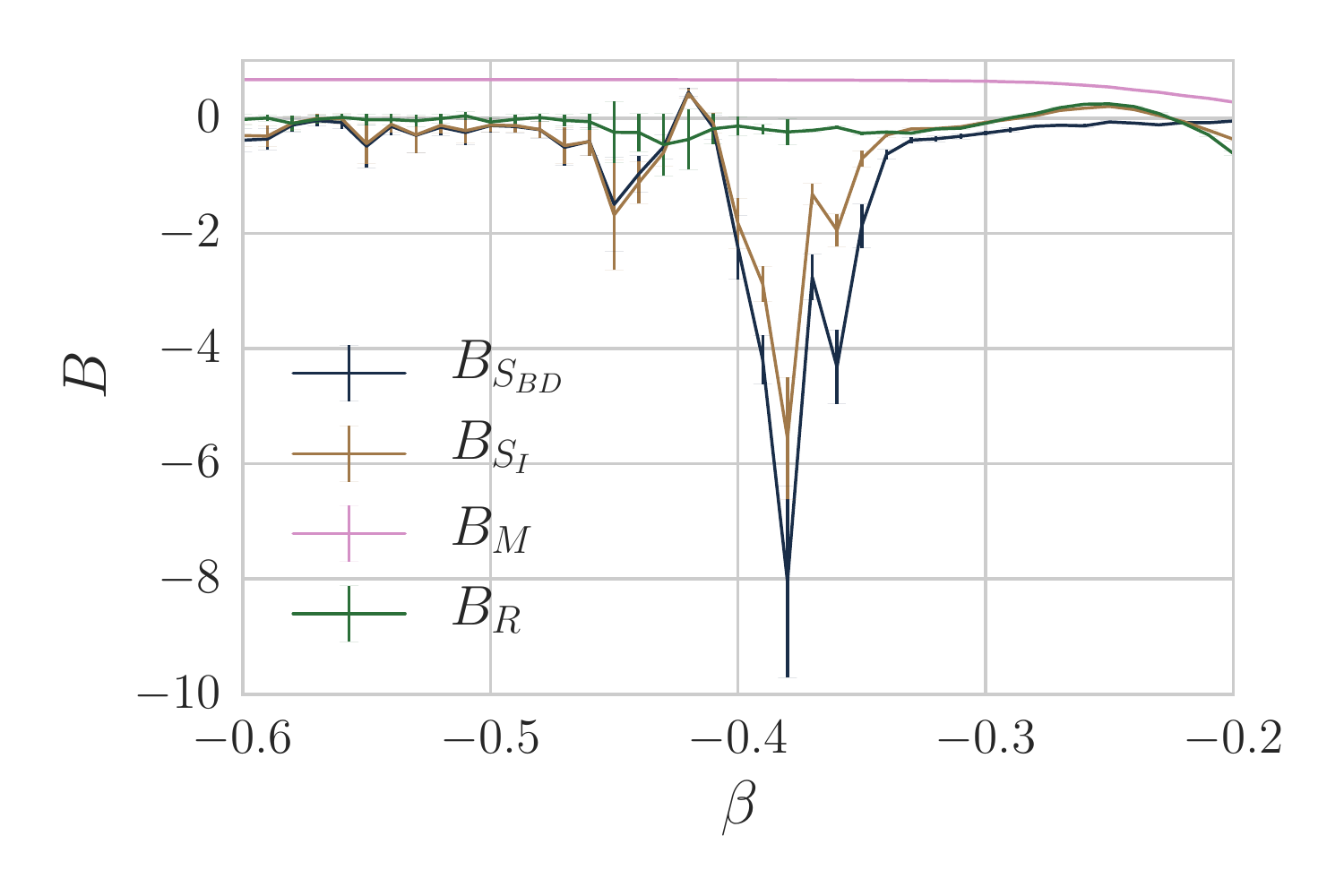}
  \caption{$j=2$ The phase transitions along this line are qualitatively very similar to those in Figure \ref{fig:b-1PTline}.}
  \label{fig:j1PTline}
\end{figure}

The causal sets along the $j=2$ line are very similar to those along the $\beta=-1$ line shown in Figure \ref{fig:b-1Causets} so we do not reproduce them here.
Overall the system undergoes the same state changes and takes on very similar values for all observables along the two orthogonal lines, this indicates that the transition between the regions remains qualitatively the same independent of the direction of approach.
We can see this by comparing Figure \ref{fig:b-1PTline} above with Figure \ref{fig:j1PTline}.
One difference is that the BD action in the random looking phase along the $j=2$ line is much lower $\approx 10$ thus hinting that the causal sets along this line are closer to truly random $2$d orders than those at $\beta=-1$.
At fixed $j=2$ the transitions happen for $\beta=-0.48$ in the actions and $\beta\approx -0.25$ in the magnetisation and correlation.
This also shows in the fourth order cumulant, although hindered somewhat by the fact that our line cut out too close to the phase transition at $\beta \approx -0.25$ for $j=2$.
We can see a rise in $B_{\Sa_I}$ and a drop in $B_M$ there, however the line ends before the behaviour stabilises.


\begin{figure}
\subfloat[][$\beta=-1$]{  \includegraphics[width=0.5\textwidth]{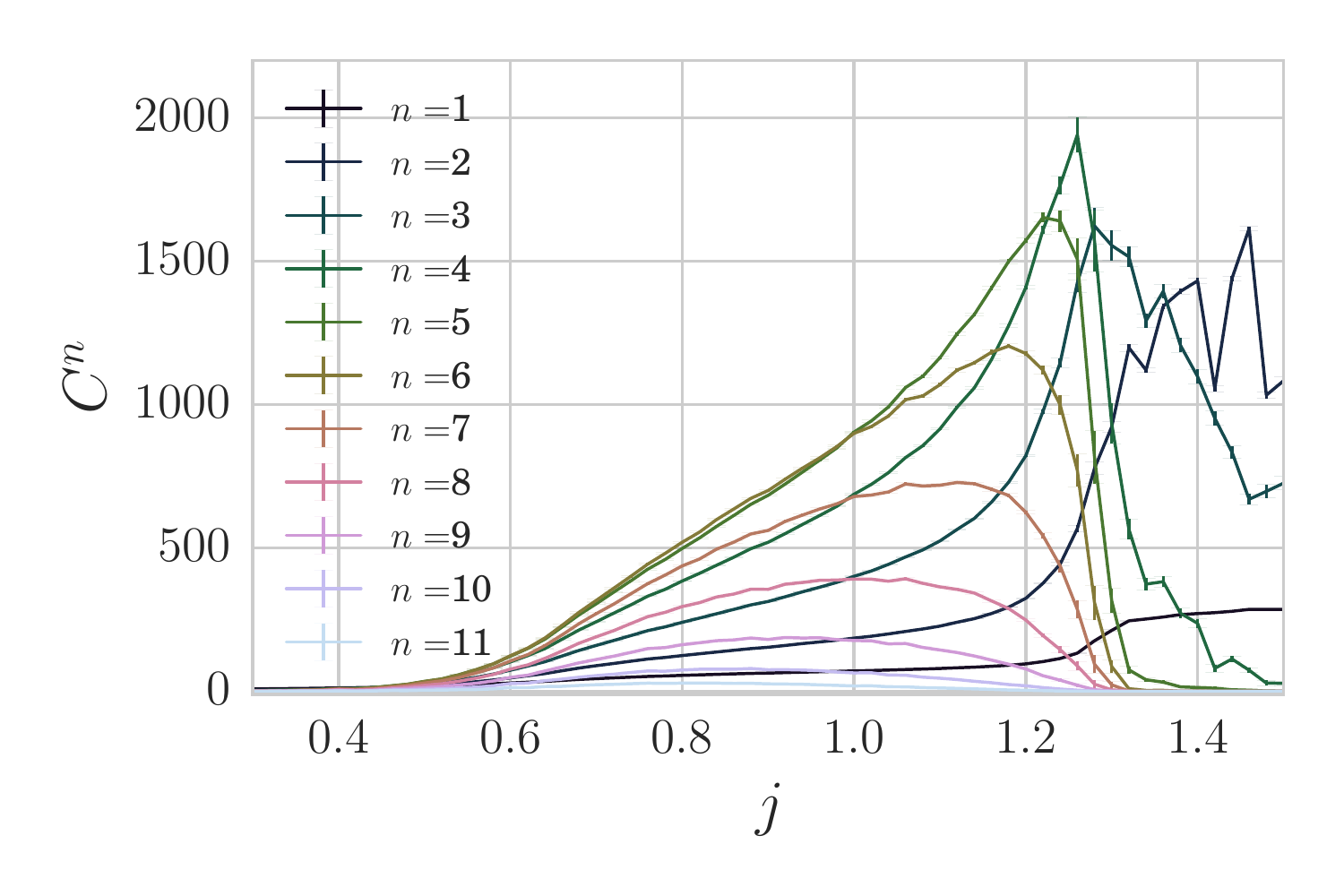}}
\subfloat[][$j=1$]{  \includegraphics[width=0.5\textwidth]{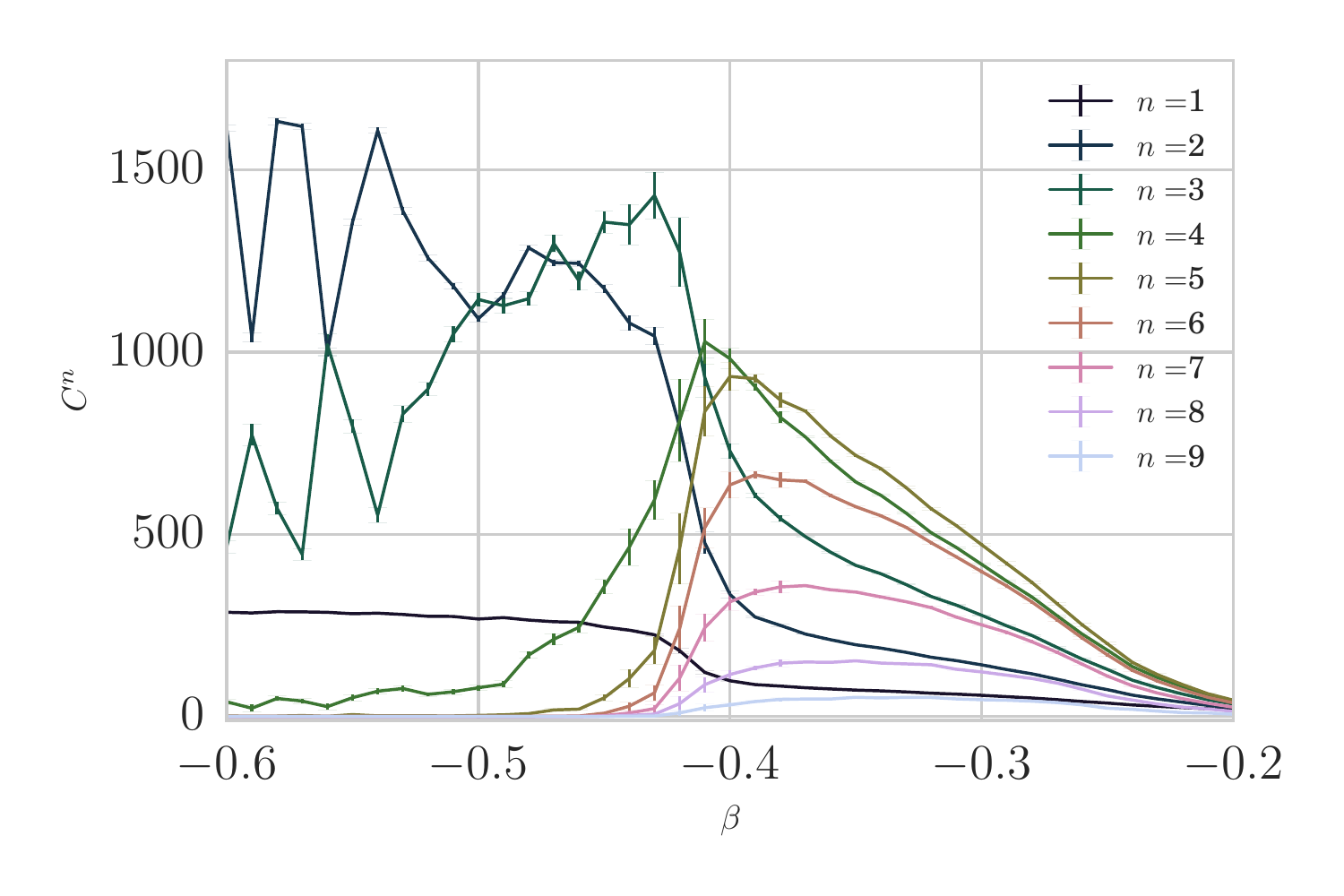}}
  \caption{The path correlators along the two lines. The behaviour of the correlators is similar, when we take into account that we approach the phase transition from different sides.}
  \label{fig:b-1j1Nchain}
\end{figure}

The difference between the phases, and similarity of the transitions along the two lines is also clear in the path correlators in Figure \ref{fig:b-1j1Nchain}.
In the disordered spin region up to $j\approx 0.3$ all correlators are $\simeq 0$, after this phase transition the correlators are non zero up to $n=10$.
There is some data for longer paths, however the errorbars on this are consistent with $0$, so $n=10$ is the longest path length for which reliable data exists.
This is not surprising, since the expected length for a longest path in a $2$d random order is $\tau\sim \sqrt{N}\approx 7$.
Then the cut-off for sensible data at $n=10$ reinforces the impression from Figure~\ref{fig:negbvis} that the causal sets in the random phase at negative $\beta$ are somewhat elongated.
After the transition to the crystalline orders at $j=1.26$ the number of paths of lengths $1$ and $2$ increases rapidly, leading to a large value for the $1,2$ paths correlators.
The values fluctuate strongly since data in this region is harder to thermalise, the Monte Carlo chain in this region tends to take very long to change between different crystal orders.
Since the number of $n$-paths depends on these orders, this leads to the strong fluctuations in the curves on this side of the phase transition.
The path correlators behave extremely similar, we just need to keep in mind that we are approaching the phase transition from  the left, low $j$ in the left hand figure and from the right, less negative $\beta$ in the right hand side.

One difference between the two states is that for $\beta=-1$ the random orders are more elongated, which is why we can there measure up to the $10$ path correlator, while we only obtain the $8$ path correlator for $j=2$.

This concludes the qualitative exploration of the different phases, we have found a total of five of them.
Along the five lines we simulated in detail we find four different types of phase transitions.
\begin{enumerate}
  \item Ising model transitions from disordered to ordered states
  \item Ising model transitions from disordered to anticorrelated states
  \item Temperature driven geometric transitions from random to crystalline states
  \item Spin driven geometric transitions from random to crystalline states
\end{enumerate}
The first and second class of transitions are almost certainly of second order, good examples for them are the two transitions at $\beta=1$.
The third type of transition is very similar to that found in~\cite{glaser_finite_2017}, a good example are the transitions at $j=-1.0$ and $j=0.9$.
The Ising spins also correlate/ anti-correlate across this transition, but the characteristics of the geometric transition do not seem to be changed by this.
The last type of transition is that described in subsection~\ref{subsec:doubletrans}.
It is the most interesting one since it points towards a strong coupling of matter to geometry.

\section{Using Mean field theory to study the phase transitions}\label{sec:meanfield}
We can use mean field theory to study those phase transitions for which the magnetisation is the order parameter.
We will first describe the results for the Ising model coupled to the fixed background causal sets, and then extend the results to the full model.

The mean field description of the lattice Ising model is more accurate in higher dimensions, because the valency of the lattice vertices increases with dimension, hence each spin has more nearest neighbours.
In manifold like causal sets the preservation of Lorentzian structure leads to non-locality in the form of high valency~\cite{bombelli_discreteness_2009}.
It is then clear that each spin in the Ising model on a random $2$d order has many nearest neighbours, which would indicate that the mean field results should be a good approximation.
The crystalline $2$d orders show even higher valency due to the layered structure, so  we also expect the mean field approximation to work for them.
The number of nearest neighbours is different for each element, and hence the effective magnetic field acting on each element is different.
We can still use the mean field approximation, but have to additionally approximate the valency of elements by the average valency $\av{z}$, which depends on the causal set the Ising model is simulated on.
In the mean field approximation we write the action in terms of the perturbation around the average magnetisation $m_i= \av{s_i}$,
\begin{align}
    \Sa_{I}
      & = j \sum_{i,k} (m_i +\delta s_i)(m_k +\delta s_k) L_{i k}\\
      \intertext{neglecting terms higher order in the perturbations $\delta s_i$ and replacing $\delta s_i= s_i-m_i$ leads to}
      &= j \left(- \sum_{i,k} m_i m_k L_{i k} + 2 \sum_{i,k} m_i s_k L_{i k} \right)
\end{align}
Next we make the simplifying assumption that all $m_i$ are equal $m$ and that each spin has valency $\av{z}$.
Then the sum $\sum_{i,k} (m_i m_k) L_{i k}$ becomes $N \av{z} m^2/2$ and the sum $\sum_{i,k} m_i s_k L_{i k}$ becomes $m \av{z} \sum_k s_k$.
While these assumptions are well justified on a regular grid, they are non-trivial on a causal set, so if this approximation holds it is an example of how well mean field theory works.
With these assumptions we find
\begin{align}
  \Sa_{I} &= j \left(- \frac{N \av{z} m^2}{2} + 2 m \av{z} \sum_k s_k \right) \;.
\end{align}
This makes it possible to calculate the partition function analytically
\begin{align}
  \mathcal{Z}_N = \sum_{\{s_i\}} e^{- \beta \Sa_{I}} = e^{ \beta j \frac{N \av{z} m^2}{2}} \left[ 2 \tanh{ \left(- 2 \beta m \av{z} j \right)} \right]^N
\end{align}
from which we can calculate the magnetisation and the phase transition point $- \beta j = 1/\av{z}$.
We can then calculate the predicted value $j$ for $\jc{-}$ using the average valency of the graph.

\begin{figure}
\subfloat[][Random orders]{\includegraphics[width=0.5\textwidth]{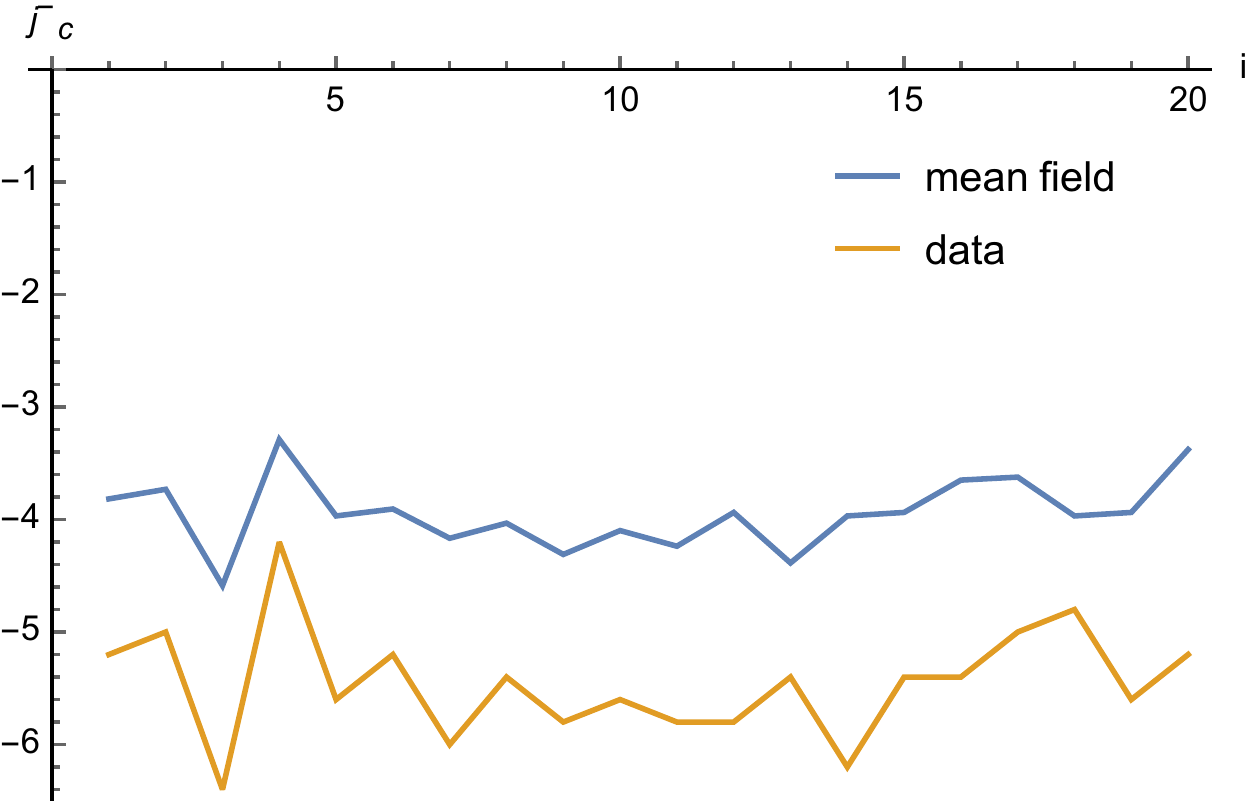}}
\subfloat[][Crystal causal sets]{\includegraphics[width=0.5\textwidth]{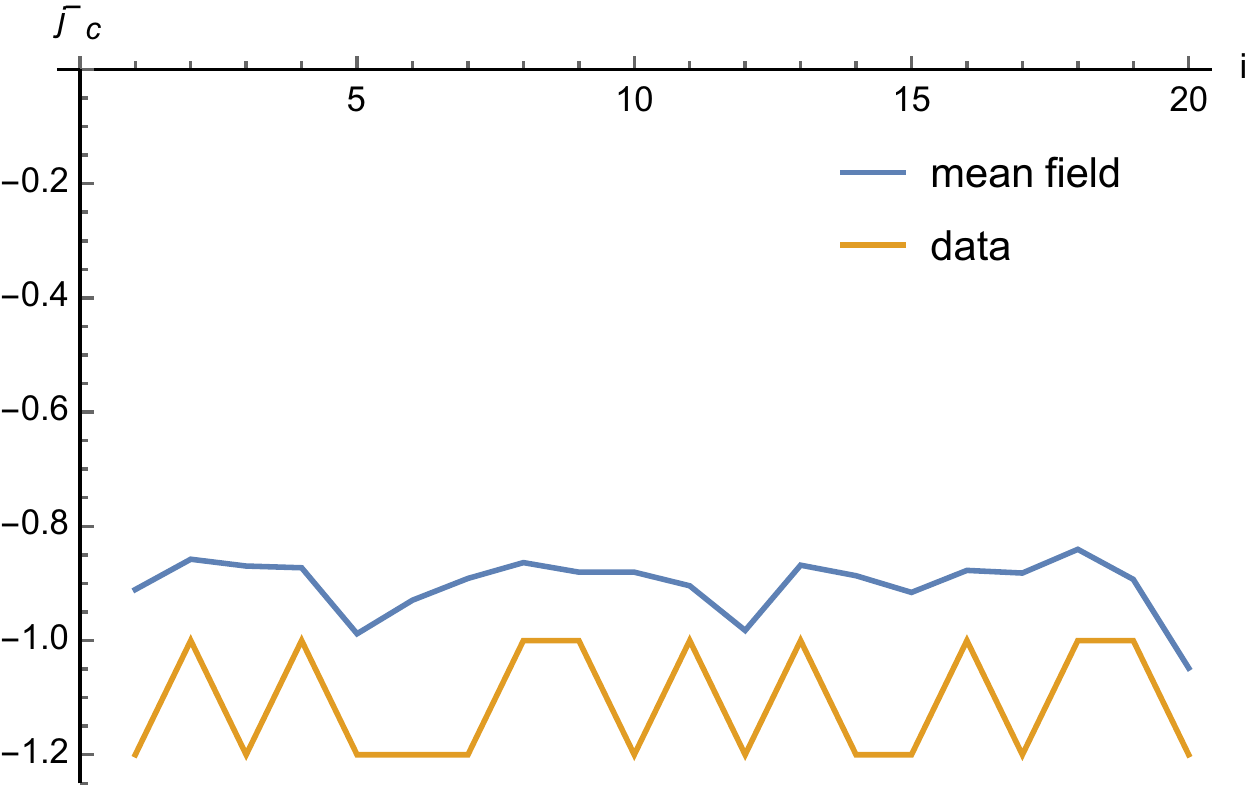}}
  \caption{Comparing value of $\jc{-}$ predicted by mean field theory to the measured values for the $20$ random and $20$ crystal orders of size $50$.}
  \label{fig:meanfieldrandom}
\end{figure}

In Figure \ref{fig:meanfieldrandom} we can see that while the predicted value and the measured value do not agree for the random $2$d orders they have a relatively consistent difference from each other of about $1.2$.
This shows that while the approximation is too rough to capture the exact location, it does qualitatively describe the system.
For the crystal orders the $\jc{-}$ curves vary much less and the transition happens at much lower $\jc{-}$, this is because the valency of nodes in the crystal orders is much higher than in the random $2$d orders.
The mean field prediction and the actual data differ by approximately $0.2$.
In both cases the mean field predicts a higher value of $\jc{-}$, so an earlier phase transition.
This indicates that the irregularities present in the causal sets delay the phase transition.
Unfortunately this method can not tell us anything about the phase transitions in the anticorrelated phase, since there the magnetisation is not the relevant observable.

\begin{figure}
\subfloat[][$\beta=1$]{\includegraphics[width=0.4\textwidth]{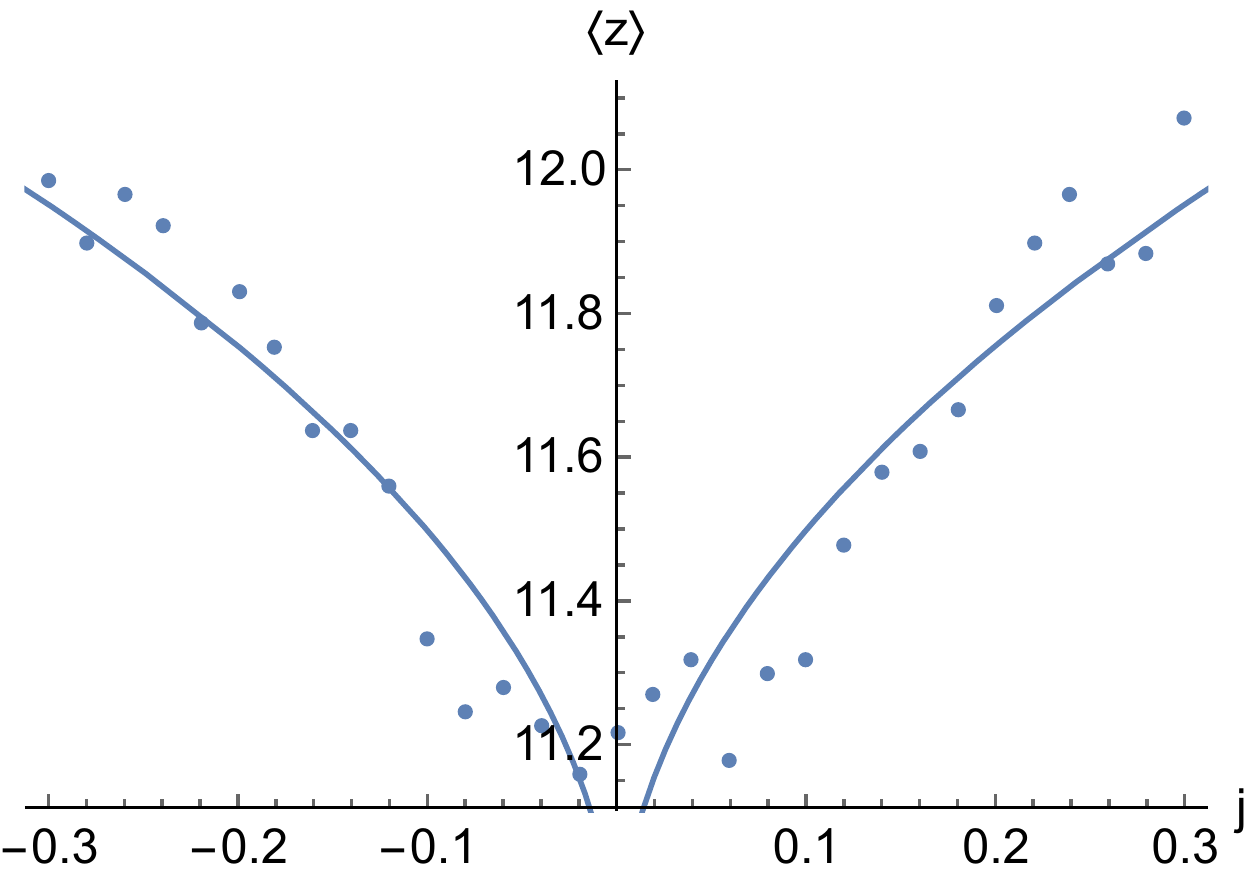}}
\subfloat[][$\beta=-1$]{\includegraphics[width=0.4\textwidth]{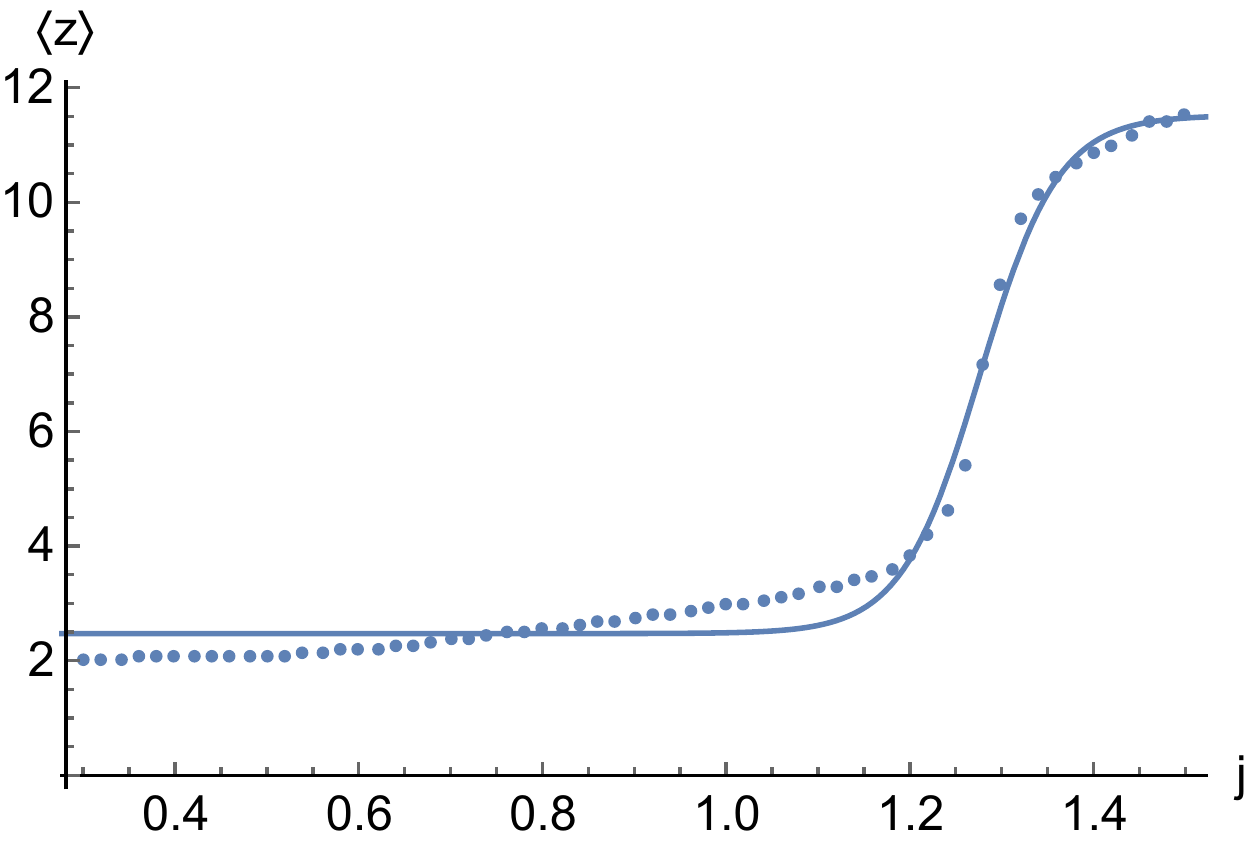}}

\subfloat[][$j=-1$]{\includegraphics[width=0.4\textwidth]{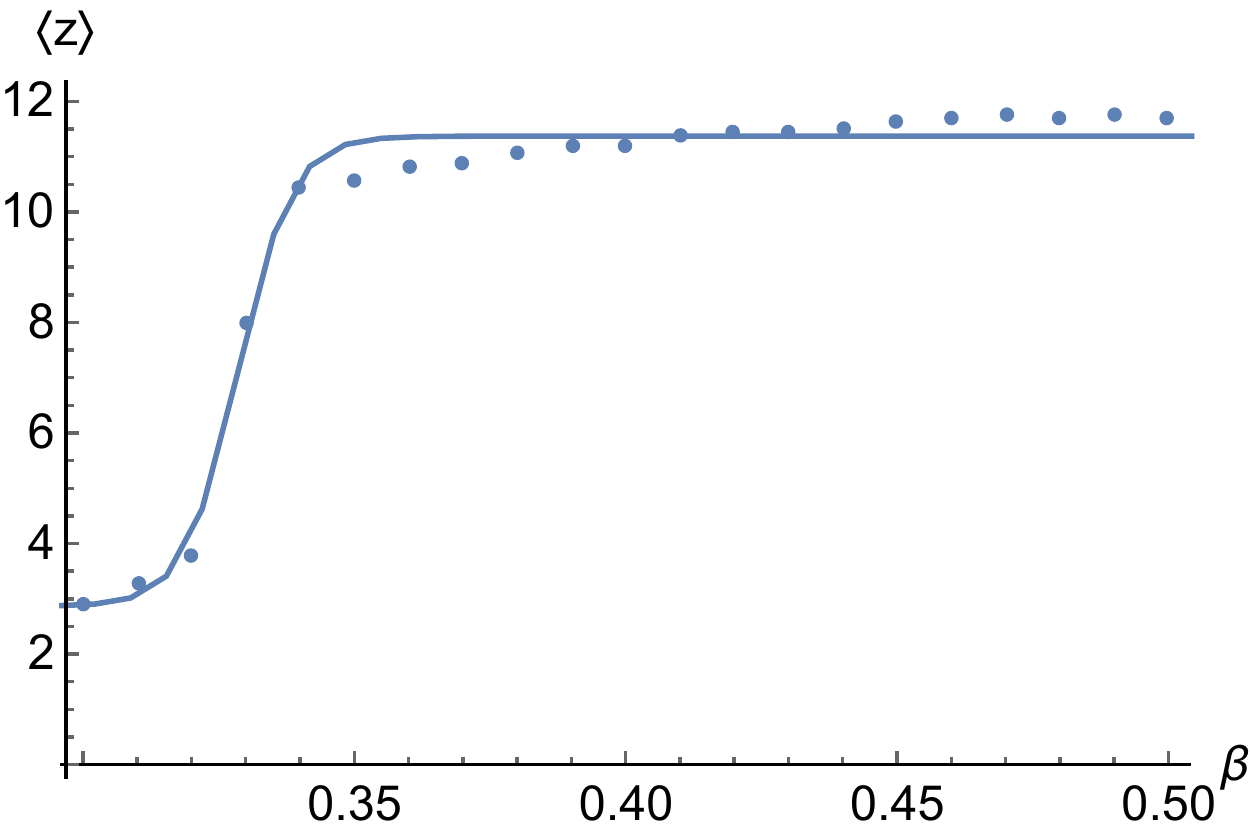}}
\subfloat[][$j=2$]{\includegraphics[width=0.4\textwidth]{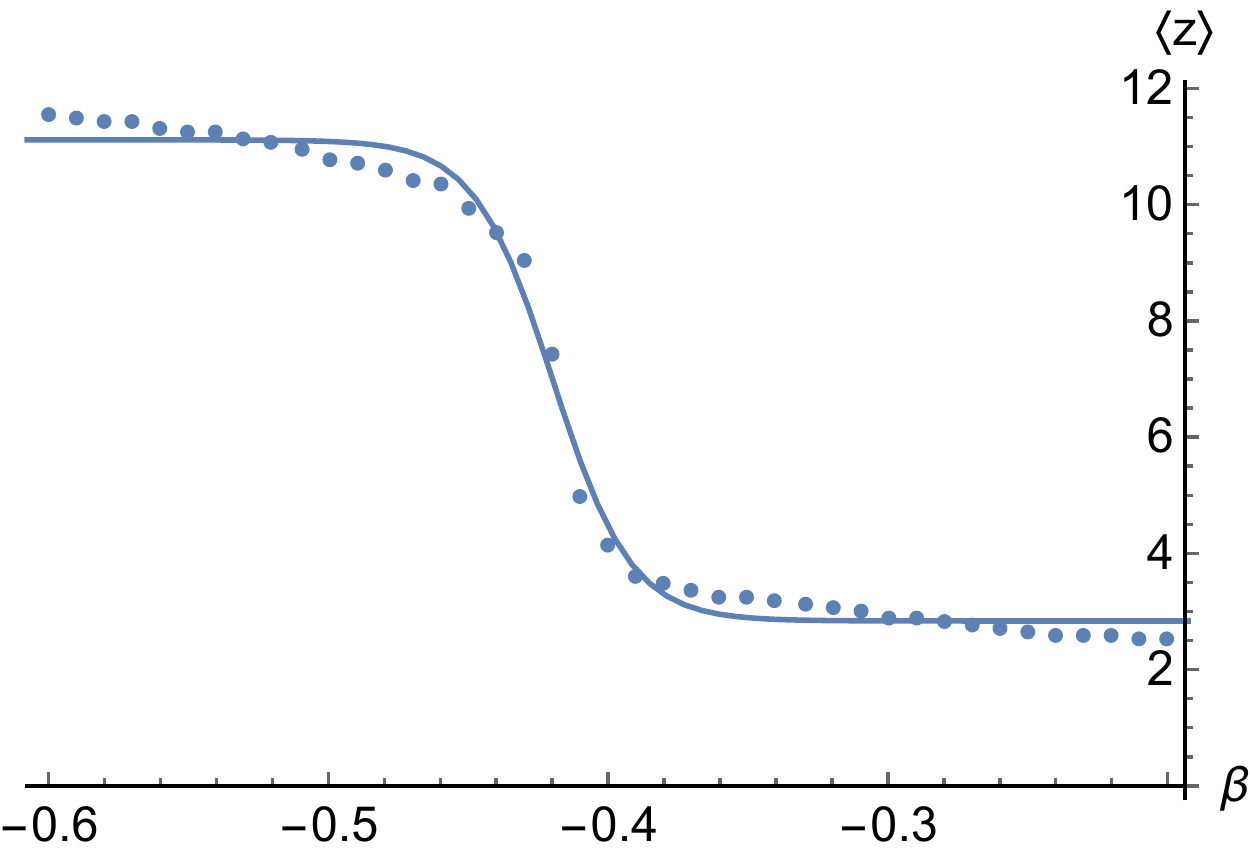}}
  \caption{The best fit functions for the average valency $\av{z}$ as a function of either $\beta$ or $j$.}
  \label{fig:valencyFits}
 \end{figure}

We can also try and use mean field theory for the phase transition when the causal set is allowed to vary.
Of course, this means that we need to take the average over the coordination number $z$ for each $\beta,j$ value.
We calculate the average over the coordination number along the lines of fixed $\beta,j$ to calculate $z(\beta,j=const.),z(\beta=const.,j)$ along the lines.
The clearest signals for an ordered to disordered transition happen at $\beta=1, j_c=0.1$ and $j=-1,\beta_c=0.33$.
The other transitions that show a jump in the magnetisation are the transitions at negative $\beta$ and positive $j$,
$j=2, \beta_c=-0.25$ and $\beta=-1, j_c=0.6$ respectively.
To calculate the critical values of $\beta, j$ we then interpolate  $z(\beta,j=const.),z(\beta=const.,j)$ through a simple function for the $\beta=-1$, and $j=-1,2$ lines
\begin{align}
  z(x)= a \tanh{\left(b(x-c) \right)} +d
\end{align}
where $x$ can be $\beta$ or $j$ and
\begin{align}
  z(x)= a \sqrt{|x|}+ b
\end{align}
for $\beta=1$, which is the only line in which we are not crossing a geometric phase transition.
The best fit functions with the data are shown in Figure~\ref{fig:valencyFits}.

Using these functions for $z$ we find that as above the phase transition should happen at $\beta j = \frac{1}{z}$, only now $z$ depends on $\beta, j$.
We can use this to predict the phase transition along the lines we have simulated.
One might rightly argue that the change in valency is a signal for the geometric phase transition, however the mean field theory is only sensitive to this geometric transition through the function $\av{z}$.

Using the functions $\av{z}$ shown above we find that for $\beta=1$ the mean field value $j_{mf}=-0.09$ which compares well with the value of $\jc{-}=-0.1$ we found in our simulations.
This is not surprising, since the geometry along this transition remains almost constant, so that the mean field theory should give very good results.

For $j=-1$ the geometry changes from the random to the crystalline orders, at $\beta_c=0.33$, which is also the location of the Ising model transition along this line.
The mean field value we find is $\beta_{mf}=0.31$ and thus in very good agreement with the simulation.
Artificially varying the geometric phase transition location, by manipulating $\av{z}(\beta)$ shows us that along this line mean field theory does not necessarily require these transitions to coincide, so their agreement in our system is a coincidence.

The other two transitions are those in which the geometric transition and the spin transition do not coincide.
For $\beta=-1$ the critical value for the magnetic transition is $\jc{-}=0.6$, while the geometric transition happens at $\jc{}=1.24$.
Using mean field theory we find $j_{mf}=0.4$ which is much closer to the magnetic transition, even though it does underestimate the critical value.
For $j=2$ the critical value for the magnetic transition is $\beta_c^{-}\approx-0.25$, with the geometric transition following at $\beta_c=-0.48$.
We find the mean field theory value to be $\beta_{mf}=-0.18$, which is again an underestimation.

It would hence seem that mean field theory can help us estimate the location of phase transitions in which the magnetisation is the order parameter.
The two transitions with the better accuracy are those in which the Ising model and the geometry are moving in the same direction, while the results for the cases in which the geometry and the Ising model work in opposite directions are somewhat less precise.
Overall this shows again that mean field theory, with suitable modifications qualitatively describes the magnetic phase transition of the Ising model coupled to random $2$d orders.

\section{Conclusion}
The work presented here is a first step in  exploring causal sets coupled to matter.
We have seen that the Ising model coupled to the $2$d orders gives rise to a rich and varied phase structure.
In particular we have seen that in the region at negative $\beta$ and positive $j$ matter can push the geometry into different states, and that geometric transitions induced in this manner are possibly of higher order.

These are of course only the first steps in combining causal set theory and matter in computer simulations.
The next step will be to explore how the path integral over the entire class of causal sets will behave when coupled to the Ising model.
In this work we focused on exploring the phase diagram of the model to gain some qualitative understanding, for future work it would be interesting to examine the phase transitions discovered here using finite size scaling as in~\cite{glaser_finite_2017}.
In such a study one should also focus on better understanding the pure causal set system in the negative $\beta$ region and the region where both $j,\beta$ are negative in which this study did not find a phase transition.
Since the observables used in this study are rather limited, and in particular ill suited to find transitions in anticorrelated behaviour it is very likely that we have overlooked a phase there.

In addition to studying the system through computer simulations we also used mean field theory to calculate the locations of magnetic phase transitions.
We found that, despite the need for an additional approximation since the coordination number for causal set elements is not constant, the results are surprisingly good.
This makes us optimistic that we can use mean field theory in future studies to estimate possible regions of interest before starting simulations.

The Ising model is a very simple matter model, which makes it a favourite for first explorations.
In the future we should also study coupling scalar fields to the causal set in the path integral, either using the d'Alembertian~\cite{sorkin_does_2007,dowker_causal_2013,glaser_closed_2014} or the Greens function~\cite{johnston_particle_2008}.

\section*{Acknowledgements}
I would like to thank Will Cunningham for the use of his code and help in modifying it to include the Ising model.
I also owe thanks to Sumati Surya, Rafael Sorkin, David Rideout and Bianca Dittrich for discussions in the early stages of this project.
I have received funding from the People Programme (Marie Curie Actions) H2020 REA grant agreement n.706349 "Renormalisation Group methods for discrete Quantum Gravity".
This research was supported in part by Perimeter Institute for Theoretical Physics which hosted an informal workshop on Causal Set theory where I presented this idea in a very early form.
Research at the Perimeter Institute is supported by the Government of Canada through the Department of Innovation, Science and Economic Development and by the Province of Ontario through the Ministry of Research and Innovation.
\bibliography{bibliography}
\end{document}